\documentclass{article}
\usepackage[utf8]{inputenc}
\usepackage{amsmath}
\usepackage{empheq}
\usepackage{amssymb}
\usepackage{wasysym}
\usepackage{multirow}
\usepackage{graphicx}
\usepackage{hyperref}
\usepackage{cleveref}
\usepackage{color}
\usepackage[square, numbers, sort&compress]{natbib}
\usepackage{soul}
\usepackage{comment}
\usepackage{cancel}
\usepackage{ulem}
\newcommand{\fg}[1]{\mbox{\pmb{$#1$}}}
\newcommand{\vep}{\varepsilon}

\newcommand{\bec}{\begin{center}}
\newcommand{\eec}{\end{center}}

\newcommand{\Q}{\boldsymbol{Q}}
\newcommand{\B}{\boldsymbol{B}}
\newcommand{\E}{\boldsymbol{E}}
\newcommand{\m}{\boldsymbol{m}}
\newcommand{\n}{\boldsymbol{n}}
\newcommand{\e}{\boldsymbol{e}}
\newcommand{\C}{\boldsymbol{C}}
\newcommand{\s}{\boldsymbol{\sigma}}
\newcommand{\I}{\boldsymbol{I}}
\newcommand{\F}{\boldsymbol{F}}
\newcommand{\ome}{\omega}
\newcommand{\om}{\boldsymbol{\omega}}
\newcommand{\T}{\boldsymbol{T}}
\newcommand{\U}{\boldsymbol{U}}

\newcommand{\R}{\boldsymbol{R}}
\newcommand{\ph}{\boldsymbol{\phi}}

\newcommand{\p}{\partial}
\newcommand{\beps}{\boldsymbol{\epsilon}}
\newcommand{\eps}{\epsilon}

\newcommand{\cd}{\cdot}
\newcommand{\be}{\boldsymbol{\beta}}
\newcommand{\la}{\boldsymbol{\lambda}}
\newcommand{\bd}{\boldsymbol{d}}
\newcommand{\bl}{\boldsymbol{l}}
\newcommand{\bw}{\boldsymbol{w}}
\newcommand{\bey}{\begin{eqnarray}}
\newcommand{\eey}{\end{eqnarray}}

\usepackage[top=2.5cm, bottom=2.5cm, left=2.5cm, right=2.5cm]{geometry} 
\usepackage    {tikz}
\usetikzlibrary{calc}
\usetikzlibrary{tikzmark}

\tikzset
{
	pics/vlines/.default=2,
	pics/vlines/.style={
		code={%
			\foreach\i in {1,...,#1}
			\draw[pic actions] (-0.05*#1+0.1*\i-0.05,-0.25) --++ (0,-0.15);
	}},
	underb/.style n args={3}{
		to path={++(#1,-0.5) --++ (0,-#3) -| ($(\tikztotarget)+(#2,-0.5)$)}
	}
}

\vspace{-1.8 in}
\title{{\bf Nonlinear elasticity of pre-stressed single crystals at high pressure and various elastic moduli}}
\date{}

\begin{document}
\large
\belowdisplayskip15pt
\maketitle


\vspace{-0.8 in}
\bec
{\large Valery I. Levitas}
\eec

\bec
Iowa State University, Departments of Aerospace Engineering and Mechanical Engineering,
 Ames, Iowa 50011, USA, vlevitas@iastate.edu;

 Ames Laboratory, Division of Materials Science and Engineering,  Ames, IA, USA
\eec
\vspace{7mm}

A general nonlinear theory for the elasticity of pre-stressed single crystals is presented. Various types of elastic moduli are defined, their importance is determined, and relationships between them are presented. In particular, $\B$ moduli are present in the relationship between the Jaumann objective time derivative of the  Cauchy stress and deformation rate and are broadly used in computational algorithms in various finite-element codes. Possible applications to simplified linear solutions for complex  nonlinear elasticity problems are outlined and illustrated for a superdislocation. The effect of finite rotations is fully taken into account and analyzed. Different types of the bulk and shear moduli under different constraints are defined and connected to the effective properties of polycrystalline aggregates.
Expressions for elastic energy and stress-strain relationships for small distortions with respect to a pre-stressed configuration are derived in detail. Under initial hydrostatic load, general consistency conditions for elastic moduli and compliances are derived that follow from the existence of the generalized tensorial equation of state under hydrostatic loading obtained from single crystal or polycrystal.   It is shown that $\B$ moduli can be found from the expression for the Gibbs energy. However, higher-order elastic moduli defined from the Gibbs energy do not have any meaning since they do not directly participate in any   known equations, like stress-strain relationships and wave propagation equation.
The deviatoric projection of $\B$ can also be found from the expression for the elastic energy for isochoric small strain increments, and the missing components of $\B$ can be found from the consistency conditions. Numerous inconsistencies and errors in the known works are analyzed.

\section{INTRODUCTION\label{Introduction}}

Elastic moduli under high pressures play a fundamental role for the solution of numerous basic and applied problems. These problems include the determination of tensorial stress-strain relationships for single crystal and polycrystalline aggregates,
which are utilized for  continuum simulations of deformation of material  under extreme dynamic~\citep{clayton2014analysis, clayton2015crystal,Jafarzadeh-Levitas-etal-Arxiv-20} or static~\citep{feng2016large, levitas2019tensorial}  loadings.
Nonlinearity of the elasticity rules lead
to  elastic instabilities
 ~\citep{wallace-72,wang1993crystal,Grimvall-etal-2012,Pokluda-etal-2015,de2017ideal}.
They cause various physical phenomena, including phase transitions (PT, i.e., crystal-crystal ~\citep{huang2003crystal, hatcher2009martensitic,Levitasetal-Instab-17,Levitasetal-PRL-17,zarkevich2018lattice}, amorphization ~\citep{binggeli1992elastic,kingma1993microstructural,brazhkin1996lattice,zhao2018shock,chen2019amorphization}, and melting ~\citep{tallon1989hierarchy,levitas2012virtual}), slip ~\citep{chowdhury2017revisit}, twinning ~\citep{ogata2005energy}, and different fracture modes  ~\citep{de2017ideal,Pokluda-etal-2015,Cernyetal-JPCM-13,tang1994lattice}.

A quite general and correct description of nonlinear elasticity in terms of elastic moduli with respect to the reference configuration  under  arbitrary  stress tensor  was presented in  ~\citep{Huang-50,Leibfried-Ludwig-61,Barron-Klein-65,wallace-67,wallace-70}. While  in  ~\citep{Huang-50,Leibfried-Ludwig-61} the presentation was heavily mixed with atomic treatment,   in 
~\citep{Barron-Klein-65} and then ~\citep{wallace-67,wallace-70} elastic moduli $\B$ are introduced that determined
relationship between small Cauchy stress increment  and strain with respect to the current configuration, which played the major role in the determination of elastic lattice instability under homogeneous perturbations in  ~\citep{wang1993crystal,Grimvall-etal-2012,Pokluda-etal-2015,de2017ideal}.

Unfortunately,  in the most popular for this topic  textbook ~\citep{wallace-72} derivation of the elastic moduli $\B$ from the Cauchy (true) stress - small strain relationship is very confusing
because it is based on small strain increment from the intermediate reference configuration with respect to stress-free configuration (instead of from the intermediate to the current configuration) and corresponding linearization. In ~\citep{Barron-Klein-65} stress-free reference configuration is not introduced at all; that is why the relationship between elastic moduli in the current configuration $\bar{\C}$ and the all-rank elastic constant tensors $\C_0^n$ of the undeformed crystal are not presented. This, however, was elaborated in ~\citep{wallace-67}.

It was suggested   in ~\citep{barsch1968second} for cubic crystals under initial hydrostatic pressure 
to use the second derivatives of the enthalpy  (instead of Helmholtz energy) with respect to the Lagrangian
strain tensor to determine elastic moduli, claiming that they play the same role under pressure as usual moduli for unstressed solid.  However, justification was not given, and as we will discuss, this is not true for the third- and higher-order elastic moduli.
This direction was further elaborated in terms of the Gibbs energy and applied for atomistic determination of elastic moduli for specific materials in ~\citep{marcus2002importance,Marcus-Qiu-09,Mosyaginetal-08,Krasilnikovetal-12,Vekilovetal-15}. However, as we will show in the paper, such elastic moduli, starting with the third order, do not have any direct physical applications.

One of the methods to determine some of the components of the tensor $\B$ is based calculation of the elastic energy for the small incremental strain $\beps$, which  is isochoric  up to terms $\beps^2$  ~\citep{Cohen-etal-1997,Steinle-Neumann-etal-PRB-99,Gulseren-Cohen-02}.
However, because justification was not given, this method was criticized in ~\citep{marcus2002importance} and ~\citep{sin2002ab}. This led to mutual critics
in  ~\citep{steinle2004comment,Marcus-Qiu-04rep}. 
It is shown in  ~\citep{steinle2004comment,Grimvall-etal-2012}  for  hexagonal crystals that for several  small   isochoric strains without rotations, the free energy indeed represents quadratic form with  $\B$ moduli. However, the general proof is lacking.
Also, there is no general consideration for exactly which components of the $\B$ moduli can be determined with such a method and how to find the remaining components.

While using the density functional theory (e.g.,  ~\citep{Mehl-etal-1990,Fastetal-95,Marcus-01}),
a linear elasticity theory that neglects the effect of initial stresses was used for calculations of elastic moduli as the coefficients in quadratic elastic energy.
However, when the same approaches have been applied  for a stressed solid, the effect of prestressing is often neglected.
For example, in ~\citep{Stixrude-Cohen-1995} these equations have been applied at high pressures but  pressure correction was not mentioned.

Elastic moduli $\B$ were   found directly from the stress-small strain relationship in
~\citep{Wangetal-08} for $\alpha$, $\omega$, and $\beta$ Zr. One of the general problems is that in most papers, 
different elastic moduli $\C$ and $\B$ are often confused. The importance of a clear definition of which elastic moduli are considered in the problem under study is demonstrated
in ~\citep{gregoryanz2000high} and the following discussion ~\citep{muser2003comment,gregoryanz2003gregoryanz}.
In ~\citep{gregoryanz2000high}, elastic moduli for $\alpha-$ quartz  were taken from the linear relationship between the stress and strain tensors, without specifying which stress and strain, and used for evaluation of lattice instability. It was assumed in ~\citep{muser2003comment} that these elastic moduli are $ \C$ and should be pressure corrected
to $ \B$ for the stability study.   Then it was explained in ~\citep{gregoryanz2003gregoryanz} that  their elastic moduli were, in fact, $ \B$, because they came directly from the stress-strain relationship, meaning Cauchy stress-small strain with respect to the deformed state. 

Also, as we will show, even the definition of the bulk modulus under nonhydrostatic loading contains ambiguities, which leads to various confusion.

We believe that it is time to revisit in a strict and consistent way, within the framework of modern continuum mechanics,  the main expressions for all reasonable elastic moduli, the relationship between them, and rigorous  determination methods.
The paper is organized as follows. Section \ref{definitions} contains finite-strain kinematics, expressions for different stresses and energy, and definitions of elastic moduli in the stress-free ($\C_0$) and arbitrary reference configurations ($\C$) for arbitrary strains, as well as elastic constants of all ranks ($\C_0^n$) in the stress-free configuration for zero strains. Relationships between all these moduli are derived. Section \ref{rate-Cauchy} presents relationships between the objective Jaumann rate of the Cauchy stress and deformation rate, which appear to be  connected by elastic moduli $\B$. The rate of the second Piola-Kirchhoff stress and Lagrangian strain are
 connected by elastic moduli $\C$.
Possible applications of the rate equations to simplified linear solutions for complex  nonlinear elasticity problems are outlined.
Various reasonable definitions of the bulk modulus and compressibility (bulk compliance) under nonhydrostatic stresses and their relationships with $\B$ moduli are presented in Section \ref{Bulk}.
Approximation for small distortions with respect to a pre-stressed intermediate configuration is described in Section \ref{small-dist}.
Elastic moduli $\tilde{\C}$ are introduced as coefficients of the quadratic elastic energy in terms of small distortions. They also directly appear in equations of motion and wave propagation. Issues related to the presence of small rotations in the elastic energy and invariance under rigid-body rotations in the current configuration  are addressed. An example  of simple shear is considered.
Section \ref{intermediate-config} presents a particular case when the intermediate configuration is under hydrostatic pressure. All equations from the previous Sections are simplified for this case, and some new relationships are derived. Consistency conditions for moduli $\B$ and corresponding compliances are derived based on the existence of the generalized equation of state under hydrostatic loading.
This is done for the data obtained for single crystal and  polycrystalline samples. Relationships are presented between bulk and shear moduli of the isotropic
polycrystal and moduli $\B$ under pressure. It is proved that the Gibbs energy is a quadratic form in small
strains with moduli $\B$. It is also proven that the elastic energy for small isochoric strains is a quadratic form with the  deviatoric projection of moduli $\B$. The rest of the components of $\B$ can be determined from the consistency conditions.
Hydrostatic loading and isotropic deformation are obtained through energy minimization in Section \ref{minimization}.
Simplifications for isotropic materials and cubic crystals are presented.
The principle of superposition for defects and inelasticity in nonlinear elasticity with application to superdislocation and promotion of phase transformations by dislocation pileups are considered in Section \ref{superposition}.
Summary of the current work is presented in Section \ref{Summarizing-remarks}.
Tensor notations used in the paper are presented in Appendix A.  Appendix B contains derivation of some equations.
Appendix C describes analysis the invariance under superposed rigid-body rotations in the current
configuration. Appendix D treats a simple shear under hydrostatic pressure.
Analysis of the previous  approaches is given in Appendix E.

\section{MAIN DEFINITIONS\label{definitions}}
\subsection{Kinematics\label{Kinematics}}

\begin{figure}[htp]
\centering
\includegraphics[width=0.45\textwidth]{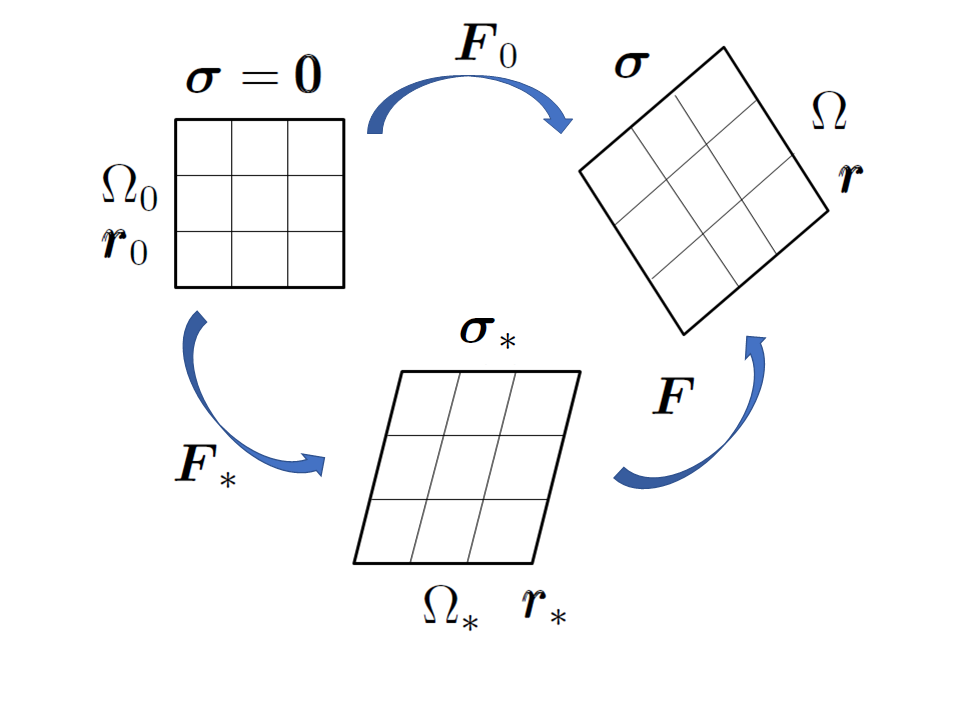}
\caption{ The reference stress-free configuration $\Omega_0$, the  deformed (current) configuration $\Omega$ under the Cauchy stress $\s$, and an arbitrary intermediate configuration $\Omega_*$ with initial stress $\s_*$.
Multiplicative decomposition of
the deformation gradient $  \F_{0}=   \F \cd \F_{*} $ is valid.
\label{kinematics}}
\end{figure}
\noindent
The  motion of an elastic body  is presented by a continuous function $  {\fg  r}={\fg  r}({\fg  r}_0, t)  $,
where $  {\fg r}_0  $ and $  {\fg r}  $ are the positions of material points in the undeformed (stress-free, $\s=\fg 0$) $  \Omega_0  $  and  the current deformed $  \Omega  $
configurations under the Cauchy (true) stress tensor $\s$, respectively; $t$ is time.
Let us consider some homogeneously deformed  intermediate configuration $  \Omega_*  $ described by position vector $  {\fg  r}_*={\fg  r}_*({\fg  r}_0, t)  $, which can be  arbitrarily chosen depending on the goals. The following multiplicative  decomposition of the total deformation gradient $ \F_{0} $ from the reference configuration $  \Omega_0  $ to the current configuration $  \Omega  $ is valid:
\begin{equation}
    \F_{0}:=  \frac{\p \fg r}{\p \fg r_0} = \frac{\p \fg r}{\p \fg r_*} \cd  \frac{\p \fg r_*}{\p \fg r_0} =  \F \cd \F_{*};
    \qquad  \F:= \frac{\p \fg r}{\p \fg r_*}; \qquad \F_{*}:=  \frac{\p \fg r_*}{\p \fg r_0},
    \label{tag500}
\end{equation}
where $ \F_{*} $ is deformation gradient describing deformation from the reference configuration $  \Omega_0  $ to the intermediate configuration $  \Omega_*  $ and $ \F $ is deformation gradient describing deformation from the intermediate configuration $  \Omega_* $  to the current configuration $ \Omega $.
In the component form
\begin{equation}
    F_{0,ij}:=  \frac{\p r_i}{\p r_{0,j}} = \frac{\p r_i}{\p r_{*k}}   \frac{\p  r_{*k}}{\p  r_{0,j}} =  F_{ik}  F_{*kj};
    \qquad  F_{ik}:= \frac{\p r_i}{\p r_{*k}}; \qquad F_{*kj}:=   \frac{\p  r_{*k}}{\p  r_{0,j}} .
    \label{tag500C}
\end{equation}
In contrast to the intermediate configurations in elastoplasticity \cite{levitas-book96,Lubarda-01} and theory of phase transformations \cite{levitas-98,Babaei-Levitas-PRL-20}, which corresponds to the stress-free state after unloading after plastic deformation and phase transformation, the intermediate configuration here can be chosen arbitrarily based on the problem in hands and convenience. For example, to derive relationships for small strains superposed on the finite strains, the intermediate configuration is chosen close to the current one; to derive rate equations, they are infinitesimally close. The intermediate configuration may correspond to the chosen pre-stressed state, particularly hydrostatically pre-stressed, to study the effect of this stress on elastic moduli or wave propagation. { Pre-stressing can be heterogeneous, e.g., due to heterogeneous loading or inelastic deformation gradient  $\F_{in}$, e.g., plastic, thermal, transformational, at others; then the decomposition ~\eqref{tag500} should be substituted with $ \F_{0}=  \F \cd \F_{*} \cd \F_{in}$ \cite{levitas-book96,Lubarda-01,levitas-98,Babaei-Levitas-PRL-20}. In general, the intermediate configuration and multiplicative decomposition can be introduced for each point locally through the field of $\fg F_*$, which is incompatible, i.e., a continuous vector function $\fg r_* ({\fg  r}_0)$  does not exist.   This is similar to incompatible plastic and transformational deformation gradients in the elastoplasticity \cite{levitas-book96,Lubarda-01} and the theory of phase transformations \cite{levitas-98,Babaei-Levitas-PRL-20}.  For example, when the hydrostatic pressure is much larger than the deviatoric stresses and the elasticity rule is presented in terms of pressure-dependent elastic moduli, then for the heterogeneously loaded body, the intermediate configuration for each point is under current local hydrostatic pressure (i.e., is obtained by local relaxation of the deviatoric stresses) and is generally incompatible.  If  the field $\F_{*}({\fg  r}_0)$ is associated with the current heterogeneously deformed configuration, it is compatible.}

Eq.~\eqref{tag500} implies for Jacobian determinants
\begin{equation}
  J_{0}:= det \F_0= J J_*; \qquad
  J:=  det \F; \qquad J_*:= det \F_{*},
    \label{tag500J}
\end{equation}
where $det $ is the determinant of a tensor, $ J_0 $,   $ J $, and $J_*$  characterize the volume-ratios for the   corresponding deformation gradients.

The Lagrangian total strain $ \E_0 $ and strains based on  $ \F_* $ and $ \F $ are defined as
\begin{eqnarray}
&& \E_0=\frac{1}{2}(\F_0^T \cd \F_0-\I);\quad
\E_*=\frac{1}{2}(\F_*^T{\cd}\F_*-\I);\quad
\E=\frac{1}{2}(\F^T \cd \F-\I); \qquad
\nonumber \\
&& \E_0=\F_*^T{\cd} \E {\cd} \F_*+\E_*,
\label{Esall}
\end{eqnarray}
where the decomposition of $ \E_0 $ in the last Eq.~\eqref{Esall} can be checked using multiplicative decomposition in Eq.~\eqref{tag500} and $\I$ is the second-order unit tensor (see Appendix A).  In the component form
\begin{eqnarray}
&& E_{0,ij}=\frac{1}{2}(F_{0,ki} F_{0,kj}-\delta_{ij});\quad
E_{*ij}=\frac{1}{2}(F_{*ki} F_{*kj}-\delta_{ij});\quad
E_{ij}=\frac{1}{2}(F_{ki} F_{kj}-\delta_{ij}); \qquad
\nonumber \\
&& E_{0,ij}=F_{*ki} E_{kl} F_{*lj} +E_{*ij}.
\label{EsallC}
\end{eqnarray}

\subsection{Energy and stresses\label{Energy-stresses}}

We will not consider temperature or entropy variation here for brevity, but they can be trivially included.
Then the  Helmholtz energy  $\psi_0$ for isothermal processes or the internal energy $U_0$ for isentropic processes, per unit volume in the reference configuration $  \Omega_0  $ and the intermediate configuration $  \Omega_*  $ are
\begin{equation}
\begin{split}
 \psi_0= \psi_0 (\E_0);     \qquad \psi =  \psi_0/ J_*=\psi (0) + \s_* :\E+ \Tilde \psi (\E),
    \label{tag504a-0}
\end{split}
\end{equation}
where $\psi (0) $ and $ \s_*$ are the energy and the second Piola-Kirchhoff stress  at $\E= \fg 0$ (see Eq.~\eqref{tag505}), and $ \Tilde \psi (\E)$ contains quadratic and higher order terms in $\E$.  The constitutive equations and, in particular, elastic moduli, will be isothermal if the  Helmholtz energy is used and isentropic (adiabatic) if the internal energy $U_0$ is utilized. In the component form, Eq.~\eqref{tag504a-0} can be presented as
\begin{equation}
	\qquad \psi =  \psi_0 / J_*=\psi  (0) + \sigma_{*ij}E_{ji}+ \Tilde \psi (E_{mn}).
 \label{psi=psi_0}
 \end{equation}

The second Piola-Kirchhoff stress in the reference $ \T_0 $ and intermediate $ \T $ configurations are defined as follows:
\begin{eqnarray}
 &&  {\T_0}=  {J_0}\ {\F_0}^{-1}\cd\s\cd {\F_0}^{T-1} = \frac{\p\psi_0}{\p\E_0};
       \qquad \T=J\F^{-1}\cd\s\cd\F^{T-1}=
       \frac{\p\psi}{\p \E}=\s_*+\frac{\p \Tilde{\psi}}{\p\E}; \quad
       \nonumber\\
    &&   \T_0= J_*{\F}^{-1}_* \cd \T \cd {\F}^{T-1}_*; \qquad  \T= J_*^{-1}{\F}_* \cd \T_0 \cd {\F}^{T}_*;
    \qquad \s_*:=  \s|_{\E= 0}=\T|_{\E=0} .
    \label{tag505}
\end{eqnarray}
It is clear that for $ \F=\I $ one has $\E=\fg 0 $ and $ \s=\T=  \s_* $, which justifies presentation Eq.~\eqref{tag504a-0} for $ \psi $. In the component form
\begin{eqnarray}
 && T_{0,ij}=  {J_0} F_{0,ik}^{-1}\sigma_{kl}F_{0,jl}^{-1} = \frac{\p\psi_0}{\p E_{0,ij}};
       \qquad  T_{ij}=  {J} F_{ik}^{-1}\sigma_{kl}F_{jl}^{-1} = \frac{\p\psi}{\p E_{ij}}=\sigma_{*ij}+\frac{\p \Tilde{\psi}}{\p E_{ij}}; \quad
       \nonumber\\
    &&   T_{0,ij}=  {J_*} F_{*ik}^{-1}T_{kl}F_{*jl}^{-1} ; \qquad   T_{ij}=  {J_*^{-1}} {F_{*ik}}T_{0,kl}{F_{*jl}}.
    \label{tag505C}
\end{eqnarray}

It follows from  Eq.~\eqref{tag505}
\begin{eqnarray}
  \s= {J_0}^{-1}{\F_0}\cd {\T_0}\cd {\F_0}^{T}= {J_0}^{-1}{\F_0}\cd \frac{\p\psi_0}{\p\E_0}\cd {\F_0}^{T}  ={J}^{-1}{\F}\cd {\T}\cd {\F}^{T}= {J}^{-1}{\F}\cd \frac{\p\psi}{\p\E}\cd {\F}^{T};
          \label{tag505s}
\end{eqnarray}
\begin{eqnarray}
  \sigma_{ij}= J_0^{-1}F_{0,ik}T_{0,kl}F_{0,jl}= J_0^{-1}F_{0,ik}\frac{\p\psi_0}{\p E_{0,kl}} F_{0,jl}
  = J^{-1}F_{ik}T_{kl}F_{jl}= J^{-1}F_{ik}\frac{\p\psi}{\p E_{kl}} F_{jl}.
          \label{tag505sC}
\end{eqnarray}

Note that while $\s_*:=  \s|_{\E= 0}=\T|_{\E=0}$ has the physical meaning of stresses, in the thermodynamic treatment  (e.g., application of thermodynamic laws to derive elasticity rules or temperature evolution equation), $\s_*$ should be treated just  as a constant in the linear term in Eq. ~\eqref{psi=psi_0} assuming the fixed intermediate configuration.   The intermediate configuration and $\s_*$ can be varied in the final equations.

\subsection{Elastic moduli in the reference and intermediate configurations  \label{elastic-moduli}}

Elastic moduli in the reference $\C_0$ and intermediate $\C$ configurations are defined as follows:
\begin{eqnarray}
 \C_0  \left( \E_0 \right):= \frac{\p \T_0}{\p \E_0 }= \frac{\p^{2} \psi_0}{\p \E_0 \p \E_0}; \quad
 \C \left( \E \right):= \frac{\p \T}{\p \E }= \frac{\p^{2} \psi}{\p \E \p \E};
    \label{tag501}
\end{eqnarray}
\begin{eqnarray}
	C_{0,ijkl}  \left( E_{0,mn} \right):= \frac{\p T_{0,ij}}{\p E_{0,kl} }= \frac{\p^{2} \psi_0}{\p E_{0,ij} \p E_{0,kl}}; \quad
	C_{ijkl} \left( E_{mn} \right):= \frac{\p T_{ij}}{\p E_{kl} }= \frac{\p^{2} \psi}{\p E_{ij} \p E_{kl}}.
 \label{tag501C}
 \end{eqnarray}
In particular, $\C \left( \E \right)$ can be evaluated at $\F=\fg I$, i.e., at $\E_0=\E_*$, when the intermediate configuration coincides with the current one, like in ~\citep{wallace-67,wallace-70}.

To derive relationships between $\C$ and $\C_0$, we first present
\begin{eqnarray}
\F_*^T{\cd} \E {\cd} \F_*= \F_*^T \stackrel{2}* \F_*^T {\cd} \E = \F_*^T \stackrel{2}* \F_*^T {\cd} \fg I^4_s \fg :\E    ;
    \label{I4E}
\end{eqnarray}
\begin{eqnarray}
	F_{*il}^T E_{lk} F_{*kj}=F_{*li} F_{*kj} E_{lk} = \frac{1}{2} (F_{*li} F_{*kj}+F_{*ki} F_{*lj})E_{lk},
	\label{dE0dEcom}
\end{eqnarray}
where we took into account symmetry of $E_{lk}$, or with the symmetrizing fourth-rank tensor,
\begin{eqnarray}
&& F_{*li} F_{*kj} E_{lk} =F_{*li} F_{*kj} I_{s,lkmn}^4 E_{nm}=
\frac{1}{2} F_{*li} F_{*kj} (\delta_{ln}\delta_{km}+\delta_{lm}\delta_{kn}) E_{nm}=
\nonumber\\
&& \frac{1}{2} (F_{*ni} F_{*mj}+F_{*mi} F_{*nj})E_{nm}=
\frac{1}{2} (F_{*li} F_{*kj}+F_{*ki} F_{*lj})E_{lk}.
	\label{dE0dEcom1}
\end{eqnarray}

 Then   we evaluate
\begin{eqnarray}
  \frac{\partial   \E_0}{ \partial \E}=\frac{\partial  (\F_*^T{\cd} \E {\cd} \F_*)}{ \partial \E}
  =\frac{\partial (  \F_*^T \stackrel{2}* \F_*^T {\cd} \fg I^4_s \fg :\E )}{ \partial \E}= \F_*^T \stackrel{2}* \F_*^T {\cd} \fg I^4_s;
    \label{dE0dE}
\end{eqnarray}
\begin{eqnarray}
  \frac{\partial   E_{0ij}}{ \partial E_{kl}}= F_{*im}^T  F_{*jn}^T  I^4_{s,nmkl}= F_{*mi}  F_{*nj}  I^4_{s,nmkl}=
  \frac{1}{2} \left(F_{*ki}  F_{*lj} + F_{*li}  F_{*kj}   \right),
    \label{dE0dEcom-2}
\end{eqnarray}
which can be directly obtained from Eq.~\eqref{dE0dEcom1}.
Note that in a similar expression in ~\citep{wallace-67,wallace-70}, symmetrization is missing; this, however, did not affect the final results.

Next, using Eqs.~\eqref{tag505} and ~\eqref{dE0dE}, we obtain
\begin{eqnarray}
     &&
        \T= J_*^{-1}{\F}_* \cd \frac{\p\psi_0}{\p\E_0} \cd {\F}^{T}_*;
     \quad
       \nonumber\\
 &&          \C= \frac{\p\T}{\p\E}= J_*^{-1}{\F}_*    \stackrel{2}* {\F}_* \cd \frac{\p^2\psi_0}{\p\E_0 \p\E}
    =
    J_*^{-1}{\F}_*    \stackrel{2}* {\F}_* \cd \frac{\p^2\psi_0}{\p\E_0 \p\E_0}  \fg: \frac{\p \E_0}{\p\E}
    =
     \nonumber\\
 &&
     J_*^{-1}{\F}_*    \stackrel{2}* {\F}_* \cd \C_0  \fg: \left(\F_*^T \stackrel{2}* \F_*^T {\cd} \fg I^4_s \right)
     =
 \left(     J_*^{-1}{\F}_*    \stackrel{2}* {\F}_* \cd \C_0  \cd \F_*^T \stackrel{2}* \F_*^T \right) \fg : \fg I^4_s
   =
 \nonumber\\
 &&
   J_*^{-1}{\F}_*    \stackrel{2}* {\F}_* \cd \C_0  \cd \F_*^T \stackrel{2}* \F_*^T =
     J_*^{-1}\F_*  \stackrel{4}* \F_*  \stackrel{3}*  {\F}_*    \stackrel{2}* {\F}_* \cd \C_0 .
    \label{tag505C-C0}
\end{eqnarray}
Similarly, in component form, we derive:
\begin{eqnarray}
     &&         C_{ijkl}= \frac{\p T_{ij}}{\p E_{kl}}= J_*^{-1}{F}_{*in}  {F}_{*jm}  \frac{\p^2\psi_0}{\p E_{0,mn} \p E_{kl}}
    =
 J_*^{-1}{F}_{*in}  {F}_{*jm}  \frac{\p^2\psi_0}{\p E_{0,mn} \p E_{0,st}} \frac{\p E_{0,st}}{\p E_{kl}}
    =
     \nonumber\\
 &&
\frac{1}{2}  J_*^{-1}{F}_{*in}  {F}_{*jm}  C_{0,mnst}  \left(F_{*ks}  F_{*lt} + F_{*ls}  F_{*kt}   \right)
     =
J_*^{-1}  F_{*lt}F_{*ks}  {F}_{*in}  {F}_{*jm}  C_{0,mnst}
,
    \label{tag505C-C0-comp}
\end{eqnarray}
where we took into account symmetry of $C_{0,mnst}$, in the given case over permutation of  $s$ and $t$.
Thus,
 \begin{eqnarray}
   \C (\E)=  J_*^{-1}\F_*  \stackrel{4}* \F_*  \stackrel{3}*  {\F}_*    \stackrel{2}* {\F}_* \cd \C_0 (\E_0);
   \quad
    C_{ijkl}(E_{ab}) =J_*^{-1} {F}_{*im} {F}_{*jn}F_{*ks} F_{*lt}  C_{0,mnst} (E_{0,ab}).
    \label{C-C0}
\end{eqnarray}
Eq.~\eqref{C-C0} gives the relationship between elastic moduli in two different configurations connected by deformation gradient ${\F}_*$.   Note that  $\C_0$ is a function of $\E_0$ and $\C$ is a function of $\E$, and can be expressed via Eq.~\eqref{C-C0} in terms of $\E_0$ and $\F_*$. They do not keep symmetry of the initial non-deformed lattice and for general $\E_0$  they have symmetry of trigonal crystal, i.e., all 21 unequal components. They have complete Voigt symmetry, i.e., are invariant with respect to exchange indices $1\leftrightarrow 2$, $3\leftrightarrow 4$, and $ (1,2)\leftrightarrow (3,4)$.

\subsection{Elastic constants of all orders in the stress-free reference configuration  \label{elastic-constants}}

To find the functional relationship $\C_0 \left( \E_0 \right)$, we need to { know the explicit expression for the energy $ \psi_0 (\E_0)$. Usually, it is defined in terms of } the elastic constants of all orders in the stress-free reference configuration.
Let us expand
\begin{eqnarray}
 && \psi_0 (\E_0)=  \psi_0 (\fg 0)+ 1/2  \E_0 \fg : \C_0^2 \fg : \E_0 + 1/(3!)  (\E_0 \fg : \C_0^3 \fg : \E_0) \fg : \E_0 +
 \nonumber\\
&& 1/(4!)  \E_0  \fg :  (\E_0 \fg : \C_0^4 \fg : \E_0) \fg : \E_0  + 1/(5!) ( \E_0  \fg :  (\E_0 \fg : \C_0^{4} \fg : \E_0) \fg : \E_0) \fg : \E_0 + ...\, ,
    \label{tag504-exp}
\end{eqnarray}
\begin{eqnarray}
&	&\psi_0 (E_{0,mn})=  \psi_0 ( 0)+ 1/2 C_{0,ijkl}^2  E_{0,ji} E_{0,lk} + 1/(3!) C_{0,ijklef}^3 E_{0,ij} E_{0,kl}E_{0,fe} +
	\nonumber\\
&&	1/(4!) C_{0,ijklefqst}^4 E_{0,ij} E_{0,kl}E_{0,fe} E_{0,ts} + 1/(5!) C_{0,ijklefqstmn}^{5} E_{0,ij} E_{0,kl}E_{0,fe} E_{0,ts}E_{0,nm} + ...\,  ,
\end{eqnarray}
where the $2n^{th}$-rank tensors $\C_0^n$ are $n$-order elastic constants of the undeformed crystal, which {\it possess symmetry of the undeformed crystal, and contain complete information about elastic energy, moduli, and stress measures at any strain} $\E_0$.

By definition,
\begin{eqnarray}
 && \C_0^2=:  \frac{\p^{2} \psi_0}{\p \E_0 \p \E_0}|_{\E_0=0}= \frac{\p \T_0}{\p \E_0 }|_{\E_0=0};
 \qquad
  \C_0^3=:  \frac{\p^{3} \psi_0}{\p \E_0 \p \E_0 \p \E_0}|_{\E_0=0}= \frac{\p^2 \T_0}{\p \E_0 \p \E_0}|_{\E_0=0};
\nonumber\\
 && \C_0^4=:  \frac{\p^{4} \psi_0}{\p \E_0 \p \E_0 \p \E_0 \p \E_0}|_{\E_0=0}= \frac{\p^3 \T_0}{\p \E_0 \p \E_0 \p \E_0}|_{\E_0=0};
\nonumber\\
 & & \C_0^{5}=:  \frac{\p^{5} \psi_0}{\p \E_0 \p \E_0\p \E_0 \p \E_0 \p \E_0}|_{\E_0=0}= \frac{\p^4 \T_0}{\p \E_0 \p \E_0 \p \E_0\p \E_0}|_{\E_0=0}.
    \label{tag504-exp-Cs}
\end{eqnarray}
\begin{eqnarray}
	&& C_{0,ijkl}^2=:  \frac{\p^{2} \psi_0}{\p E_{0,ij} \p E_{0,kl}}|_{E_{0,cd}=0}= \frac{\p T_{0,ij}}{\p E_{0,kl} }|_{E_{0,cd}=0};
\label{t04-exp-Cs}\\
	\qquad
	&& C_{0,ijklef}^3: = \frac{\p^{3} \psi_0}{\p E_{0,ij} \p E_{0,kl} \p E_{0,ef}}|_{E_{0,cd}=0}= \frac{\p^2 T_{0,ij}}{\p E_{0,kl} \p E_{0,ef}}|_{E_{0,cd}=0};
	\nonumber\\
	&& C_{0,ijklefst}^4=:  \frac{\p^{4} \psi_0}{\p E_{0,ij} \p E_{0,kl} \p E_{0,ef} \p E_{0,st}}|_{E_{0,cd}=0}= \frac{\p^3 T_{0,ij}}{\p E_{0,kl} \p E_{0,ef} \p E_{0,st}}|_{E_{0,cd}=0};
\nonumber\\
	&& C_{0,ijklefstmn}^{5}=:  \frac{\p^{4} \psi_0}{\p E_{0,ij} \p E_{0,kl} \p E_{0,ef} \p E_{0,st}E_{0,mn}}|_{E_{0,cd}=0}= \frac{\p^3 T_{0,ij}}{\p E_{0,kl} \p E_{0,ef} \p E_{0,st} \p E_{0,mn} }|_{E_{0,cd}=0}.
\nonumber
\end{eqnarray}
Examples of elastic moduli are given up to third order in  ~\citep{zhao2007first,lopuszynski2007ab,Cao-etal-PRL-18},  fourth order  in  ~\citep{wang2009ab,telichko2017diamond} at small strains (e.g., 0.02-0.03), and up to fifth order for finite strains (0.37 for Si I) in ~\citep{Chen-etal-NPJCM-20}.

The knowledge of the above elastic constants allows one to determine elastic moduli $\C_0  \left( \E_0 \right)$
according to  Eq.~\eqref{tag501},
\begin{eqnarray}
 \C_0  \left( \E_0 \right)= \C_0^2  + \C_0^3 \fg :  \E_0 + 1/2  (\C_0^4 \fg : \E_0 ) \fg :  \E_0+
1/(3!)  ( (\C_0^{5} \fg : \E_0) \fg : \E_0 )\fg : \E_0+ ...\, ;
    \label{tag501CE}
\end{eqnarray}
\begin{eqnarray}
&&	C_{0,ijkl}  \left( E_{0,ab} \right)= C_{0,ijkl}^2 +   C_{0,ijklef}^3 E_{0,fe} +1/2 C_{0,ijklefst}^4 E_{0,fe} E_{0,ts} +
	\nonumber\\
&	&1/(3!) C_{0,ijklefstmn}^{5}E_{0,fe} E_{0,ts}E_{0,nm} + ...\,  .
 \label{tag501CE-co}
 \end{eqnarray}
Eq.~\eqref{C-C0} can be utilized to determine moduli $\C(\E)$ for an arbitrary intermediate configuration.
In particular, they can be evaluated at $\F=\fg I$, i.e., at $\E_0=\E_*$, when the intermediate configuration coincides with the current one, like in ~\citep{wallace-67,wallace-70}. For finite deviation of the intermediate configuration from the current one, one may need higher order elastic moduli in the intermediate configuration to evaluate energy and stresses.
That is why the choice of the intermediate configuration that coincides with the current one is the most convenient and popular.

\section{RATE OF THE CAUCHY STRESS VERSUS DEFORMATION RATE  \label{rate-Cauchy}}

\subsection{Kinematics\label{sub-kinematic}}

Let $\fg v =\frac{\p}{\p t}{\fg  r}({\fg  r}_0, t) = \dot{\fg  r}({\fg  r}_0, t)   $ be the material velocity vector and $t$ is time. Using the invertibility of the
relationship $  {\fg  r}={\fg  r}({\fg  r}_0, t)  $, i. e., $  {\fg  r}_0={\fg  r}_0({\fg  r}, t)  $, velocity can be expressed as a function of  $  {\fg  r}$ and $t$, $  {\fg v}={\fg  v}({\fg  r}, t)  $, i. e., in spatial (Eulerian) presentation.
Let us define the velocity gradients in configurations $\Omega_0 $ and $\Omega$, respectively:
\bey
&&{ \partial \fg v \over \partial \fg r_0 }={ \partial  \over \partial \fg r_0 }{ \partial \fg r \over \partial t }=
{\partial \over \partial t}{ \partial \fg r \over \partial \fg r_0}={\partial \over \partial t}\fg F=\dot{\fg F};
\nonumber\\
&& { \partial v_i \over \partial r_{0,j} }\, \fg e_i \fg e_j={ \partial  \over \partial r_{0,j} }  { \partial r_i \over \partial t }\, \fg e_i \fg e_j= { \partial  \over \partial t}  { \partial r_i \over \partial r_{0,j}}\, \fg e_i \fg e_j= { \partial  \over \partial t}  F_{0,ij}\, \fg e_i \fg e_j =\dot{F}_{0,ij}\, \fg e_i \fg e_j;
 \label{11-13}
\eey
\bey
\fg l={ \partial \fg v \over \partial \fg r }= { \partial \fg v \over \partial \fg r_0 } \cdot { \partial \fg r_0 \over \partial \fg r }= \dot {\fg F} \cdot \fg F^{-1}; \qquad \fg l={ \partial v_i \over \partial r_j }\fg e_i \fg e_j=
{ \partial v_i \over \partial r_{0,k} }\, { \partial r_{0,k} \over \partial r_j }\, \fg e_i \fg e_j
= \dot{F}_{ik} F_{kj}^{-1}
\, \fg e_i \fg e_j.
\label{11-14}
\eey
We took into account the permutability of the spatial and time derivative in the material description in terms of $({\fg  r}_0, t)$.
As it follows from Eq.~\eqref{11-14}, this is not the case for variable reference configuration, i.e., in spatial presentation
in terms of $({\fg  r}, t)$.

The velocity gradient can be decomposed into the symmetric deformation rate $\bd$ and the antisymmetric spin tensor
$\bw$:
\bey
&& \fg l=\bd+\bw; \qquad   \bd=\bl_s= \left(\dot {\fg F} \cdot \fg F^{-1}\right)_s  ; \qquad \bw=\bl_{as}=  \left(\dot {\fg F} \cdot \fg F^{-1}\right)_{as};
\nonumber\\
&&
l_{ij}=d _{ij} + w_{ij}; \qquad  d _{ij}= \left(\dot{F}_{ik} F_{kj}^{-1}\right)_s; \qquad w_{ij}= \left(\dot{F}_{ik} F_{kj}^{-1}\right)_{as}.
\label{11-14-de}
\eey
All these tensors are defined in the current configuration and independent of the reference configuration.
We will use known relationship between $\dot {\fg E}$ and $\fg d$ \cite{Lurie-90,levitas-book96}:
\begin{eqnarray}
\fg d= \fg F^{-1T} \cdot \dot {\fg E}\cdot \fg F^{-1} ; \qquad \dot {\fg E}=\fg F^T \cdot \fg d \cdot \fg F .
\label{11-23}
\end{eqnarray}

\subsection{Stress rate - deformation rate relationships\label{stress-strain}}

{  In computational mechanics, it is traditional to find the relationship between some objective time derivative of the Cauchy stress and deformation rate for different material models \cite{Simo-Hughes-98}.  The coefficients in such relationships are used as an input in various computational codes (e.g., ABAQUS or DealII). We will do this here for nonlinear elasticity.}
The time-derivative of the constitutive equation $\T= \frac{\p\psi}{\p \E}$ for an arbitrarily  chosen fixed intermediate configuration (including stress-free configuration) yields:
\begin{eqnarray}
    \dot{\T}=\C:\dot{\E};
\qquad
	\dot{T_{ij}}=C_{ijkl}\dot{E}_{lk}.
    \label{b1}
\end{eqnarray}
Note that because the intermediate configuration is fixed, there are no terms with $\dot{\s}_*$ and $\dot{\F}_*$.
It is derived in Appendix B that
\begin{eqnarray}
    \dot{\s}+\s(\bd:\I)-\bl\cd\s-\s\cd\bl^{T}=\frac{1}{J}(\F\stackrel{2}*\F\cd\C\cd\F^{T}\stackrel{2}*\F^T):\bd =\bar{ \C} :\bd,
    \label{b5}
\end{eqnarray}
where 
\begin{eqnarray}
\bar{\C}=\frac{1}{J}\F\stackrel{2}*\F\cd\C\cd\F^{T}\stackrel{2}*\F^T
=\frac{1}{J}\F\stackrel{4}*\F\stackrel{3}*\F\stackrel{2}*\F\cd\C
=
\frac{1}{J_0}\F_0\stackrel{4}*\F_0\stackrel{3}*\F_0\stackrel{2}*\F_0 \cd \C_0.
   \label{barC}
\end{eqnarray}
Eq.~\eqref{C-C0}  and multiplicative decomposition ~\eqref{tag500} were utilized.
The tensor $\bar{\C}$ is the tensor of elastic moduli in the current configuration, which is independent of the choice of the reference configuration.  Note that the intermediate configuration was utilized for generality; it is not necessary for the derivation of Eq.~\eqref{b5}.

 In component notations,
\begin{eqnarray}
&&
\dot{\sigma_{ij}}+\sigma_{ij}(d_{kl}\delta_{lk})-l_{il}\sigma_{lj}-\sigma_{im} l_{jm}=\frac{1}{J}F_{lt}F_{ks}  {F}_{in}  {F}_{jm}  C_{mnst} d_{lk} =\bar{ C}_{ijkl} d_{lk},
\nonumber\\
& & \bar{C}_{ijkl} =J^{-1}  F_{lt}F_{ks}  {F}_{in}  {F}_{jm}  C_{mnst}
  =J_0^{-1}  F_{0,lt}F_{0,ks}  {F}_{0,in}  {F}_{0,jm}  C_{0,mnst}.
\label{b5com}
\end{eqnarray}
The left-hand side (LHS) of  Eq.~\eqref{b5} is known as the Truesdell objective rate of the Cauchy stress
\begin{eqnarray}
 \stackrel{\nabla}\s_{Tr}   := \dot{\s}+\s(\bd:\I)-\bl\cd\s-\s\cd\bl^{T}=\bar{ \C} :\bd,
    \label{Tr}
    \end{eqnarray}
which gives one more physical interpretation of the  tensor of elastic moduli in the current configuration $\bar{\C}$.
Using the definition of the Jaumann objective derivative $\stackrel{\nabla}\s_{J} :=\dot\s-\bw\cd\s-\s\cd\bw^{T}$ in Eq.~\eqref{b5}, it can be expressed as
\begin{eqnarray}
    \stackrel{\nabla}\s_{J}=\dot{\s}-\bw\cd\s-\s\cd\bw^{T}=
    \bar{\C}:\bd-\s\I:\bd+\bd\cd\s+\s\cd\bd;
    \label{b6}
\end{eqnarray}
\begin{eqnarray}
	\stackrel{\nabla}\sigma_{J,ij}=\dot{\sigma}_{ij}-w_{ik}\sigma_{kj}-\sigma_{ik}w_{jk}=
	\bar{C}_{ijkl}d_{lk}-\sigma_{ij}\delta_{mn}d_{nm}+d_{ik}\sigma_{kj}+\sigma_{ik}d_{kj}.
   \label{b6-c}
   \end{eqnarray}
Using derivations in Appendix B, Eq.~\eqref{b6}  can be transformed to 
\begin{eqnarray}
    \stackrel{\nabla}\s_{J}=\dot{\s}-\bw\cd\s-\s\cd\bw^{T}=
    (\bar{\C}-\s\I+\s\cd\I_s^{4}+(\I_s^{4}\cd\s)^{T}):\bd = \boldsymbol{B}:\bd,
    \label{mb7}
\end{eqnarray}
\begin{eqnarray}
\B := \bar{\C}-\s\I+\s\cd\I_s^{4}+(\I_s^{4}\cd\s)^{T};
 \label{mB}
\end{eqnarray}
\begin{eqnarray}
    B_{ijkl} := \bar{C}_{ijkl} - \sigma_{ij}\delta_{kl} +\frac{1}{2} (\sigma_{il}\delta_{jk} + \sigma_{ik}\delta_{jl} +
     \sigma_{kj}\delta_{li}+  \sigma_{lj}\delta_{ki}).
      \label{mB-co}
     \end{eqnarray}
     Thus, the relationship between the Jaumann derivative of the Cauchy stress and deformation rate is described by the effective elastic moduli tensor $\B$. { The tensor $\B$ was introduced in ~\citep{Barron-Klein-65} as the coefficients in the relationship between stresses and small strains  of the pre-deformed crystals and
 in ~\citep{wallace-67,wallace-70} as the derivative of the Cauchy stress with respect to small strain increment from the current configuration; thus, Eq.~\eqref{mB} is an alternative definition convenient for numerical realization in the modern finite-element codes (see Section \ref{Time integration}).} It is easy to check directly (and since $\B$ connects two symmetric tensors) that $B_{ijkl}$ is symmetric with respect to permutations $i\leftrightarrow j$ and $k \leftrightarrow l$, but is generally not symmetric in $(i,j) \leftrightarrow (k,l)$. The effective elastic moduli tensor $\B$ is the main elastic moduli tensor used for the evaluation of crystal lattice instability
      ~\citep{wallace-72,Grimvall-etal-2012,de2017ideal,Pokluda-etal-2015,wang1993crystal}.
      Here, it is introduced through the objective stress rate vs. deformation rate relationship.

Using transformations from Appendix B, Eq.~\eqref{mb7} can also be presented as
\begin{eqnarray}
    \dot{\s}= \B:\bd+\left[(\I_{as}^{4}\cd\s)^{T} - \s\cd\I_{as}^{4} \right]:\bw;
    \label{b10}
\end{eqnarray}
\begin{eqnarray}
    \dot{\sigma}_{ij}= B_{ijkl}d_{lk}+
  \frac{1}{2}  ( \sigma_{kj}\delta_{li} - \sigma_{jl} \delta_{ik}  -\sigma_{il}\delta_{jk}+ \sigma_{ik} \delta_{jl}) w_{lk} .
    \label{b10com}
\end{eqnarray}

\subsection{Time integration\label{Time integration}}

Let us  introduce the distortion tensor $ \be $ with respect to  the intermediate configuration $  \Omega_* $, as well as its symmetric $ \beps $ and antisymmetric $ \om $ parts:
\begin{equation}
    \be:= \F-\I = \beps + \om; \qquad \beps :=  (\be)_s; \quad \om:=  (\be)_{as}.
    \label{tag508a-0}
\end{equation}
For infinitesimal distortions, i.e., infinitesimal difference between the current and intermediate configurations, $|\be|\ll 1$, the velocity gradient is $\fg l \simeq \dot{\be}$, the deformation rate is
$\bd \simeq \dot{\beps}$, and spin is $\bw\simeq \dot{\om}$. Then Eq.~\eqref{b10com}
can be integrated for small time increment:
\begin{eqnarray}
  {\s}(t+\Delta t) = \s (t)+ \B: \dot{\beps} \Delta t+\left[(\I_{as}^{4}\cd\s)^{T} - \s\cd\I_{as}^{4} \right]: \dot{\om} \Delta t;
    \label{b10-int}
\end{eqnarray}
\begin{eqnarray}
   {\sigma}_{ij}(t+\Delta t)=  {\sigma}_{ij}(t)+  B_{ijkl}\dot{\epsilon}_{lk}  \Delta t +
  \frac{1}{2}  ( \sigma_{kj}\delta_{li} - \sigma_{jl} \delta_{ik}  -\sigma_{il}\delta_{jk}+ \sigma_{ik} \delta_{jl}) \dot{\omega}_{lk} \Delta t.
    \label{b10com-int}
\end{eqnarray}
Eq.~\eqref{b10com-int} is used in numerical algorithms for solution of boundary-value problems. Many finite-element programs (e.g., ABAQUS or DealII) use the fourth-rank tensor relating objective stress rate and deformation rate as the input from the user-developed material models.
While we presented the simplest explicit integration, any implicit integration scheme (e.g., predictor-corrector, backward Euler,  Crank-Nicolson, etc.) can be applied in the traditional way.

If an  intermediate configuration $\Omega_*$ is updated, then small strain $\beps $ and rotations $\om$ are evaluated with respect to updated configuration $\Omega_*$ without remembering previous values of $\beps$ and  $\om$, Eq.~\eqref{b10-int}
can be presented in terms of  small strains $\beps=\Delta \beps=\dot{\beps} \Delta t$ and rotations
$\om= \Delta \om = \dot{\om} \Delta t$
\begin{eqnarray}
  {\s}(t+\Delta t) = \s (t)+ \B: \beps +\left[(\I_{as}^{4}\cd\s)^{T} - \s\cd\I_{as}^{4} \right]: \om,
    \label{b10-int-1}
\end{eqnarray}
\begin{eqnarray}
   {\sigma}_{ij}(t+\Delta t)=  {\sigma}_{ij}(t)+  B_{ijkl} \epsilon _{lk}  +
  \frac{1}{2}  ( \sigma_{kj}\delta_{li} - \sigma_{jl} \delta_{ik}  -\sigma_{il}\delta_{jk}+ \sigma_{ik} \delta_{jl}) \omega _{lk} .
    \label{b10com-int-1}
\end{eqnarray}

\section{BULK MODULUS AND COMPRESSIBILITY \label{Bulk}}

While it is well known that the bulk modulus $K:= -V\frac{\p p}{\p V}$, where $p=- {1 \over 3 }\s \fg : \fg I$ is the pressure and
$V$ is the deformed volume of the system, we need to understand how this follows from definitions used above and which elastic  moduli does $K$ correspond to: $\C_0$,  $\bar{\C}$, or $\fg B$? For this purpose, we need to give strict definitions of the bulk modulus and compressibility for the general  stress-strain state.
Also, the importance of mentioning the exact components of the strain or stress tensors that are fixed in the chosen definition of the bulk moduli and compressibility for the general stress-strain state is illustrated.

\subsection{Hydrostatic loading of material with elastic energy depending on volume\label{hydro}}

For liquids and gases, since under hydrostatic loading stress work per unit reference volume $V_0$ or more generally, $V_*$ (in the intermediate configuration) is $-V_*^{-1} p d V= -p d J$, it then follows from
\begin{equation}
-V_*^{-1} p d V=  - p d J= d \psi \quad  \rightarrow \quad p=-\frac{\p \psi }{\p J}= -{V_*}\frac{\p \psi }{\p V},
    \label{press}
\end{equation}
where we remind that  $ \psi$ is the elastic energy per unit volume $V_*$ in the intermediate configuration $\Omega_*$.
Then the bulk modulus in the intermediate configuration ${K}_*$ can be defined as
\begin{equation}
   {K}_*:=-\frac{\p p}{\p J}= \frac{\p^2{\psi}}{\p {J}^2}= - V_* \frac{\p p}{\p V}= V_*^2 \frac{\p^2{\psi}}{\p V^2}.
    \label{tagK*}
\end{equation}
We use subscript $*$ to distinguish this modulus from the moduli $\C$ in the intermediate configuration because for  ${K}_*$ the derivative is calculated for the Cauchy stress, while for $\C$ it was evaluated for the second Piola-Kirchhoff stress.
When the intermediate configuration coincides with the current one,
then $V_*=V$, elastic energy is defined per unit current volume and is designated as $\psi_c$, and
\begin{equation}
p= -{V}\frac{\p \psi_c }{\p V}; \qquad    {K}=- V \frac{\p p}{\p V}= V^2 \frac{\p^2{\psi_c}}{\p V^2}= -  \frac{\p p}{\p lnV},
    \label{tagK}
\end{equation}
i.e., the classical definition is obtained. The following alternative expressions for $K$ are valid for any
{\it fixed}  intermediate configuration:
\begin{equation}
   {K}=- \frac{V}{V_*} \frac{\p p}{\p (V/V_*)}=-J \frac{\p p}{\p J}=J \frac{\p^2 \psi}{\p J^2} =- \frac{\p p}{\p \vep_0} = J K_* ;
   \qquad \vep_0: ={\ln J},
       \label{tagKalt}
\end{equation}
where $\vep_0$ is the logarithmic volumetric strain.
Comparisons of Eqs.~\eqref{tagK} and ~\eqref{tagKalt} leads to $d lnV=d\vep_0=d \ln\frac{V}{V_0}= \frac{d V}{V_0}\frac{V_0}{V}$, which  means that the increment of the logarithmic strain is independent of the reference configuration (volume).

Compressibility or bulk compliance is determined as
\begin{equation}
  k=\frac{1}{K}=- \frac{1}{V} \frac{\p V}{\p p}=  -  \frac{\p lnV}{\p p}=- \frac{\p \vep_0}{\p p}=-\frac{1}{J} \frac{\p J}{\p p}  .
    \label{tag-k}
\end{equation}

Note that all the above equations are strict if $\vep_0$ is the only strain-related variable; otherwise, derivatives should be calculated
at fixed other strain-related  variables or deviatoric stress, which will be discussed below.


\subsection{General elastic material\label{general-bulk}}

{\it Bulk modulus. }Let us consider general stress-strain state and find explicit connection between $K$ and  $\B$. We decompose the Cauchy stress and the deformation rate into spherical and deviatoric ($ \s_{dev}$ and $\bd_{dev}$) parts
\begin{equation}
\s=-p \fg I + \s_{dev}; \qquad \bd=\frac{1}{3} \dot{\vep}_0 \fg I +\bd_{dev}; \quad   \s_{dev} \fg : \fg I=\bd_{dev} \fg : \fg I=\fg 0,
       \label{sph-dev}
\end{equation}
since $\bd \fg : \fg I= \dot{J}/ J = \dot{\ln J}=\dot{\vep}_0 $ \cite{Lurie-90,levitas-book96}.
Substituting
 Eq.~\eqref{sph-dev}  in Eq.~\eqref{mb7}, we obtain
\begin{eqnarray}
    \stackrel{\nabla}\s_{J}=-\dot{p}\fg  I + \dot{\s}_{dev}  -\bw\cd \s_{dev} -\s_{dev} \cd\bw^{T}
    = \B :({1}/{3}\,  \dot{\vep}_0 \fg I +\bd_{dev}) ,
    \label{mb7dec}
\end{eqnarray}
where we took into account that $\bw\cd \fg I+\fg I \cd\bw^{T}=\fg 0$. Calculating trace of Eq.~\eqref{mb7dec},
we derive
\begin{eqnarray}
 - \dot{p}
    ={1}/{9}\, \dot{\vep}_0 \fg I \fg : \B \fg :  \fg I    + {1}/{3}\, \fg I \fg : \B \fg : \bd_{dev}   ={1}/{9}\, \dot{\vep}_0 \fg I \fg : \B \fg :  \fg I    + {1}/{3}\, \fg I \fg : \B \fg : \dot{\beps}_{dev}.
    \label{dotp}
\end{eqnarray}
One can define different bulk moduli using Eq.~\eqref{tagKalt}, i.e., $- \frac{\p p}{\p \vep_0}$, under various constraints on the strain or stress states.  One of the natural definitions is to fix deviatoric strain; then  from  Eq. ~\eqref{dotp} we obtain:
\begin{eqnarray}
K_V=- \frac{\p p}{\p \vep_0}|_{{\beps}_{dev}} =1/9\, \fg I \fg : \B \fg :  \fg I = 1/9\,  B_{iijj}= 1/9\, ( \bar{C}_{iijj}+ 3 p).
    \label{K-B}
\end{eqnarray}
Subscript $V$ is for Voigt, because relationship ~\eqref{K-B} between $K_V$ and $\B$ corresponds to the bulk modulus of the polycrystalline aggregate based on the Voigt averaging (i.e., for the same strain in all crystals ~\citep{Voigt-28,Hill-52,Mura-87}). The last equality was obtained using relationship ~\eqref{mB-co} between the tensors  $\B$ and $\bar{\C}$ for an arbitrary $\s$.
Thus, $K_V$ corresponds to the moduli $\fg B$, which is clear since it connects components of the Jaumann rate of the Cauchy stress and the deformation rate.

Note that generally, the bulk modulus in Eq. ~\eqref{K-B} can be determined for arbitrary strain or stress states (which contribute to the definition of $\B$); the small strain increment should be pure volumetric (isotropic) only. In particular, the current  state can be an isotropically
strained material with respect to stress-free state or material under hydrostatic pressure $p$, which produces a generally anisotropic strain.
In particular, let one impose in atomistic simulations $\F=J^{1/3} \I$ and calculate energy $\psi(J)$.
Then from the power balance
\bey
\s: \bd dt = -p  d{\vep}_0 +  \s_{dev}:\bd_{dev} dt = J^{-1} d\psi,
 \label{power-0}
\eey
where we took into account Eq.~\eqref{sph-dev}, $\bd_{dev}= \fg 0$ implies
\bey
 -p  d{\vep}_0 =-p J^{-1} dJ = J^{-1} d\psi (J) \quad \rightarrow \quad p=-\frac{\p \psi (J)}{\p J}|_{\F=J^{1/3} \I}.
 \label{power-0-a}
\eey
Then Eqs. ~\eqref{K-B} and  ~\eqref{power-0-a} imply
 that the same relationships like in Section  \ref{hydro} are applicable:
\bey
 && {K}_V=- \frac{\p p}{\p \vep_0}|_{\F=J^{1/3} \I}=
-J \frac{\p p}{\p J}|_{\F=J^{1/3} \I}=J \frac{\p^2 \psi}{\p J^2}|_{\F=J^{1/3} \I}=
\nonumber\\
&& - V \frac{\p p}{\p V}|_{\F=J^{1/3} \I}= V^2 \frac{\p^2{\psi_c}}{\p V^2}|_{\F=J^{1/3} \I}.
       \label{K-nondev}
\eey

{\it Compressibility.}
Let us introduce the fourth-rank compliance tensor $\la$, inverse to $\B$, by relationship $\la  : \B=\fg I^4_s$  ($\lambda_{ijmn} B_{nmlk}=\frac{1}{2} (\delta_{il}\delta_{jk}+\delta_{ik}\delta_{jl})$).  Then producing double contraction of $\la$ and both sides of Eq.~\eqref{mb7dec}, we obtain
\begin{eqnarray}
  {1}/{3}\,  \dot{\vep}_0 \fg I +\bd_{dev}  =
    \la \fg :\left( -\dot{p}\fg  I +    \stackrel{\nabla}\s_{dev,J} \right).
    \label{mb7dec+}
\end{eqnarray}
Trace of Eq.~\eqref{mb7dec+} has the form
\begin{eqnarray}
 \dot{\vep}_0  =
 - \fg I \fg :   \la \fg : \fg  I  \dot{p}+  \fg I \fg :   \la \fg :   \stackrel{\nabla}\s_{dev,J}.
    \label{mb7dec+tr}
\end{eqnarray}
Similar to the bulk moduli, one can define different compressibilities  using Eq.~\eqref{tag-k}, i.e.,
$- \frac{\p \vep_0}{\p p} $, under various constraints on the strain or stress states.  One of the natural definitions is to impose $ \stackrel{\nabla}\s_{dev,J}=\fg 0$, i.e., to fix deviatoric stress $\s_{dev}$ and rotations $\om$; then  from  Eq. ~\eqref{mb7dec+tr} we  obtain the bulk compliance (compressibility):
\begin{eqnarray}
k_R= - \frac{\p \vep_0}{\p p }|_{\s_{dev}} =  \fg I \fg :   \la \fg : \fg  I  = \lambda_{iijj}.
    \label{k-S}
\end{eqnarray}
Here subscript $R$  is for Reuss because relationship ~\eqref{k-S} between $k_R$ and $\la$ corresponds to the compressibility of the polycrystalline aggregate based on the Reuss averaging (i.e., for the same stress in all crystals) ~\citep{Reuss-29,Mura-87}; subscript $\om$ for fixed rotation is omitted for brevity but should always be kept in mind.
Similar to the bulk modulus,  generally, the compressibility in Eq. ~\eqref{k-S} can be determined for arbitrary strain or stress states (which contribute to the definition of $\la$); the small stress increment should be pure hydrostatic only. In particular, the current  state can be an isotropically
strained material with respect to the stress-free state or material under hydrostatic pressure $p$, which generally produces anisotropic strain.

Note that generally
$K_V \neq  {1}/{ k_R}$,
because derivatives $\frac{\p p}{\p \vep_0}|_{{\beps}_{dev}}$ and $ \frac{\p \vep_0}{\p p }|_{\s_{dev}}$ are evaluated while fixing different parameters that do not correspond to each other. It is known that $K_V=1/k_R$ for cubic crystals and isotropic materials
only, because for them fixation of small deviatoric  strain and stress is equivalent.

One can define alternative bulk modulus $K_R$ and compressibility $k_V$ by equations
\begin{eqnarray}
K_R :=- \frac{\p p}{\p \vep_0}|_{\s_{dev}} =  \frac{1}{ k_R}=  \frac{1}{ \lambda_{iijj}}; \qquad k_V:= - \frac{\p \vep_0}{\p p }|_{\beps_{dev}}=  \frac{1}{ K_V}=  \frac{9}{ B_{iijj}}.
    \label{mixed-k-S}
\end{eqnarray}
In particular, if for the entire straining $\s_{dev}=\fg 0$ and loading is hydrostatic, then
energy can be calculated for hydrostatic loading as  $\psi (J)$, and
Eq.  ~\eqref{power-0-a} implies
\bey
-p J^{-1} dJ = J^{-1} d\psi (J) \quad \rightarrow \quad p=-\frac{\p \psi (J)}{\p J}|_{\s_{dev}=  0}.
 \label{power-0-b}
\eey
Then again, the same relationships like in Section  \ref{hydro} are applicable:
\bey
{K}_R=- \frac{\p p}{\p \vep_0}|_{\s_{dev}= 0}=
-J \frac{\p p}{\p J}|_{\s_{dev}=  0}=J \frac{\p^2 \psi}{\p J^2}|_{\s_{dev}=  0}=
- V \frac{\p p}{\p V}|_{\s_{dev}=  0}= V^2 \frac{\p^2{\psi_c}}{\p V^2}|_{\s_{dev}=0}.
       \label{K-hydros}
\eey
To summarize, there is no unique bulk modulus and compressibility under general stress-strain states; they can be defined by derivatives
$- \frac{\p p}{\p \vep_0}$ and $- \frac{\p \vep_0}{\p p}$, respectively, under various constraints on the strain or stress states, which should be very clearly stated. For one of them, $K_V=1/k_V$, we fix deviatoric strain, and for another, $K_R=1/k_R$, we fix deviatoric stress. Generally, the bulk modulus and compressibility  can be determined for arbitrary strain or stress states (which contribute to the definition of $\B$).  For $K_V=1/k_V$ small strain increment should be pure volumetric (isotropic) only and
for $K_R=1/k_R$ small stress increment should be pure hydrostatic pressure only. If the entire loading corresponds to isotropic straining or  hydrostatic loading, then two different energy functions $\psi (J)$ can be defined, and bulk modulus for both loadings can be in addition defined in terms of the second derivative of energy. While there is no averaging in the definition of the bulk modulus and compressibility for a single crystal, the expression for $K_V=1/k_V$ in terms of components $B_{ij}$ coincides with the Voigt average for polycrystal, and  expression for $k_R=1/K_R$ in terms of components $\lambda_{ij}$ coincides with the Reuss average. Neglecting the difference between the two types of bulk moduli/compressibilities may lead to quantitative misinterpretation of experiments and qualitative contradictions, which will be illustrated in Section \ref{Consist-bulk}.

\section{APPROXIMATION FOR SMALL DISTORTIONS WITH RESPECT TO AN INTERMEDIATE CONFIGURATION\label{small-dist}}
\subsection{Elastic energy\label{elastic energy}}

Using Eqs.~\eqref{Esall} and ~\eqref{tag508a-0},  for the general case of finite distortions $\be
= \F-\I = \beps + \om$
\begin{eqnarray}
\E=\frac{1}{2}(\F^T \cd \F-\I)= \beps +\frac{1}{2} \be^T \cd \be =
 \beps+ \frac{1}{2} \left( \beps \cd \beps + 2(\beps \cd \om)_s + \om^T \cd \om \right). \qquad
\label{Esall-1}
\end{eqnarray}
The   energy per unit volume in the intermediate configuration $  \Omega_*  $ in the quadratic in $\E$ approximation is
\begin{equation}
\begin{split}
\psi =  \psi_0 /J_*=\psi  (0) + \s_* :\E+\frac{1}{2}\E:\C:\E. \,
    \label{tag504a-in}
\end{split}
\end{equation}
The corresponding second Piola-Kirchhoff stress is
\begin{equation}
    \T=\s_* +\C:\E.
    \label{tag504-1d}
\end{equation}
For small distortions $\be$, substituting Eq.~\eqref{Esall-1}
in Eq.~\eqref{tag504a-in}, we obtain in quadratic in distortions approximation
\begin{eqnarray}
  \psi & =\psi  (0) + \s_* : \left(  \beps +\frac{1}{2} \be^T \cd \be \right) +\frac{1}{2} \be^T:\C:\be^T;
    \label{tag504a-T}
\end{eqnarray}
\begin{eqnarray}
  \psi & =\psi  (0) + \sigma_{*ij}  \left( \epsilon_{ji} +\frac{1}{2} \beta_{kj} \beta_{ki} \right) +\frac{1}{2} {C}_{ijkl} \beta_{ij} \beta_{kl},
    \label{tag504a-T-com}
\end{eqnarray}
where $\frac{1}{2} \be^T:\C:\be^T= \frac{1}{2} \beps:\C:\beps$ due to symmetry of $\C$, but we will keep the expression with $\be^T$ for a while in order to express the final quadratic form  like $\frac{1}{2} \tilde{C}_{ijkl} \beta_{ij} \beta_{kl}$ with moduli $\tilde{C}_{ijkl} $ to be determined.

The  second Piola-Kirchhoff stress in linear in distortions approximation reduces to
\begin{equation}
    \T=\s_* +\C:\beps .
    \label{tag504-1ds}
\end{equation}
We transform
\begin{eqnarray}
 \s_* :  \be^T \cd \be = \be \cd \s_* :  \be^T=  \be^T \fg : \left( \fg I^4_t \cd\s_* \right) :  \be^T;
    \label{tag504a-Tt}
\end{eqnarray}
 \begin{eqnarray}
\beta_{ij}  I^4_{t,ijkl}  \sigma_{*lm} \beta_{km}= \sigma_{*lm} \delta_{ik} \delta_{jl}\beta_{ij}\beta_{km}=
 \sigma_{*jm} \delta_{ik} \beta_{ij}\beta_{km}
 =
   \sigma_{*jl}\delta_{ik} \beta_{ij} \beta_{kl}.
    \label{tag504a-TI}
\end{eqnarray}
or more straightforwardly
$ \sigma_{*ij}\beta_{kj} \beta_{ki}=    \sigma_{*ij} \delta_{mk}\beta_{kj} \beta_{mi} =  \sigma_{*lj} \delta_{ki} \beta_{ij} \beta_{kl}.$
Then,
\begin{eqnarray}
 & &\psi =\psi  (0) + \s_* : \beps   + \frac{1}{2}  \be^T: \tilde{\C}: \be^T=
 \psi  (0) +  \sigma_{*ij} \epsilon_{ji}   + \frac{1}{2}  \tilde{C}_{ijkl} \beta_{ij} \beta_{kl};
      \nonumber\\
 && \tilde{\C}:= \C +  \fg I^4_t \cd \s_*; \qquad  \tilde{C}_{ijkl}:=  C_{ijkl}+  \sigma_{*lj} \delta_{ki}.
    \label{tag504a-Ta}
\end{eqnarray}
The tensor $\tilde{C}_{ijkl}$ is not symmetric with respect to $i \leftrightarrow j$  and $k \leftrightarrow l$ and does not possess the Voigt  symmetry, but it is
symmetric in exchange between  pairs $ij$ and $kl$.
 Eq.~\eqref{tag504a-Ta} was suggested in ~\citep{Huang-50}. For the intermediate configuration infinitesimally close to the current one, we have $\bar{\C} \simeq \C$ and $\s_* \simeq \s$. Then it follows from the comparison of   Eqs.~\eqref{mB}  and ~\eqref{tag504a-Ta} that

 \begin{eqnarray}
\B := \tilde{\C}-  \I^4_t \cd \s -\s\I + \s\cd\I_{s}^{4}+(\I_s^{4}\cd\s)^{T}=  \tilde{\C} -\s\I + \s\cd\I_{s}^{4}+(\I_{as}^{4}\cd\s)^{T};
 \label{mB-t}
\end{eqnarray}
\begin{eqnarray}
    B_{ijkl} := \tilde{C}_{ijkl} - \sigma_{ij}\delta_{kl} +\frac{1}{2} (\sigma_{il}\delta_{jk} + \sigma_{ik}\delta_{jl} +
     \sigma_{kj}\delta_{li}-  \sigma_{lj}\delta_{ki}).
      \label{mB-co-t}
     \end{eqnarray}
Note that if higher-order terms in $\E$ will be utilized in Eq.~\eqref{tag504a-in}, stress $\s_*$ will not contribute to the higher-order terms in $\be$, because $\E$ does not contain higher than the second-order terms in $\be$.
 Decomposing $\be$ into $\beps$ and $\om$ in Eq.~\eqref{Esall-1}, we obtain 
 (see Appendix B)
\begin{eqnarray}
 & &\psi =\psi  (0) + \s_* : \beps   + \frac{1}{2}
  \beps :  {\C}^{\eps \eps} : \beps 
  +
    \om \fg  :   {\C}^{ \om \eps} \fg :  \beps +  \frac{1}{2}   \om :   {\C}^{ \om \om}:  \om^T;
\label{504a-Ta}\\
&& \C^{\eps \eps} :={\C} + \fg I^4_s  \cd \s_* :  \fg I^4_s; \qquad    
{\C}^{ \om \eps}:=  \fg I^4_{as}  \cd \s_* :  \fg I^4_s  ;
\qquad {\C}^{ \om \om}:= \fg I^4_{as}   \cd \s_* : \fg I^4_{as};
  \nonumber
\end{eqnarray}
\begin{eqnarray}
 && \psi =\psi  (0) + \sigma_{*ij} \epsilon_{ji}    + \frac{1}{2}
C^{\eps \eps}_{ijkl}  \eps_{ij}  \eps_{kl} 
 +
 C^{ \ome\eps}_{ijkl}   \eps_{kl}  \ome_{ji}  +  \frac{1}{2}   C^{ \ome\ome}_{ijkl}   \ome_{kl}  \ome_{ji};
\nonumber\\
& &C^{\eps \eps}_{ijkl}  := C_{ijkl} +\frac{1}{4}
(\sigma_{*lj} \delta_{ki} +\sigma_{*li} \delta_{kj} + \sigma_{*jk} \delta_{li}+\sigma_{*ik} \delta_{lj});
  \nonumber\\
& &C^{ \ome\eps}_{ijkl} := \frac{1}{4}
(\sigma_{*li} \delta_{kj} -\sigma_{*lj} \delta_{ki} - \sigma_{*jk} \delta_{li}+\sigma_{*ik} \delta_{lj});
  \nonumber\\
 & & C^{ \ome\ome}_{ijkl}:=\frac{1}{4}
(\sigma_{*li} \delta_{kj}-\sigma_{*lj} \delta_{ki} + \sigma_{*jk} \delta_{li}-\sigma_{*ik} \delta_{lj}).
 \label{504a-TaC}
  \end{eqnarray}
  The tensor ${\C}^{\eps \eps}$ possesses full Voigt symmetry. The tensor $C^{ \ome\eps}_{ijkl}$ is symmetric in $k \leftrightarrow l$ and antisymmetric in $i \leftrightarrow j$. The tensor $ C^{ \ome\ome}_{ijkl}$ is antisymmetric in $i \leftrightarrow j$ and  $k \leftrightarrow l$, and invariant under exchange of pairs $(ij) \leftrightarrow (lk)$,
 $(kl) \leftrightarrow (ji)$,   $(ij) \leftrightarrow (kl)$.


A delicate moment in consistent keeping in Eq.~\eqref{tag504a-T} all quadratic in distortions terms is that in the linear in $ \E $ term, we keep all quadratic terms, but in quadratic in $ \E $ term, we retain the linear term $ \E= \beps $ only. This modifies elastic moduli $ \C $ by $ \fg I^4_s  \cd \s_* :  \fg I^4_s$ and also shows that the  energy depends on the small rotations $ \om $. Small rotations may be related to $ \beps $ for specific loadings (e.g., for simple shear), thus also changing elastic moduli.

As a general conclusion, it follows from Eqs.~\eqref{tag504-1ds}, ~\eqref{b10-int-1},
and ~\eqref{504a-Ta} that
\begin{eqnarray}
&& \T\neq  \frac{\p \psi}{\p \beps}; \qquad    \C \neq  \frac{\p^2 \psi}{\p \beps \p \beps};
    \qquad
 \s \neq  \frac{\p \psi}{\p \beps};
   \qquad   \s \neq  \frac{1}{J}\frac{\p \psi}{\p \beps};
   \qquad
  \B \neq  \frac{\p^2 \psi}{\p \beps \p \beps}
,
    \label{noteq}
\end{eqnarray}
 despite that this is expected from small strain theory.

In many works (e.g., ~\citep{Mehl-etal-1990,Fastetal-95,Marcus-01}), where the  effect of initial stresses and rotations is neglected,  the energy under the small strain increment in $\Omega_*$ is approximated like in traditional small strain theory, i.e.
\begin{eqnarray}
\psi - \psi  (0)\simeq  \frac{1}{2} \beps \fg : \C \fg : \beps.
\label{psi-wr}
\end{eqnarray}
In atomistic simulations, the energy is calculated for different distortions $\be$ and then is approximated by quadratic function Eq.~\eqref{psi-wr}. If these evaluations are performed under initial stresses and with small rotations, then all neglected terms in Eq. \eqref{504a-Ta} are attributed to $\C$ and produce significant error in $\C$. This will be shown in the examples below.

 \subsection{Equation of motion and wave propagation in the pre-stressed solids\label{wave-ch}}
It is  shown, e.g.,  in  ~\citep{Huang-50} that the Lagrangian equations of motion is
 \begin{eqnarray}
\rho { \frac{\p^2  u_i}{\p t^2}}= \tilde{C}_{ijkl}^s \frac{\p^2  u_k}{\p r_{*j} \p r_{*l}}; \qquad
 \tilde{C}_{ijkl}^s:=\frac{1}{2} ( \tilde{C}_{ijkl}+ \tilde{C}_{ilkj}),
  \label{motion}
\end{eqnarray}
 i. e., it utilizes the same tensor $ \tilde{\C}$ like in  Eq.~\eqref{tag504a-Ta} for the energy but symmetrized with respect to $j$ and $l$, because the second derivative in Eq.~\eqref{motion} is symmetric with respect to $j$ and $l$. Here $u_i$ are components of the displacement vector $\fg u = \fg r - \fg r_*$. A similar equation of motion was derived in
 ~\citep{wallace-67} in a different way. While ~\citep{Huang-50} gave a pure mechanical interpretation of $\tilde{\C}$, ~\citep{wallace-67} has generalized it for thermoelasticity.

 Eq.~\eqref{motion} results in the following plane wave propagation equation ~\citep{Huang-50,wallace-67}
 \begin{eqnarray}
\rho v_w^2  u_i = \tilde{C}_{ijkl}^s k_j k_l u_k =L_{ik }u_k; \qquad  L_{ik }:= \tilde{C}_{ijkl}^s k_j k_l,
  \label{wave}
\end{eqnarray}
where $v_w$ is the wave velocity, $k_i$ are the components of the unit vector in the direction of wave propagation, and $L_{ik }$ is the propagation matrix. As one of the known stability conditions, the propagation matrix $L_{ik }$ should be a positive definite for all possible $k_i$ to guarantee possibility of wave propagation in any direction, which shows an additional importance of the tensor $\tilde{\C}$. Eq.~\eqref{wave} shows the  possibility to determine
 the tensor $\tilde{\C}^s$  from the wave propagation experiment, while the antisymmetric in $j$ and $l$ part of $\tilde{C}_{ijkl}$
can be determined from the expression for elastic energy only.

\subsection{Simple shear\label{sec-shear}}

\begin{figure}[htp]
\centering
\includegraphics[width=0.45\textwidth]{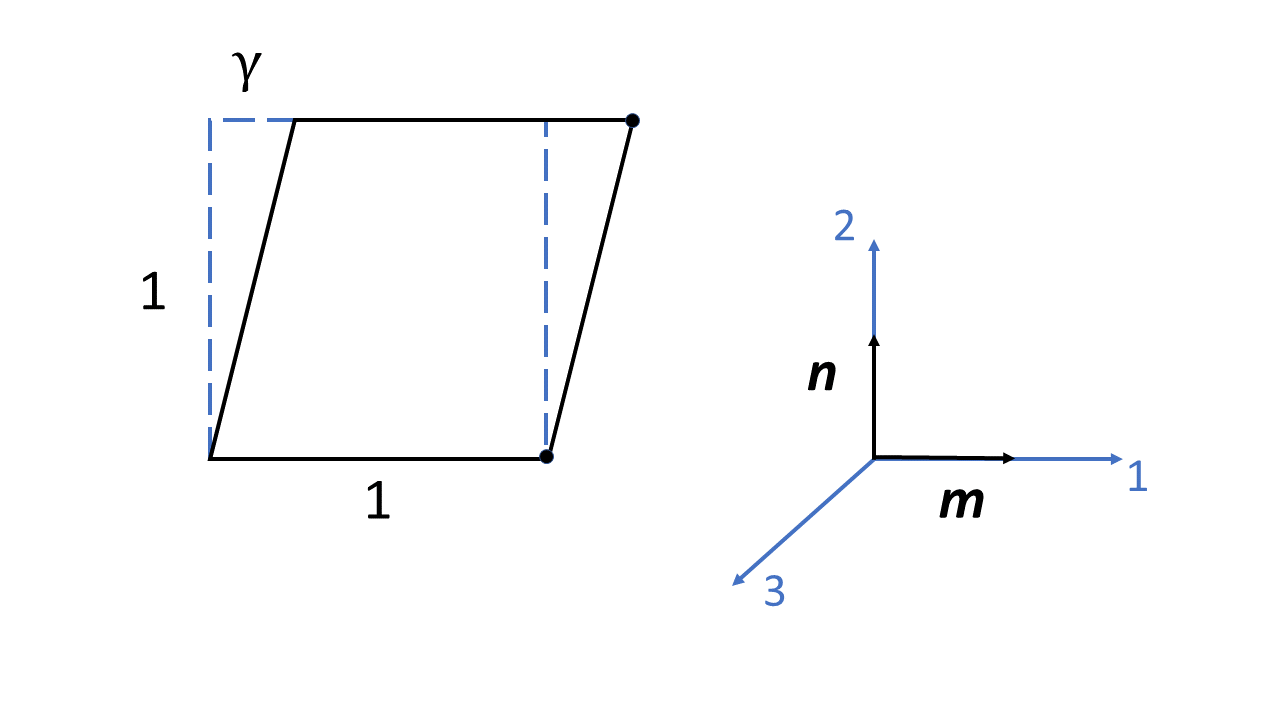}
\caption{ Simple shear $ \F=\fg I+\gamma \m \n$ with the shear strain $\gamma$ at the plane with the unit normal $\n$ along the unit direction $\m$. Crystal lattice of any symmetry can be oriented arbitrarily.
\label{shear}}
\end{figure}
\noindent
Let us consider  a simple shear as an example:
\begin{eqnarray}
  &&
 \F- \fg I= \be=\gamma \m \n =
  \gamma   \begin{pmatrix}
  0 & 1 & 0 \\
  0 & 0 & 0 \\
  0 & 0 & 0 \\
 \end{pmatrix};
 \nonumber\\
&&
\E=\frac{1}{2} (\F^T \cd \F -\fg I)= \frac{1}{2} \gamma  (\m\n + \n\m +\gamma \n\n) =
 \frac{1}{2} \gamma   \begin{pmatrix}
  0 & 1 & 0 \\
  1 & \gamma & 0 \\
  0 & 0 & 0 \\
 \end{pmatrix}.
    \label{tag409sg}
\end{eqnarray}
Here, $\n$ is the unit normal to the shear plane, $\m$ is the unit vector in the shear direction, $\gamma$ is the shear strain,
and the matrices are presented in the coordinate system shown in Fig. \ref{shear}.
 Combining all terms from Appendix B, we obtain from Eq.~\eqref{504a-TaC} for the energy
\begin{eqnarray}
 & &\psi =\psi  (0) + \sigma_{*12} \epsilon_{12}    + \frac{1}{2}
\underbrace{(C_{1212}+ \frac{1}{4}   ( \sigma_{*11}+ \sigma_{*22}))}_{C_{1212}^{\eps\eps}}\gamma^2
 +
\underbrace{\frac{1}{4}   ( \sigma_{*22}- \sigma_{*11})}_{C^{ \ome\eps}_{2121}}\gamma^2
  +
  \nonumber\\
  & &\frac{1}{2} \gamma^2 \underbrace{\frac{1}{4}   ( \sigma_{*22}+ \sigma_{*11})}_{C^{ \ome\ome}_{1221} }
=\psi  (0) + \sigma_{*12} \epsilon_{12}    + \frac{1}{2} \underbrace{ (C_{1212}+\sigma_{*22})}_{C_{\psi}}\gamma^2
.
 \label{504a-TaCsh}
  \end{eqnarray}
The last equation in Eq.~\eqref{504a-TaCsh} can be obtained more quickly by substituting $\be$ from Eq.~\eqref{tag409sg} in Eq.~\eqref{tag504a-T}.
Note that from Eq.~\eqref{mB-co}
\begin{eqnarray}
    B_{1212} = {C}_{1212}  +\frac{1}{2} (\sigma_{*11}+  \sigma_{*22})=   C_{1212}^{\eps\eps} + C^{ \ome\ome}_{1221} \neq  C_{1212}^{\eps\eps}   \neq  C_{\psi}.
      \label{mB-co-sh}
     \end{eqnarray}

For the straining with $\be=\beps = \frac{1}{2} \gamma (\m \n +\n\m)$ and $\om=\fg 0$ (rotation-free shear),  the terms with  $C^{ \ome\eps}_{1212}$ and $C^{ \ome\ome}_{1212}$ disappear from the expression for $\psi$ and one obtains
\begin{eqnarray}
 & &\psi =\psi  (0) + \sigma_{*12} \epsilon_{12}    + \frac{1}{2}
\underbrace{(C_{1212}+ \frac{1}{4}   ( \sigma_{*11}+ \sigma_{*22}))}_{C_{1212}^{\eps\eps}}\gamma^2
 .
 \label{TaCsh}
  \end{eqnarray}
  Again, $  B_{1212}  \neq  C_{1212}^{\eps\eps} $.

The above example explicitly shows the effect of initial stress and small rotations on the energy and determination of the elastic moduli based on energy. If neglected, they produce errors in the determination of the elastic moduli. Thus, by comparing Eq.~\eqref{504a-TaCsh}  with the simplest theory in Eq.~\eqref{psi-wr}, we see that $C_{1212}$ is corrected by the term $\sigma_{*22}$.  However, when rotations are absent and Eq.~\eqref{TaCsh} is valid,
$C_{1212}$ is corrected by the term $  \frac{1}{4}   ( \sigma_{*11}+ \sigma_{*22})$.




\section{RELATIONSHIPS FOR THE INTERMEDIATE CONFIGURATION UNDER HYDROSTATIC PRESSURE\label{intermediate-config}}

In the treatment that follows, the deformed intermediate configuration under hydrostatic pressure $p$ is considered, i.e.,
\begin{equation}
\s_*= - p \I; \qquad  \sigma_{*ij}= -p \delta_{ij}.
\label{eq-p}
\end{equation}
This substitution can be made for all equations of the previous sections. However, this particular case also allows some qualitatively new results, which are not valid for the general preliminary stress tensor.

For single crystals, 
the symmetric deformation gradient tensor $ \F_* $ that produces  the intermediate configuration is described by the experimentally determined function
\begin{equation}
\F_*= \F_* (p),
\label{Esall-1s}
\end{equation}
which we will call a generalized equation of state.
While in the  experiment, depending on the way the crystal is fixed, generally nonsymmetric $\bar{\F}_*$ may be obtained and can be decomposed into the symmetric $\F_*$ and the orthogonal $\fg R$ tensors characterizing rigid-body rotation using polar decomposition  $\bar{\F}_*= \fg R\cd \F_*$; rotation should be excluded because it can be made arbitrary by rotating an observer. Relationship Eq.~\eqref{Esall-1s}, in particular, results in the equation of state
\begin{equation}
J_*=det \F_*= J_* (p).
\label{EOS}
\end{equation}
For the cubic crystal and isotropic polycrystalline aggregate, Eq.~\eqref{EOS} is the only scalar equation that follows from Eq.~\eqref{Esall-1s}. For lower lattice symmetries, the number of independent scalar equations in Eq.~\eqref{Esall-1s} is equal to the number of nonzero components of the transformational deformation gradient that transform the cubic lattice into the lattice of interest during martensitic phase transformation. The transformation matrices for such transformations are collected, e. g., in   ~\citep{zanziotto,Bhattacharya-book}.  For example, for tetragonal and hexagonal lattices, there are two independent equations for $F_{*11}=F_{*22}$ and $F_{*33}$; for orthorhombic lattices and monoclinic lattices (with the axis of monoclinic symmetry corresponding to $<100>_{cubic}$ direction) there are four independent equations ($F_{*13}=F_{*23}=0$ only); for
monoclinic lattices (with the axis of monoclinic symmetry corresponding to $<110>_{cubic}$ direction) and triclinic lattices, there are all six independent equations.

The elastic energies per unit volume in the reference configuration $  \Omega_0  $ and the intermediate configuration $  \Omega_*  $ are
\begin{equation}
\begin{split}
 \psi_0= \psi_0 (\E_0);     \qquad \psi= \psi_0 /J_*= \psi  (0) -p\I:\E+ \Tilde \psi (\E).
    \label{tag504a-0-p}
\end{split}
\end{equation}
Similar to the general initial stress, while   $p$ has the physical meaning of pressure, in the thermodynamic treatment  (e.g., application of thermodynamic laws to derive elasticity rules or temperature evolution equation), $p$ should be treated just  as a constant in the linear term in Eq. ~\eqref{tag504a-0-p} assuming the fixed intermediate configuration.   The intermediate configuration and $p$ can be varied in the final equations. In particular, in the thermodynamic treatment, pressure-dependent elastic moduli should not be differentiated with respect to pressure.

\subsection{Stresses and stress rate\label{stress-rate}}
The second Piola-Kirchhoff stress in the reference $ \T_0 $ and intermediate $ \T $ configurations are defined as follows:
\begin{eqnarray}
 &&  {\T_0}=  {J_0}\ {\F_0}^{-1}\cd \s \cd {\F_0}^{T-1} = \frac{\p\psi}{\p\E_0};
       \qquad \T=J\F^{-1}\cd\s\cd\F^{T-1}=
       \frac{\p\psi}{\p \E}=-p\I+  \frac{\p \Tilde{\psi}}{\p\E}; \quad
       \nonumber\\
    &&   \T_0= J_*{\F}^{-1}_* \cd \T \cd {\F}^{T-1}_*; \qquad  \T= J_*^{-1}{\F}_* \cd \T_0 \cd {\F}^{T}_*,
    \label{tag505-p}
\end{eqnarray}
where $ \s $ is Cauchy (true) stress. It is clear that for $ \F=\I $ one has $ \s=\T= -p \I $, which justifies presentation Eq.~\eqref{tag504a-0-p} for $ \psi $.
In the component form
\begin{eqnarray}
 && T_{0,ij}=  {J_0} F_{0,ik}^{-1}\sigma_{kl}F_{0,jl}^{-1} = \frac{\p\psi_0}{\p E_{0,ij}};
       \qquad  T_{ij}=  {J} F_{ik}^{-1}\sigma_{kl}F_{jl}^{-1} = \frac{\p\psi}{\p E_{ij}}=-p \delta_{ij}+\frac{\p \Tilde{\psi}}{\p E_{ij}}; \quad
       \nonumber\\
    &&   T_{*ij}=  {J_*} F_{*ik}^{-1}T_{kl}F_{*jl}^{-1} ; \qquad   T_{ij}=  {J_*^{-1}} {F_{*ik}}T_{0,kl}{F_{*jl}}.
    \label{tag505C-p}
\end{eqnarray}
It follows from  Eq.~\eqref{tag505-p}
\begin{eqnarray}
 &&  \s= {J_0}^{-1}{\F_0}\cd {\T_0}\cd {\F_0}^{T}= {J_0}^{-1}{\F_0}\cd \frac{\p\psi_0}{\p\E_0}\cd {\F_0}^{T}  \quad
 \rightarrow
 \nonumber\\
&&  @ \F=\I: \qquad   -p \I = {J_*}^{-1}{\F}_{*}\cd \frac{\p  {\psi_0}}{\p  {\E_0}}\mid_{\E_0=\E_*}  \cd {\F^T}_* .
         \label{tag505s-p}
\end{eqnarray}
For general anisotropic crystals Eq.~\eqref{tag505s-p} is the inverse of   Eq.~\eqref{Esall-1s}, i.e., the generalized equation of state.
In the component form
\begin{eqnarray}
 && \sigma_{ij}= J_0^{-1}F_{0,ik}T_{0,kl}F_{0,jl}= J_0^{-1}F_{0,ik}\frac{\p\psi_0}{\p E_{0,kl}} F_{0,jl}
     \rightarrow
 \nonumber\\
&&   @ F_{ij}=\delta_{ij}: \qquad   -p  \delta_{ij} = J_*^{-1}{F}_{*ik} {F}_{*jl} \frac{\p  \psi_0}{\p  E_{0,kl}}\mid_{E_{0,ab}=E_{*ab}} .
          \label{tag505sC-p}
\end{eqnarray}
Also,
\begin{eqnarray}
 &&  \s= {J}^{-1}{\F}\cd {\T}\cd {\F}^{T}= {J}^{-1}{\F}\cd \frac{\p\psi}{\p\E}\cd {\F}^{T}
=
  -p {J}^{-1}{\F}\cd {\F}^{T}   + {J}^{-1}{\F}\cd \frac{\p \tilde{\psi}}{\p\E}\cd {\F}^{T}.
         \label{tag505si-p}
\end{eqnarray}
\begin{eqnarray}
  \sigma_{ij}=  J^{-1}F_{ik}T_{kl}F_{jl}= J^{-1}F_{ik}\frac{\p\psi}{\p E_{kl}} F_{jl}
  =
 -  p J^{-1}F_{ik} F_{jk} + J^{-1}F_{ik}\frac{\p \tilde\psi}{\p E_{kl}} F_{jl}
  .
          \label{tag505sC-p1}
\end{eqnarray}
All equations from Section \ref{elastic-moduli} for elastic moduli remain unchanged.
All equations from Section \ref{rate-Cauchy} down to Eq. ~\eqref{Tr}  remain unchanged because they do not use an intermediate configuration. If we consider in Eqs. ~\eqref{b6} and ~\eqref{b6-c} stress in the current configuration $\s=-p \fg I$, these equations simplify to
\begin{eqnarray}
 & &  \stackrel{\nabla}\s_{J}=\dot{\s}=
    \bar{\C}:\bd+ p (\I \I:\bd- 2  \bd )= (  \bar{\C}+ p (\I \I - 2 \I^4_s)   :\bd =  \B : \bd;
        \nonumber\\
 & &  \dot{\s}= \B : \bd;  \qquad \B:= \bar{\C}+ p (\I \I - 2 \I^4_s);
    \label{b6-p}
\end{eqnarray}
\begin{eqnarray}
&	&\stackrel{\nabla}\sigma_{ij}=\dot{\sigma}_{ij}=
	\bar{C}_{ijkl}d_{lk}+p (\delta_{ij}\delta_{mn}d_{nm}- 2 d_{ij})=
( \bar{C}_{ijkl}d_{lk}+p (\delta_{ij}  \delta_{kl}-   \delta_{ik}\delta_{jl}-  \delta_{il}\delta_{jk} )  d_{kl};     \nonumber\\
 & &\dot{\sigma}_{ij}= B_{ijkl} d_{kl};
 \qquad B_{ijkl}:= \bar{C}_{ijkl}d_{lk}+p (\delta_{ij}  \delta_{kl}-   \delta_{ik}\delta_{jl}-   \delta_{il}\delta_{jk} ).
   \label{b6-c-p}
   \end{eqnarray}
We took into account that $-(\bw\cd\s+\s\cd\bw^{T})= p(\bw+\bw^{T} )=0$, i.e., the spin tensor does not contribute to the stress rate for initially hydrostatically loaded crystal.  Based on the definition, we see that in contrast to the general stress state at time $t$, for the initial hydrostatic loading tensor $B_{ijkl} = B_{klij}$. Thus, for this case, the  tensor $\B$ has full Voigt symmetry. Also, hydrostatic pressure does not change the symmetry of the stress-free crystal lattice, which significantly simplifies analysis.
Eq.~\eqref{b6-p}
can be integrated for small time increment:
\begin{eqnarray}
  {\s}(t+\Delta t) = -p (t) \I + \B: \dot{\beps} \Delta t;
  \quad
   {\sigma}_{ij}(t+\Delta t)=  -p (t) \delta_{ij}+  B_{ijkl}\dot{\epsilon}_{lk}  \Delta t.
    \label{b10-int-p}
\end{eqnarray}
The application of this equation is limited to the case when, for the next time step, stress at the previous time step is hydrostatic. This is the case if the entire loading is hydrostatic, which is of limited interest.
However, based on $ {\s}(t+\Delta t)$, one can update pressure $p(t+\Delta t) $ and using Eq.~\eqref{Esall-1s},  update $\F_* (t+\Delta t)$ and the intermediate configuration. This configuration under updated pressure can be used as
a new intermediate configuration $\Omega_*$ for the next time step and strain increment $\dot{\beps} \Delta t$ should be calculated with respect to this configuration.

If an  intermediate configuration $\Omega_*$ is updated, then strain $\beps $ is evaluated with respect to updated configuration $\Omega_*$ without remembering previous values of $\beps$, Eq.~\eqref{b10-int-p}
can be presented in terms of  $\beps=\Delta \beps=\dot{\beps} \Delta t$
\begin{eqnarray}
  {\s} = -p (\F_*) \I + \B: \beps; \qquad     \beps= \F_*^{-1T}(p){\cd}(\E_0- \E_*(p)) {\cd} \F_*^{-1}(p);
    \label{b10-int-p-a}
\end{eqnarray}
\begin{eqnarray}
   {\sigma}_{ij}=  -p (\F_{*mn}) \delta_{ij}+  B_{ijkl}\epsilon_{lk};
   \qquad
   \epsilon_{lk} =    F_{*ik}^{-1}(p)  (  E_{0,ij}-E_{*ij}(p)) F_{*jl}^{-1}(p)
,
    \label{b10com-int-pa}
\end{eqnarray}
where Eqs.~\eqref{Esall} and ~\eqref{EsallC} were  utilized. Both equations are nonlinearly connected, since $p$ depends on $\F_*$ in the first equation and $\F_*$ depends on $p$ in the second equation, and should be solved iteratively, if $\E_0$ is prescribed.
Of course, one can use Eq.~\eqref{b10-int} with general $\s$ dependent  expression for the tensor $\B$ in Eq.~\eqref{mB}. However, the advantage of utilizing Eq.~\eqref{b10-int-p}
is that
  the tensor $\B$ has full Voigt symmetry and the symmetry of the stress-free crystal lattice, which significantly simplifies analysis.

\subsection{Consistency conditions  utilizing single crystal data  \label{Consistency}}
Here, we will find which  constraints on the elastic compliances and moduli are imposed from the experimental generalized equation of state ~\eqref{Esall-1s} under hydrostatic loading, which is usually obtained using an x-ray diffraction measurement of the change of lattice parameters under hydrostatic pressure; we will call them  the consistency conditions.
We will consider two cases: (a) When the generalized equation of state is obtained from the experiment on a single crystal and (b)
when it is extracted from the polycrystalline sample.


One of the universal consistency conditions for any crystal symmetry can be immediately found from the equation of state ~\eqref{EOS} when the intermediate configuration coincides with the current one: it defines the bulk modulus $K$ and compliance $k$, which are connected to elastic moduli and compliances by  Eq.~\eqref{K-B} and Eq.~\eqref{k-S}, respectively:
\begin{eqnarray}
&&   {K}_V=-J_* \frac{\p p}{\p J_*}|_{\beps_{dev}}= - \frac{\p p}{\p \ln J_*}|_{\beps_{dev}} =1/9\, \fg I \fg : \B \fg :  \fg I = 1/9\,  B_{iijj};
     \label{EOS-con-1}
\end{eqnarray}
\begin{eqnarray}
 &&  k_R= - \frac{\p \ln J_*}{\p p } |_{\s_{dev}}=  \fg I \fg :   \la \fg : \fg  I  = \lambda_{iijj}.
       \label{EOS-con}
\end{eqnarray}
Eq.~\eqref{EOS-con-1} or ~\eqref{EOS-con} is routinely used to check correctness of the determined elastic moduli or helps to find $n$ elastic moduli from $n-1$ experiments or atomistic simulations that do not involve $K$ or $k$.
Note that Eqs.~\eqref{EOS-con-1} and ~\eqref{EOS-con} are not equivalent.

Below we will derive all consistency conditions related to the generalized equation of state ~\eqref{Esall-1s} under hydrostatic loading.
When $\s$ represents hydrostatic state of stress, Eq.~\eqref{b6-p} reduces to the following equation:
\begin{eqnarray}
    -\dot{p} \I=  \B(p):\bd
     \label{tag510-1b}.
\end{eqnarray}
\subsubsection{Consistency conditions for elastic compliances\label{compliances}}
Inverting Eq.~\eqref{tag510-1b} by double contraction with elastic compliance $\la  (  p)$  ($\la  (  p) : \B(p) =\I^4_s$) gives:
\begin{eqnarray}
 \bd =  - \la  (  p) :  \I \dot{p}.
      \label{tag510-1c}
\end{eqnarray}
From the other side, it follows from Eq.~\eqref{Esall-1s}
\begin{equation}
 \bd_*: =  (\dot{\fg F_*} {\fg \cdot}{\fg F_*}^{- 1})_{s}=  \left(\frac{d \F_*(p)}{d p} {\fg \cdot}{\fg F}_*^{- 1} \right)_{s} \dot{p}.
\label{Esall-1s-a}
\end{equation}
Since under hydrostatic loading $ \bd = \bd_{*}$, by comparing Eqs. (\ref{tag510-1c}) and Eq. (\ref{Esall-1s-a}) , we obtain
\begin{equation}
 \la  (  p) :  \I =-  \left(\frac{d \F_*(p)}{d p} {\fg \cdot}{\fg F}_{*}^{- 1} \right)_{s}  \quad \rightarrow \quad
 \lambda_{ijkk}=-  \left(\frac{d {F_{*}}_{im}(p)}{d p} {F_{*}}_{mj}^{- 1} \right)_{s}
 .
\label{Esall-1s-la}
\end{equation}
Eq. (\ref{Esall-1s-la}), which we call a consistency condition for elastic compliances, imposes linear constraints on the elastic compliances coming from the experimental generalized equation of state ~\eqref{Esall-1s} under hydrostatic loading. The number of constraints is equal to the number
of independent strain components that appear under hydrostatic pressure for a given symmetry of a lattice. It  equals to the number of independent lattice parameters or components of the transformation strain tensor for phase transformation from the cubic to the given lattice.
Thus, it is two for hexagonal, trigonal, rhombohedral, and tetragonal lattices, three for orthorhombic, four for monoclinic,  and six for triclinic.

In particular, the trace of Eq. (\ref{Esall-1s-la}) in combination with the definition (\ref{k-S})  of  the bulk compliance $k_R$  leads to
\begin{eqnarray}
k_R=\lambda_{iijj}= \I : \la  (  p) :  \I =-  \frac{d \F_*(p)}{d p} :{\fg F}_{*}^{- 1} =- \frac{d {F_{*}}_{im}(p)}{d p} {F_{*}}_{mi}^{- 1}  .
    \label{k-S-cmp}
\end{eqnarray}
Since $dJ_*/J_*= d\ln J_* = d\F_* :{\fg F}_{*}^{- 1}= d {F_{*}}_{im} {F_{*}}_{mi}^{- 1} $ \cite{Lurie-90,levitas-book96}, Eq. (\ref{k-S-cmp}) is equivalent to Eq.~\eqref{EOS-con}.
For example, if $\F_*$ is a diagonal tensor (for orthorhombic lattices), then Eq. (\ref{Esall-1s-la}) takes the following  form
\begin{eqnarray}
 &&\lambda_{11kk}=-   \frac{1}{{F_{*}}_{11}}\frac{d {F_{*}}_{11}(p)}{d p}=
  -   \frac{d \ln \left[{F_{*}}_{11}(p)\right]}{d p} ;
 \quad
  \lambda_{22kk}=-   \frac{1}{{F_{*}}_{22}}\frac{d {F_{*}}_{22}(p)}{d p}=
  -   \frac{d \ln \left[{F_{*}}_{22}(p)\right]}{d p} ;
  \nonumber \\
 &&  \lambda_{33kk}=-   \frac{1}{{F_{*}}_{33}}\frac{d {F_{*}}_{33}(p)}{d p}=
  -   \frac{d \ln \left[{F_{*}}_{33}(p)\right]}{d p}
 .
\label{Esall-1s-la-1}
\end{eqnarray}
Recall, that  $\lambda_{11kk}$, $\lambda_{22kk}$, and $\lambda_{33kk}$ are linear compliances along the orthorhombic axes ~\citep{Nye-85}.
For hexagonal and tetragonal systems $ {F_{*}}_{11}= {F_{*}}_{22}$ and $\lambda_{11kk}=\lambda_{22kk}$, and nontrivial Eqs. (\ref{Esall-1s-la-1}) in more explicit form are
\begin{equation}
 \lambda_{1111}+ \lambda_{1122}+ \lambda_{1133}=-   \frac{d \ln \left[{F_{*}}_{11}(p)\right]}{d p} ;
  \quad
   \lambda_{ 33 11}+   \lambda_{3322}+  \lambda_{3333}=-    \frac{d \ln \left[{F_{*}}_{33}(p)\right]}{d p}
 .
\label{Esall-1s-la-1a}
\end{equation}
With Voigt designation, Eq. (\ref{Esall-1s-la-1a}) looks like
\begin{equation}
 \lambda_{11 }+ \lambda_{12}+ \lambda_{13}=-   \frac{d \ln \left[{F_{*}}_{1}(p)\right]}{d p} ;
  \quad
 2  \lambda_{13}+  \lambda_{33}=-    \frac{d \ln \left[{F_{*}}_{3}(p)\right]}{d p}
 ,
\label{Esall-1s-la-1a2}
\end{equation}
where symmetry $ \lambda_{32}= \lambda_{23}$ and $ \lambda_{23}= \lambda_{13}$ is taken into account.

\subsubsection{Consistency condition for elastic moduli\label{consistency-moduli}}
Let substitute in  Eq. (\ref{tag510-1b})
deformation rate $\bd_*$ for hydrostatic loading from  Eq. (\ref{Esall-1s-a}):  
\begin{eqnarray}
&&-\dot{p} \I=  \B (p):  \left(\frac{d \F_{*}(p)}{d p} {\fg \cdot}{\fg F}_{*}^{- 1} \right)_{s} \dot{p}  \quad \rightarrow \quad \B (p):  \left(\frac{d \F_*(p)}{d p} {\fg \cdot}{\fg F}_{*}^{- 1} \right)_s =- \I \quad \rightarrow \quad
\nonumber\\
&& B_{ijkl}(p)   \left(\frac{d {F_{*}}_{lm}(p)}{d p} {F_{*}}^{- 1}_{mk} \right)_s =-\delta_{ij},
      \label{tag510-1bd}
\end{eqnarray}
where symmetrization can be omitted due to symmetry of $B_{ijkl}$ in $k$ and $l$.
Eq. (\ref{tag510-1bd}) represents the desired linear constraints on elastic moduli.

As an example, for orthorhombic lattices Eq. (\ref{tag510-1bd}) simplifies to
\begin{eqnarray}
\B (p):  \frac{d \ln \left[ \F_*(p)\right]}{d p}=-\I;
 \qquad
B_{ijkk}(p)  \frac{d \ln \left[ {F_{*}}_{kk}(p)\right]}{d p}= - \delta_{ij} .
        \label{tag510-1bd-or}
\end{eqnarray}
For hexagonal and tetragonal crystals Eq. (\ref{tag510-1bd-or}) reduces to
\begin{eqnarray}
 [B_{11}(p) + B_{12}(p))]\frac{d \ln \left[ {F_{*}}_{11}(p)\right]}{d p}+ B_{13}(p)  \frac{d \ln \left[ {F_{*}}_{33}(p)\right]}{d p}=-1;
  \nonumber\\
2 B_{13}(p)  \frac{d \ln \left[ {F_{*}}_{11}(p)\right]}{d p}+ B_{33}(p)  \frac{d \ln \left[ {F_{*}}_{33}(p)\right]}{d p}=-1.
         \label{tag510-1bd-hx}
\end{eqnarray}
The difference between consistency conditions for compliance and moduli is that the former constrains the sum of some compliances, but the latter
constrains the weighted sum of elastic moduli. 

Let us express for comparison the constraints on compliances  Eq. (\ref{Esall-1s-la-1a2}) for hexagonal and tetragonal lattices in terms of elastic moduli.
We recollect relationships between $ \lambda_{ij}$ and $B_{ij}$ for hexagonal and tetragonal (classes $4mm$, $\bar{4}2m$, $42m$, and $4/mmm$) systems ~\citep{Nye-85}:
\begin{eqnarray}
&& \lambda_{11}+\lambda_{12}=B_{33}/B; \quad \lambda_{13}=-B_{13}/B; \quad
\lambda_{33}=(B_{11}+B_{12})/B; \quad
\nonumber\\
&& B=B_{33}(B_{11}+B_{12})-2B_{13}^2.
      \label{C-lam}
\end{eqnarray}
Substituting them in Eq. (\ref{Esall-1s-la-1a2}), we obtain consistency conditions for compliances expressed in terms of elastic moduli:
\begin{equation}
\frac{B_{33}-B_{13}}{B}=-   \frac{d \ln \left[{F_{*}}_{1}(p)\right]}{d p} ;
  \quad
\frac{B_{11}+B_{12}-2B_{13}}{B} =-    \frac{d \ln \left[{F_{*}}_{3}(p)\right]}{d p}
 .
\label{Esall-1s-C}
\end{equation}
They are  equivalent to Eq. (\ref{tag510-1bd-hx}) and can be obtained from Eq. (\ref{tag510-1bd-hx}) by solving them for
$ \frac{d \ln \left[{F_{*}}_{1}(p)\right]}{d p}$ and $ \frac{d \ln \left[{F_{*}}_{3}(p)\right]}{d p}$.
Due to non-linearity of these constraints in terms of  $B_{ij}$, their application is less convenient  than in Eq. (\ref{tag510-1bd-hx}) or Eq. (\ref{Esall-1s-la-1a2}) for $\lambda_{ij}$.

{\it Equivalence of the consistency conditions} for elastic moduli and compliances can be proven in the general case. Indeed, producing double contraction of  both sides of Eq. (\ref{Esall-1s-la}) with $\B$ we obtain
\bey
&& \B: \la  (  p) :  \I =\I^4_s : \I = \I =    - \B: \left(\frac{d \F_*(p)}{d p} {\fg \cdot}{\fg F}_{*}^{- 1} \right)_{s} ;
\nonumber\\
&& B_{ijab}  \lambda_{abkk}=I^4_{s,ijkk}= \delta_{ij} =  -  B_{ijab} \left(\frac{d {F_{*}}_{am}(p)}{d p} {F_{*}}_{mb}^{- 1} \right)_{s},
\label{El-1s-la}
\eey
which coincides with  Eq. (\ref{tag510-1bd}). We can also present Eq. (\ref{Esall-1s-la}) in terms of elastic moduli
\begin{equation}
 \B^{-1}  (  p) :  \I =-  \left(\frac{d \F_*(p)}{d p} {\fg \cdot}{\fg F}_{*}^{- 1} \right)_{s};  \qquad
B^{-1}_{ijkk}=-  \left(\frac{d {F_{*}}_{im}(p)}{d p} {F_{*}}_{mj}^{- 1} \right)_{s}
\label{all-1s-la}
\end{equation}
and Eq. (\ref{tag510-1bd}) in terms of  elastic compliances
\begin{eqnarray}
 \la^{-1} (p):  \left(\frac{d \F_*(p)}{d p} {\fg \cdot}{\fg F}_{*}^{- 1} \right)_s =- \I;  \qquad   \lambda^{-1}_{ijkl}(p)   \left(\frac{d {F_{*}}_{lm}(p)}{d p} {F_{*}}^{- 1}_{mk} \right)_s =-\delta_{ij}.
      \label{10-1bd}
\end{eqnarray}
Due to inversion, these are nonlinear constraints on the elastic moduli and compliances, which are less convenient than the original linear constraints (\ref{Esall-1s-la}) and (\ref{tag510-1bd}).

\subsubsection{Constraint involving the bulk modulus\label{Consist-bulk}}
As a particular case of Eq. (\ref{tag510-1bd}), let us try to derive the constraint related to the definition of the bulk modulus $K_V$ the way we did for bulk compliance, i.e. by finding trace of the consistency condition.  First, with the help of the definition of the deformation rate  (\ref{11-14-de}), when the current configuration coincides with the intermediate configuration,
$\bd_*= \left(\dot {\fg F}_* \cdot \fg F^{-1}_*\right)_s$ and its decomposition (\ref{sph-dev})
into spherical and deviatoric parts $\bd_*=\frac{1}{3} \dot{\vep}_{0*} \fg I +\bd_{dev*}$, we obtain
\begin{eqnarray}
&&  \left(\frac{d \F_*(p)}{d p} {\fg \cdot}{\fg F}_{*}^{- 1} \right)_s
= \bd_* \frac{dt}{dp}=(\frac{1}{3} \dot{\vep}_{0*} \fg I +\bd_{dev*})\frac{dt}{dp};
\nonumber\\
&&   \left(\frac{d {F_{*}}_{im}(p)}{d p} {F_{*}}^{- 1}_{mj} \right)_s=  d_{*ij} \frac{dt}{dp}=(\frac{1}{3} \dot{\vep}_{0*} \delta_{ij} +d_{dev*,ij})\frac{dt}{dp}.
      \label{10-1bd-r}
\end{eqnarray}

Then finding the trace of Eq. (\ref{tag510-1bd}), we derive
\begin{eqnarray}
&&- 3= \I : \B (p):  \left(\frac{1}{3} \dot{\vep}_{0*} \fg I +\bd_{dev*}\right)\frac{dt}{dp}
=\left(3 K_V \dot{\vep}_{0*}  +\I : \B (p): \bd_{dev*}\right)\frac{dt}{dp}
=
\nonumber\\
&&  3K_V \frac{d {\vep}_{0*}}{d p}|_{\s_{dev}}   +\I : \B (p):  \left(\frac{d \F_*(p)}{d p} {\fg \cdot}{\fg F}_{*}^{- 1} \right)_{dev} ;
\nonumber\\
&&
-3=    B_{iikl}(p) \left(\frac{1}{3} \dot{\vep}_{*0} \delta_{kl} +d_{dev*,kl} \right)\frac{dt}{dp}
=
\left(3 K_V \dot{\vep}_{0*}  +B_{iikl}(p)  d_{dev*,kl}\right)\frac{dt}{dp}=
\nonumber\\
&&
3 K_V \frac{d {\vep}_{0*}}{d p}|_{\s_{dev}} + B_{iikl}(p)
 \left(\frac{d {F_{*}}_{km}(p)}{d p} {F_{*}}^{- 1}_{ml} \right)_{dev}.
      \label{10-1bd-d}
\end{eqnarray}
We used definition  Eq. (\ref{K-B}) of the bulk modulus $K_V= 1/9\, \fg I \fg : \B \fg :  \fg I =1/9\,  B_{iijj}$ and also that pressure derivatives are evaluated at fixed $\s_{dev}$ and $\om$, like in the initial  Eq. (\ref{tag510-1bd}) .
With  $\vep_{0*}: ={\ln J_*}=det {\fg F}_{*}$ from Eq. (\ref{tagKalt}) for the current configuration coinciding with the intermediate configuration, Eq. (\ref{10-1bd-d}) can be presented in the form
\begin{eqnarray}
&&
K_V \frac{d \ln J_*}{dp}|_{\s_{dev}}  + \frac{1}{3}\I : \B (p):  \left(\frac{d \F_*(p)}{d p} {\fg \cdot}{\fg F}_{*}^{- 1} \right)_{dev}
=- 1 ;   
\nonumber\\
&&
K_V \frac{d \ln J_*}{dp}|_{\s_{dev}}+  \frac{1}{3}B_{iikl}(p) \left(\frac{d {F_{*}}_{km}(p)}{d p} {F_{*}}^{- 1}_{ml} \right)_{dev}
=- 1
.
      \label{10-1bd-r1}
\end{eqnarray}
Eq. (\ref{10-1bd-r1}) is clearly not equivalent to the constraint (\ref{EOS-con-1}) on the bulk modulus $K_V$, because it  explicitly depends on the deviatoric part
$\left(\frac{d \F_*(p)}{d p} {\fg \cdot}{\fg F}_{*}^{- 1} \right)_{dev} $.
However, this is noncontradictory because careful utilization of the definitions leads to
\begin{eqnarray}
&&
K_V \frac{d \ln J_*}{dp}|_{\s_{dev}}  = - \frac{dp}{d \ln J_*}|_{\beps_{dev}} \frac{d \ln J_*}{dp}|_{\s_{dev}}
\neq - 1
,
      \label{9-1bd-r}
\end{eqnarray}
because derivatives are evaluated for different and nonequivalent constraints. Thus,  Eq. (\ref{10-1bd-r1})  represents noncontradictory
constraint on  $\I : \B (p):  \left(\frac{d \F_*(p)}{d p} {\fg \cdot}{\fg F}_{*}^{- 1} \right)_{dev}$, which is part of  the general constraints  (\ref{tag510-1bd}).

If we would neglect this difference and  insert  $K_V \frac{d \ln J_*}{dp} = - \frac{dp}{d \ln J_*} \frac{d \ln J_*}{dp}=-1$ in   Eq. (\ref{10-1bd-r1}) , we would obtain
\begin{eqnarray}
&&
\I : \B (p):  \left(\frac{d \F_*(p)}{d p} {\fg \cdot}{\fg F}_{*}^{- 1} \right)_{dev}
=  B_{iikl}(p) \left(\frac{d {F_{*}}_{km}(p)}{d p} {F_{*}}^{- 1}_{ml} \right)_{dev}
=0   \rightarrow {\rm nonsense} ,
      \label{10-1bd-r2}
\end{eqnarray}
because it does not follow from the general constraints  (\ref{tag510-1bd}). This became more evident, e.g., for hexagonal  and tetragonal
crystals, for which $(\ln F_{*11})_{dev}=(\ln F_{*22})_{dev}=-\frac{1}{2} (\ln F_{*33})_{dev}$, and (\ref{10-1bd-r2})  simplifies to
  \begin{eqnarray}
&&
B_{iikl}(p) \left(\frac{d {F_{*}}_{km}(p)}{d p} {F_{*}}^{- 1}_{ml} \right)_{dev}
=B_{iikk}(p) \left(\frac{d \ln [{F_{*}}_{kk}(p)]}{d p} \right)_{dev}=
\nonumber\\
&&( B_{ii11}(p) +  B_{ii22}(p) - 2B_{ii33}(p))   \left(\frac{d \ln [{F_{*}}_{11}(p)]}{d p} \right)_{dev}
=
\nonumber\\
&&
2 ( B_{11}(p) +  B_{12}(p) - B_{13}(p)- B_{33}(p))   \left(\frac{d \ln [{F_{*}}_{11}(p)]}{d p} \right)_{dev}
=0
\rightarrow
\nonumber\\
&&
 B_{11}(p) +  B_{12}(p) - B_{13}(p)- B_{33}(p) =0 \rightarrow {\rm nonsense}
,
      \label{10-1bd-r5}
\end{eqnarray}
because there is no relationship between these 4 elastic moduli for tetragonal and hexagonal lattices.
Thus,  constraint (\ref{EOS-con-1}) on the elastic moduli expressed in terms of bulk modulus  $K_V= 1/9\, \fg I \fg : \B \fg :  \fg I =1/9\,  B_{iijj}$ cannot be directly derived from the trace ~\eqref{dotp} of the general  constraint(\ref{mb7dec}).  This is in contrast to the constraint (\ref{EOS-con}) on elastic compliances expressed in terms of bulk compliance $k_R= \fg I \fg :   \la \fg : \fg  I  = \lambda_{iijj}$, because it is strictly derived in Eq. (\ref{k-S-cmp})  from the trace of the general consistency condition for elastic compliances Eq. (\ref{Esall-1s-la}).

The difference between these results for bulk modulus and compressibility  are in general equations ~\eqref{dotp} and ~\eqref{mb7dec+tr} and the fact that consistency conditions are based on the relationship between hydrostatic pressure $p$ and some deformation gradient $\F_*$, which includes anisotropic strain.
For hydrostatic loading  Eq.~\eqref{mb7dec+tr} gives
\begin{eqnarray}
 \dot{\vep}_0  =
 - \fg I \fg :   \la \fg : \fg  I  \dot{p}= k_R  \dot{p} \rightarrow  k_R= - \fg I \fg :   \la \fg : \fg  I .
    \label{mb7dec+tr-1}
\end{eqnarray}
However,  for hydrostatic loading  we cannot set in Eq.~\eqref{dotp}
$\dot{\beps}_{dev}=0$ and receive $K_V={1}/{9}\,  \fg I \fg : \B \fg :  \fg I $; instead, we obtain the more general Eq.~\eqref{10-1bd-r1}.
Expression $K_V={1}/{9}\,  \fg I \fg : \B \fg :  \fg I $ is derived independently in Eq. (\ref{EOS-con-1})  under nonhydrostatic stresses but fixed  ${\beps}_{dev}$.

Thus, either of constraints on bulk moduli,  $K_V={1}/{9}\,  \fg I \fg : \B \fg :  \fg I $ or $K_R =  \frac{1}{ k_R}=  \frac{1}{ \lambda_{iijj}}$ can be applied as one of the equations for finding all elastic moduli, provided that $K_V$ and $K_R$ are correctly determined from the experiments or atomic simulations. Modulus $K_V$ is easier to calculate by prescribing  isotropic expansion to lattice, calculating energy, the stress tensor and pressure and using ${K}_V=- V \frac{\p p}{\p V}|_{{\beps}_{dev}}= V^2 \frac{\p^2{\psi_c}}{\p V^2}|_{\F=J^{1/3} \I}= -  \frac{\p p}{\p {\vep}_0 }|_{{\beps}_{dev}}$. Calculation of $K_R$ should be produced under hydrostatic loading, i.e., lattice parameters under fixed volume should be varied to relax all deviatoric stresses, which requires much more calculations.  Then ${K}_R=- V \frac{\p p}{\p V}|_{{\s}_{dev}}= V^2 \frac{\p^2{\psi_c}}{\p V^2}|_{{\s}_{dev}=0}= -  \frac{\p p}{\p {\vep}_0 }|_{{\s}_{dev}}$.  Also, $K_V$ imposes linear constraint on $\B$ components, but $K_R$  represents a nonlinear constraint. Both these results demonstrate some advantage of using $K_V$-related constraint in comparison with $K_V$-related constraint. However, relaxation of deviatoric stresses for each pressure should be performed in any case to find all other elastic moduli under pressure.


\subsection{Consistency conditions utilizing data from polycrystalline sample \label{polycrystalline}}

{\it Utilizing function $\F_* (p)$}. While for single crystal the loading is clear, i.e., pressure is applied, and deformation is measured, for polycrystalline aggregates, due to different orientation and interaction of crystals, there is a distribution of stress and strains.
For applied  pressure, $p$, in gas or liquid, each crystal of a polycrystalline aggregate possesses its own stress tensor $\s$, which causes own $\F_*$ in the crystallographic axes of each crystal.
X-ray diffraction measures averaged over the aggregate change in lattice parameters.
That is, in a thought experiment, all crystals are rotated to the same orientation of their cells, like for a single crystal,  then each component of  $\F_*$ is averaged  over the aggregate and is related to the averaged pressure in a polycrystal, which is equal to pressure $p$ in a liquid. Then dependence $\F_* (p)$ in terms of averaged $\F_*$ and $p$ can be used in equations of Section
\ref{Consistency}
the same way as for a single crystal.

This approach is simple, straightforward, and differs from the case when one uses some
effective properties of the polycrystalline aggregate instead of  $\F_* (p)$, e.g., effective bulk and shear moduli for isotropic polycrystal in terms of elastic moduli  of a single crystal.

{\it Utilizing effective elastic properties. }
There are numerous methods of determining the effective elastic properties of heterogeneous materials.
The simplest way to determine the effective elastic properties is to assume that all crystals have the same strain or distortion (Voigt approximation ~\citep{Voigt-28,Hill-52,Mura-87}), for which averaging is performed for elastic moduli $\B_V$ and the same stress (Reuss approximation ~\citep{Reuss-29,Hill-52,Mura-87}), for which averaging is performed for
 elastic compliances $\la_R$. For isotropic polycrystals, the  effective bulk modulus $K_p$ and shear modulus $\mu_p$  are bounded by
\begin{eqnarray}
&&
K_R= 1/ k_R \leq K_p \leq K_V; \qquad    \mu_R= 1/ s_R \leq \mu_p \leq \mu_V
,
      \label{bound}
\end{eqnarray}
where $s$ is shear compliance.  According to ~\citep{Hill-52}, a good approximation is
\begin{eqnarray}
&&
K_p=K_H=\frac{1}{2} (K_V +K_R); \qquad   \mu_p= \mu_H= \frac{1}{2}( \mu_V+ \mu_R).
      \label{bound-1}
\end{eqnarray}
It is clear that these equations are valid for any pressure.
For cubic crystals only $K_H=K_V =K_R$. For effective anisotropic (textured) polycrystalline aggregate, similar lower Reuss and upper Voigt bounds on elastic moduli can be formulated in an energetic sense,
 and one of the possible Hill-type averaging is
\begin{eqnarray}
&&
\B_p(p)=\frac{1}{2} (\B_V(p) +\B_R(p)); \qquad   \B_R(p)= \la^{-1}_R(p),
      \label{bound-2}
\end{eqnarray}
see e.g., ~\citep{Morris-70}.
For triclinic crystals
\begin{eqnarray}
&&   {K}_V =1/9\,  B_{iijj}=1/9\, (B_{11}+B_{22}+ B_{33}+2(B_{12}+B_{23}+B_{13}));
  \nonumber\\
 &&  k_R=  \lambda_{iijj}= \lambda_{11}+ \lambda_{22}+  \lambda_{33}+2( \lambda_{12}+ \lambda_{23}+ \lambda_{13}).
   \nonumber\\
   &&   \mu_V= 1/15\, (B_{11}+B_{22}+ B_{33}-(B_{12}+B_{23}+B_{13})+3(B_{44}+B_{55}+B_{66}));
  \nonumber\\
 &&  15/ \mu_R= 4 (\lambda_{11}+\lambda_{22}+ \lambda_{33}) - 4(\lambda_{12}+\lambda_{23}+\lambda_{13})+3(\lambda_{44}+\lambda_{55}+\lambda_{66}).
    \label{EOS-con-a}
\end{eqnarray}
While for single crystal we do not produce any averaging, Eqs.~\eqref{EOS-con-1} and  ~\eqref{EOS-con} for single crystals coincide  with Eq.~\eqref{EOS-con-a} for $K_V$ and $k_R$  for polycrystalline aggregate.
This is because it is easy to show that $K_V$ and $k_R$ in Eqs.~\eqref{EOS-con-1} and  ~\eqref{EOS-con} for arbitrary crystals are independent of crystal orientation.
That is why we added subscripts $V$ and  $R$ for single crystals to underline that  $K_V$ is based on the Voigt   approximation, and $k_R$ is based on the Reuss approximation.

For higher-symmetry crystals, Eqs.~\eqref{EOS-con-a} trivially simplifies.
For example, for tetragonal and hexagonal single crystals, we obtain from  Eq.~\eqref{EOS-con-a} for bulk modulus and bulk compliance
\begin{eqnarray}
&&   {K}_V= 1/9\, (2B_{11}+2B_{12}+4B_{13}+ B_{33});
  \nonumber\\
 &&  k_R=  2\lambda_{11}+2\lambda_{12}+4\lambda_{13}+ \lambda_{33}=\frac{B_{11}+B_{12}-4B_{13}+ 2B_{33}}{B_{33}(B_{11}+B_{12})-2B_{13}^2} \rightarrow
 \nonumber\\
 && K_R=\frac{1}{k_R}=\frac{B_{33}(B_{11}+B_{12})-2B_{13}^2}{B_{11}+B_{12}-4B_{13}+ 2B_{33}},
       \label{EOS-con-a-tet}
\end{eqnarray}
where Eq.~\eqref{C-lam} was used.
 Using expressions for $K_V$,  $K_R$, $\mu_V$, and $\mu_R$, one can find expressions for $K_H$ and $\mu_H$, and then by equaling it to experimental value $K_p$, obtain the desired two constraint equations on elastic moduli of single crystals.
Reuss, Voigt, and Hill averages for bulk or/and shear moduli were arbitrarily  and routinely  used in different papers as one or two  of the equations for determination of all elastic moduli, and of course, results depend on which approximation for $K_p$ and $\mu_p$ is chosen.
Thus, in contrast to a more detailed and precise approach based on the experimental function $\F_* (p)$, approaches utilizing effective elastic properties of  polycrystalline sample for finding  elastic moduli of single crystals always include an error related to the chosen theory for effective elastic moduli. Thus, the approach based on $\F_* (p)$ is much more preferred. Then  Eq.~\eqref{EOS-con-a}  can be used for comparison of theoretically predicted elastic moduli of polycrystal at different pressures with experiment.

On the other hand, Eq.~\eqref{EOS-con-a} can be used to find pressure-dependent elastic moduli for polycrystalline aggregates.
They also can be used to calibrate some nonlinear isotropic potentials, like Murnaghan  potential (\cite{Murnaghan-51,Lurie-90}),
which is broadly used for simulation of deformation processes under high pressure (e.g., ~\citep{feng2016large, levitas2019tensorial}).
Murnaghan potential possesses 5 elastic constants, four of which, bulk and shear moduli and their pressure derivatives, can be found
from elastic moduli of single crystals and their pressure derivative.

\subsection{Expressions for small distortions in the intermediate configuration\label{small-dist-p}}

\subsubsection{Expression for the elastic energy\label{expr-elastic}}

Limiting ourselves to quadratic terms in $ \E $ in $ \psi $, we obtain for the   energy per unit volume in the intermediate configuration $  \Omega_*  $ from Eq.~\eqref{tag504a-in}:
\begin{equation}
    \psi= \psi_0 /J_*=\psi  (0)  -p\I:\E+ \frac{1}{2}\E:\C:\E.
    \label{tag504}
\end{equation}
Utilizing
\begin{equation}
\F= \I +  \be =\I + \beps + \om; \qquad \beps :=  (\be)_s; \quad \om:=  (\be)_a
    \label{tag508a}
\end{equation}
and  Eq.~\eqref{Esall-1},
\begin{equation}
\E= \beps +\frac{1}{2} \be^T \cd \be =
 \beps+ \frac{1}{2} \left( \beps \cd \beps + 2(\beps \cd \om)_s + \om^T \cd \om \right),
    \label{tag508b}
\end{equation}
we evaluate
\begin{equation}
\I : \E= \beps : \I +\frac{1}{2} \be^T : \be =
 \eps_0+ \frac{1}{2} \left( \beps:\beps  + \om^T:\om \right); \quad \eps_0:= \beps :\I.
    \label{tag508c}
\end{equation}
For small strains and rotations with respect to the intermediate configuration $ \Omega_* $,
substituting Eq.~\eqref{tag508b} and~\eqref{tag508c} in the expression for the elastic  energy
in Eq.~\eqref{tag504}, we obtain in quadratic in distortions approximation
\begin{eqnarray}
 && \psi-\psi  (0) =-p[\beps :\I  + \frac{1}{2} \be^T : \be ] +\frac{1}{2} \be^T :\C:\be^T  = -p \beps :\I +  \frac{1}{2}  \be^T: \tilde{\C}: \be^T=
   \nonumber\\
    &&  - p  \epsilon_{kk}   + \frac{1}{2}  \tilde{C}_{ijkl} \beta_{ij} \beta_{kl};
    \nonumber\\
  & & \tilde{\C}:= \C -p   \fg I^4_t ; \qquad  \tilde{C}_{ijkl}:=  C_{ijkl}- p \delta_{jl} \delta_{ik}.
     \label{tag504a-0-q}
\end{eqnarray}
Note that if higher order terms in $\E$ will be utilized in Eq.~\eqref{tag504}, pressure $p$ will not contribute to the higher order terms in $\be$, because $\E$ does not contain higher than the second-order terms in $\be$.

After decomposition of  $\be$ into $\beps$ and $\om$ in Eq.~\eqref{tag504a-0-q}, we obtain more detailed expression

\begin{eqnarray}
 && \psi-\psi  (0) =-p[\beps :\I  + \frac{1}{2} \left( \beps : \beps  + \om^T: \om \right)] +\frac{1}{2} \beps:\C:\beps
    \nonumber\\
  & &= -p \left( \beps :\I  + \frac{1}{2} \om^T : \om \right) +\frac{1}{2} \beps:(\C-p \I^4_s):\beps,
    \label{tag504a}
\end{eqnarray}

\begin{eqnarray}
&&  \psi -\psi  (0)=-p[\epsilon_{kk}  + \frac{1}{2} \left( \epsilon_{ij}\epsilon_{ij}  +\ome_{ij}\ome_{ij} \right)] +\frac{1}{2} {C}_{ijkl} \epsilon_{ij} \epsilon_{kl}
    \nonumber\\
  & &= -p \left( \epsilon_{kk}    + \frac{1}{2} \ome_{ij}\ome_{ij} \right) +\frac{1}{2} [{C}_{ijkl}-\frac{1}{2} p  \left(\delta_{ik}  \delta_{jl} +
\delta_{il}  \delta_{jk}\right)]  \epsilon_{ij} \epsilon_{kl} .
    \label{tag504a-c}
\end{eqnarray}
A delicate moment in consistent keeping in Eq.~\eqref{tag504a} all quadratic in distortions terms is that in linear in $ \E $ term we keep all quadratic terms but in quadratic in $ \E $ term we retain linear term $ \E= \beps $ only. This modifies elastic moduli $ \C $ by $-p \I^4_s$ and also shows that the elastic energy depends on the small rotations $ \om $. Small rotations may be related to $ \beps $ for specific loadings (e.g., for simple shear), thus also changing elastic moduli.
Also, distortion $\be$ or strain $\beps$ may include higher order components, (e.g., $\beps = \frac{1}{2} \gamma (\m \n +\n\m +\frac{\gamma }{1-\gamma^2} \fg k \fg k) $ (which, as we will discuss later, represents at $\om=\fg 0$ a isochoric rotation-free shear). Here, $\fg k$ is the unit vector orthogonal to $\n$ and $\m$. Then the quadratic part of $\beps :\I$,
namely, $\frac{\gamma^2 }{1-\gamma^2}\simeq \gamma^2 $  will effectively contribute to elastic moduli. We will discuss this later in more detail. The same is generally true for the general stress tensor $\s_*$ in the intermediate configuration.

Comparing with  Eq.~\eqref{504a-Ta} for the general stress tensor $\s_*$ in the intermediate configuration,
\begin{eqnarray}
 & \psi =\psi  (0) + \s_* : \beps   + \frac{1}{2}
  \beps :  {\C}^{\eps \eps} : \beps 
  +
    \om \fg  :   {\C}^{ \om \eps} \fg :  \beps +  \frac{1}{2}   \om :   {\C}^{ \om \om}:  \om^T,
\label{504a-Ta-rep}
\end{eqnarray}
we conclude that
\begin{eqnarray}
{\C}^{\eps \eps}= \C-p \I^4_s; \qquad
{\C}^{ \om \eps} = \fg 0; \qquad
{\C}^{ \om \om}= -p \fg I^4_{as};
  \label{4a-T}
\end{eqnarray}

\begin{eqnarray}
{C}^{\eps \eps}_{ijkl} ={C}_{ijkl}-\frac{1}{2} p  \left(\delta_{ik}  \delta_{jl} +
\delta_{il}  \delta_{jk}\right); \qquad
{C}^{ \ome \eps}_{ijkl} =  0; \qquad
{C}^{ \ome \ome}_{ijkl} =  \frac{1}{2} p  \left(\delta_{ik}  \delta_{jl} -
\delta_{il}  \delta_{jk}\right).
  \label{4a-T-c}
\end{eqnarray}


\subsubsection{Expressions for stresses\label{express-stress}}

Corresponding to Eq.~\eqref{tag504} second Piola-Kirchhoff and Cauchy stresses are
\begin{equation}
    \T=-p\I+\C:\E;
    \label{tag504-1}
\end{equation}
\begin{equation}
\begin{split}
 \s=J^{-1}(-p \F\cd \F^T+ \F\cd (\C:\E)\cd \F^T).
 \label{tag507}
\end{split}
\end{equation}
Using polar decomposition of  $  \F=   \R \cd   \U$ into the orthogonal  rotation tensor $  \R$ and the symmetric right stretch tensor $ \U$,  we obtain from Eq. ~\eqref{tag507}
\bey
 \s=J^{-1}\R\cd [-p \U\cd \U+ \U\cd (\C:\E)\cd \U]\cd \R^T
 =
 J^{-1} \R\cd [-p  (2\E+ \I)  + \U \cd (\C:\E) \cd \U] \cd \R^T.
 \label{tag507-1}
\eey
The linear in distortions expression for stress $ \T $ is obtained from Eq.~\eqref{tag504-1}:
\begin{equation}
    \T= -p\I+ \C:\beps;  \qquad T_{ij}= -p \delta_{ij} +{C}_{ijkl}  \epsilon_{kl}.
    \label{tag504-1a}
\end{equation}

 As a general conclusion, it follows from Eqs.~\eqref{tag504a}
and ~\eqref{tag504-1a} that $ \T\neq  \frac{\p \psi}{\p \beps} $ and $ \C \neq  \frac{\p^2 \psi}{\p \beps \p \beps} $, despite that this is expected in the small strain theory. Since in many atomistic simulations elastic moduli at any pressure are determined as coefficients of  the quadratic form in terms of $ \beps $, this may lead to significant errors.
Of  course, if one drops the terms $ \frac{1}{2}\left( \beps : \beps  + \om^T: \om \right) $
in Eq.~\eqref{tag504a} for $ \psi $, then $ \T=  \frac{\p \psi}{\p \beps} $ and $ \C=  \frac{\p^2 \T}{\p \beps \p \beps}$ hold.
However, this would be an inconsistent step, since (a) we need to keep these terms when we will derive the elastic moduli in the current configuration and (b) in atomistic calculations its contribution  to the energy cannot be eliminated.

Let us elaborate Eq.~\eqref{tag507-1} for the  Cauchy stress $\s$ for small distortions.
In the linear approximation in distortions, we obtain
\begin{equation}
    J= 1 +(\I : \be) =1+ \eps_0; \qquad J^{-1}=1-\I:\beps =1 - \eps_0,
    \label{tag508e}
\end{equation}
since $(1+ \eps_0)(1- \eps_0)= 1- \eps_0^2 \simeq 1$.
Utilizing Eqs.~\eqref{tag508a} and ~\eqref{tag508b}  in Eq.~\eqref{tag507-1} and limiting ourselves by the linear in distortions terms in  $1/J= 1-\I:\beps$, $\E=\beps$ and $\U=\I+\beps$, $\R= \I+\om$ (because the second-order terms will be neglected in  expression for $\s$),  we derive:
\begin{equation}
    \s= (1-\I:\beps)( \I+\om) \cd [-p(\I+\beps)^{2}+(\I+\beps)\cd(\C:\beps)\cd(\I+\beps)] \cd ( \I+\om^T).
    \label{tag509}
\end{equation}
Then, neglecting quadratic and higher order terms in $\be$ in Eq.~\eqref{tag509}, we obtain:
\begin{eqnarray}
 && \s=-p(\I+ 2\beps- \I\I : \beps)+ \C:\beps =
       -p\I+\left(\C+p(\I\I-2\I_{s}^4)\right):\beps;
       \nonumber\\
&& \s=  -p\I+ \tilde{\B}:\beps ; \qquad \tilde{\B}:= \C+p(\I\I-2\I_{s}^4).
      \label{tag510}
\end{eqnarray}
\begin{eqnarray}
&	&{\sigma}_{ij}= -p(\delta_{ij}+ 2\epsilon_{ij} - \delta_{ij}\epsilon_{mm}) + {C}_{ijkl}\epsilon_{kl}
=
-p \delta_{ij}+ \left({C}_{ijkl} +p(\delta_{ij}  \delta_{kl}-   \delta_{ik}\delta_{jl}-  \delta_{il}\delta_{jk})\right)  \epsilon_{kl};
\nonumber\\
 & &{\sigma}_{ij}=-p \delta_{ij}+  \tilde{B}_{ijkl} \epsilon_{kl};
 \qquad  \tilde{B}_{ijkl}:=  {C}_{ijkl}+p (\delta_{ij}  \delta_{kl}-   \delta_{ik}\delta_{jl}-   \delta_{il}\delta_{jk} ).
   \label{tag510-c}
   \end{eqnarray}
   In the limit when the intermediate configuration coincides with the current one,
Eq.~\eqref{tag510} is equivalent to the rate equation  ~\eqref{b6-p}, $\C=\bar{\C}$, and $ \tilde{\B}=\B$, or its integrated form Eq.~\eqref{b10-int-p}. Since below we will consider small difference between the  intermediate and  current configurations, we will neglect differences between  $ \tilde{\B}$ and $\B$.
{ While sometimes the elastic moduli ${B}_{ijkl}$  are called the elastic constants ~\citep{Soderlindetal-96,Cohen-etal-1997,Steinle-Neumann-etal-PRB-99,Marcus-01,Gulseren-Cohen-02,Marcus-Qiu-09}, they are not, because they are functions of pressure or volume.
}

Equations similar to Eq.~\eqref{tag510-c}  were presented in  ~\citep{Birch-47} with elastic moduli depending on volumetric strain instead of pressure, which are, of course, connected by an equation of state.
Explicitly   Eq.~\eqref{tag510-c} appeared in ~\citep{Barron-Klein-65} and then rederived in ~\citep{wallace-67} as a particular
of  Eq.~\eqref{b10com-int-1} for the general  initial stress $\s$.
Comparing Eqs.~\eqref{tag504a-0} and ~\eqref{tag510-c} we obtain
\begin{eqnarray}
\B = \tilde{\C}+p ( \I\I-\I^{4}); \qquad
    B_{ijkl} = \tilde{C}_{ijkl} + p ( \delta_{ij}\delta_{kl}- \delta_{il}\delta_{jk}).
      \label{mB-co-t-p}
     \end{eqnarray}
The tensor $\tilde{C}_{ijkl}$  is not symmetric with respect to $i \leftrightarrow j$  and $k \leftrightarrow l$ and does not possess the Voigt  symmetry, however, it is symmetric with respect to $ (i,j)\leftrightarrow (k,l)$. At the same time,
$  B_{ijkl}$ does  possess the Voigt  symmetry ($i\leftrightarrow j$, $k\leftrightarrow l$, and $ (i,j)\leftrightarrow (k,l)$), but
is not symmetric with respect to $j \leftrightarrow l$ and $i \leftrightarrow k$.  Symmetrizing Eq. ~\eqref{mB-co-t-p} with respect to $j \leftrightarrow l$, we obtain
\begin{eqnarray}
\tilde{C}_{ijkl}^s=B_{ijkl}^s := \frac{1}{2}(B_{ijkl}+B_{ilkj}).
      \label{mB-co-t-p-s}
     \end{eqnarray}


Similar to the general case, for hydrostatically pre-stressed solid,  it follows from Eqs.~\eqref{tag504a-c}, ~\eqref{tag504-1a},
and ~\eqref{tag510} that
\begin{eqnarray}
&& \T\neq  \frac{\p \psi}{\p \beps}; \qquad    \C \neq  \frac{\p^2 \psi}{\p \beps \p \beps};
    \qquad
 \s \neq  \frac{\p \psi}{\p \beps};
    \qquad   \s \neq  \frac{1}{J}\frac{\p \psi}{\p \beps};
   \qquad
  \B \neq  \frac{\p^2 \psi}{\p \beps \p \beps}
,
    \label{noteq-p}
\end{eqnarray}
 despite that this is expected from small strain theory.

\subsubsection{Equation of motion and wave propagation  under pressure\label{wave-p-ch}}
The equations of motion ~\eqref{motion} with allowing for Eq.~\eqref{mB-co-t-p-s}
 \begin{eqnarray}
\rho \frac{\p^2  u_i}{\p^2 t}= B_{ijkl}^s \frac{\p^2  u_k}{\p r_{*j} \p r_{*l}}.
  \label{motion-p}
\end{eqnarray}
 The corresponding plane wave propagation equation is
 \begin{eqnarray}
\rho v_w^2  u_i =B_{ijkl}^s  k_j k_l u_k =L_{ik }u_k; \qquad  L_{ik }:= B_{ijkl}^s k_j k_l,
  \label{wave-p}
\end{eqnarray}
 As one of the known stability conditions, the propagation matrix $L_{ik }$ should be positive definite for all possible $k_i$ to guarantee possibility of wave propagation in any direction, which shows an addition importance of the symmetrized tensors $\tilde{\C}^s$ and tensor $\B^s$, which coincide when initial stress reduces to the hydrostatic pressure. Eq.~\eqref{wave-p} show possibility to determine
the symmetrized tensors $\tilde{\C}^s$ and $\B^s$ from the wave propagation experiment. The antisymmetric part of  $\tilde{\C}$ can be determined from the elastic energy only; the antisymmetric part of  $\B$ can be determined from the stress-strain relationship only.


\subsubsection{Gibbs energy\label{Gibbs energy}}

 First, neglecting higher than the second-degree terms in distortions $ \be$, we obtain for the Jacobian determinant
 (\cite{Lurie-90}):
\begin{equation}
    J= 1 +\I : \be + \frac{1}{2} (\I : \be)^2- \frac{1}{2} \be:\be=1+ \eps_0 + \frac{1}{2} \eps_0^2 -\frac{1}{2} \left( \beps : \beps + \om : \om \right).
    \label{tag508d}
\end{equation}
Let us calculate the Gibbs energy $G:= \psi  + p J$ per unit volume of the intermediate configuration utilizing Eq.~\eqref{tag504a} for $ \psi $:
\begin{eqnarray}
  && G-\psi  (0):= \psi -\psi  (0) + p J
  =
 -p(\cancel{\beps :\I}  + \frac{1}{2} \be^T:\be) +\frac{1}{2} \beps:\C:\beps
  +
  \nonumber\\
  &&
  p [1 +\cancel{\I : \beps} + \frac{1}{2} (\I : \beps)^2- \frac{1}{2} \be:\be]
  =
 \nonumber\\
&&  p + \frac{1}{2} \beps: \underbrace{\left(\C+p(\I\I-2\I_{s}^4)\right)}_{\B}:\beps =   p + \frac{1}{2} \beps: \B:\beps \rightarrow
    \nonumber\\
&& G= \psi + p J
  = \psi  (0) + p + \frac{1}{2} \beps: \B:\beps ,
   \label{tag514}
\end{eqnarray}
since $ \be : \be + \be^T : \be  =  2 \beps : \be= 2 \beps : ( \om+ \beps)=  2\beps : \beps $ and $ \beps : \om=0 $ as the double contraction of the symmetric and antisymmetric tensors. Thus, the terms with small rotations $ \om $ in the expression ~\eqref{tag504a} for $ \psi $ is compensated in and disappears from Eq. ~\eqref{tag514} for the Gibbs energy  $ G $.
This confirms that it could not be neglected in Eq.~\eqref{tag504a} for $\psi$.
While Gibbs energy is defined at prescribed $\theta$, in a similar way at prescribed entropy we can define the enthalpy  $U+ p J$, which will possess the same expression as $G$, since we do not distinguish here between isothermal and adiabatic processes.

{The most important conclusion from  Eqs.~\eqref{tag510} and~\eqref{tag514} is that
\begin{equation}
\s +p\I=  \frac{\p G }{\p \beps }; \qquad     \B=  \frac{\p \s}{\p \beps}=\frac{\p ^2 G }{\p \beps \p \beps}.
      \label{tag515}
\end{equation}
Thus, the effective elastic moduli in the current configuration $ \B $, defined as derivative of the  Cauchy stress with respect to  small strain $ \beps $ superposed in the intermediate configuration  at pressure $ p $, can be obtained as the second derivative of the Gibbs energy  per unit volume of the intermediate configuration with respect to small strains $ \beps $. Relationship between  superposed Cauchy stress  $ \s_R  +p\I $, the Gibbs energy $ G $, small strains $ \beps $, and the effective elastic moduli in current configuration  $ \B $ are similar to those in the traditional linear elasticity if   $ G $
is used instead of an elastic  energy. Non-traditional point in such a presentation, which was criticized  in literature~\citep{steinle2004comment}, is that usually the Gibbs energy is expressed in terms of stresses and used to find strains. Despite this, it is strictly justified here that the Gibbs rather than the elastic energy
should be used to define the superposed Cauchy stress and the effective elastic moduli $ \B $ in the current configuration.
This is generally not surprising, because in contrast to classical  thermodynamics, modern nonlinear continuum theory ~\citep{Murnaghan-51,Barron-Klein-65,wallace-67,wallace-70,Lurie-90}, considering
thermodynamic functions with respect to arbitrary pre-stressed reference state, always mix variables, e.g., use stress $\s_*$ (pressure $p$) and strain $\E$ in elastic (internal or free) energy $\psi$ in Eqs.~\eqref{tag504a-0} and~\eqref{tag504a-0-p}. Gibbs energy in nonlinear elasticity is almost never expressed in terms of stresses, since it is practically impossible to inverse stress versus strain relationship.

Next, using  Eq.~\eqref{K-B} and Eq.~\eqref{tag515}, we obtain for bulk modulus
\begin{eqnarray}
K_V=1/9\, \fg I \fg : \B \fg :  \fg I = 1/9\, \fg I \fg : \frac{\p ^2 G }{\p \beps \p \beps} \fg :  \fg I =1/9\, \frac{\p ^2 G }{\p \epsilon_{ii} \p \epsilon_{jj}} = \frac{\p ^2 G }{\p \epsilon_0^2},
    \label{K-B-1-0}
\end{eqnarray}
since
\begin{eqnarray}
\beps=\frac{1}{3}\epsilon_0\I +\beps_{dev} \rightarrow \frac{d\beps }{d \epsilon_0}=\frac{1}{3}\I  \rightarrow
\frac{\p ^2 G }{\p \epsilon_0^2}= \frac{d\beps }{d \epsilon_0} : \frac{\p ^2 G }{\p \beps \p \beps}: \frac{d\beps }{d \epsilon_0}=\frac{1}{9}  \I: \frac{\p ^2 G }{\p \beps \p \beps}: \I.
    \label{K-B-2-0}
\end{eqnarray}
Thus, the bulk modulus $K_V$ is the second derivative of the Gibbs energy with respect to small volumetric strain for pure isotropic strain increment. This contrasts with the statement in ~\citep{marcus2002importance} that the bulk modulus $K$, unlike the $\B$, is a second derivative of energy $\psi$, rather than of $G$.

 }

\subsubsection{Gibbs energy for finite strain \label{Gibbs energy-f}}

Let us derive expression for the Gibbs energy for finite strain $\E$ in quadratic and in $\E$ approximation. First,
for finite $\E$, the Jacobian determinant
 (\cite{Lurie-90}, Eq. (1.8.1)):
\begin{equation}
    J= \left(1 +2I_1 (\E) + 4 I_2(\E)+ 8 I_3(\E)\right)^{1/2},
     \label{tag508d-f}
\end{equation}
where $I_{i}(\E)$ are invariants of $\E$:
\begin{eqnarray}
  && I_1 (\E)= \I:\E= E_{ii}; \qquad
 I_2(\E) =\frac{1}{2} ((\I:\E)^2 - \E:\E)= \frac{1}{2}((E_{ii})^2- E_{ij} E_{ij}) ;
 \nonumber\\
  &&
  I_3(\E) =   \frac{1}{6} \left((\I:\E)^3 -3 (\I:\E) \E:\E +2  \E: \E \cdot\E  \right)
  =
   \nonumber\\
  &&
  \frac{1}{6} \left((E_{ii})^3-3 E_{ii} E_{ij} E_{ij}+2    E_{ij} E_{jk}E_{ki} \right).
   \label{inv}
\end{eqnarray}
For quadratic in $\E$ approximation $I_3$ can be neglected and
\begin{eqnarray}
 &&  J\simeq  1 +I_1 (\E) + 2 I_2(\E)- \frac{1}{2} (I_1 (\E) + \cancel {2 I_2(\E)})^2 \simeq
    \nonumber\\
  &&
   1+  \I:\E +\frac{1}{2} (\I:\E)^2  - \E:\E= 1 + E_{ii} +\frac{1}{2} (E_{ii})^2- E_{ij} E_{ij}.
   \label{J-2}
\end{eqnarray}
Let us calculate the Gibbs energy $G:= \psi  + p J$ per unit volume of the intermediate configuration utilizing Eq.~\eqref{tag504} for $ \psi $:
\begin{eqnarray}
  && G-\psi  (0):= \psi -\psi  (0) + p J
  =
\cancel{ -p\I:\E}+ \frac{1}{2}\E:\C:\E
  +
  \nonumber\\
  &&
  p [1 +\cancel{\I:\E} +\frac{1}{2} (\I:\E)^2  - \E:\E]
  =
 \nonumber\\
 &&  p + \frac{1}{2} \E: \underbrace{\left(\C+p(\I\I-2\I_{s}^4)\right)}_{\B}:\E =   p + \frac{1}{2} \E: \B:\E \rightarrow
    \nonumber\\
&& G= \psi + p J
  = \psi (0) + p + \frac{1}{2} \E: \B:\E ,
   \label{tag514-f}
\end{eqnarray}
i.e., $G$ has the same expression for finite $\E$ as for small strains, involving elasticity moduli $\B$.
The function $U+p(J-1)$ close to the enthalpy, similar to the  Gibbs energy, 
was defined for finite strain in ~\citep{barsch1968second}
up to  fourth-degree in $\E$ approximation, and explicit expressions for $\B$ and higher-order elastic moduli  were given for cubic crystals.
However, practical application of Eq. ~\eqref{tag514-f} for determination of elastic moduli (this was the main goal in ~\citep{barsch1968second}) even up to second-degree in $\E$ approximation is limited to the small strain.
First, since in atomistic simulations one can find $\B$ from $G$ for small $\E \simeq \beps$, then why one would do this for finite strain? Second, in  ~\citep{barsch1968second} and following papers ~\citep{Vekilovetal-15,Mosyaginetal-08,Krasilnikovetal-12}, relation between $G$ and stress $\s$ was not given. In fact,
for finite components of $\E$ not negligible in comparison with $1$,
\begin{equation}
\s +p\I \neq  \frac{\p G }{\p \E } = \B: \E,
     \label{tag515-f}
\end{equation}
because for linear $\T-\E$ relationship $\s - \E$ relationship contains non-negligible nonlinear terms.

\subsubsection{Elastic energy for isochoric deformation with respect to the intermediate configuration \label{isochoric}}
For this particular case, $J=det (\I+\be)=1$, and it follows from Eq.~\eqref{tag514}
\begin{eqnarray}
&& \psi_{j=1} - \psi  (0)  :=G_{j=1} - \psi  (0) -p =
  \frac{1}{2} \beps_{j=1}: \B :\beps_{j=1} ,
   \label{tag514-psi}
\end{eqnarray}
where subscript $j=1$ means that parameter is determined at $J=1$.

Let us consider examples of one-parametric  isochoric distortions. One of examples is  simple shear $ \F=\fg I+\gamma \m \n$. However, rotation-free shear, $ \F=\fg I+\frac{1}{2} \gamma (\m \n+\n\m)$, $\om=\fg 0$, is not isochoric because $det(\F)=1-\frac{1}{4} \gamma^2$. The straightforward way to correct this is to use  the tensor $\F_{is}= J(\F)^{-1/3} \F$: $det(\F_{is})= J(\F)^{-1} det (\F) =1$.
In particular, for the rotation-free shear,
$ \F=(1 - \frac{1}{4}\gamma^2)^{-1/3}(\fg I+\frac{1}{2} \gamma (\m \n+\n\m))$, then
\begin{eqnarray}
  &&
 \beps=\F- \fg I= \be =
 \begin{pmatrix}
  \frac{1}{(1 - \frac{1}{4}\gamma^2)^{1/3}}-1&  \frac{\frac{1}{2} \gamma }{(1 - \frac{1}{4}\gamma^2)^{1/3}}& 0 \\
   \frac{\frac{1}{2} \gamma }{(1 - \frac{1}{4}\gamma^2)^{1/3}} & \frac{1}{(1 - \frac{1}{4}\gamma^2)^{1/3}}- 1 & 0 \\
  0 & 0 &   \frac{1}{(1 - \frac{1}{4}\gamma^2)^{1/3}}-1 \\
 \end{pmatrix}
 \simeq
 \nonumber\\
&& \frac{1}{2}  \begin{pmatrix}
  \frac{\gamma^2}{6}&    \gamma  & 0 \\
  \gamma   & \frac{\gamma^2}{6} & 0 \\
  0 & 0 &  \frac{\gamma^2}{6} \\
 \end{pmatrix} + O(\gamma^4); \qquad det(\I+\beps)=1+ \frac{\gamma^6}{1728}\simeq 1 .
   \label{matr-she}
   \end{eqnarray}
Simpler option is to use expression for finite strain $\E$ for simple shear from Eq.~\eqref{tag409sg}
\begin{eqnarray}
  &&
 \beps= \be =\frac{1}{2} \gamma (\m \n+\n\m+\frac{1}{2}\gamma \n\n)=
 \frac{1}{2} \begin{pmatrix}
0&    \gamma  & 0 \\
  \gamma   &\frac{1}{2} \gamma^2 & 0 \\
  0 & 0 &  0 \\
 \end{pmatrix};
\qquad
det(\I+\beps)=1 .
   \label{matr-she1}
   \end{eqnarray}
This and other correction options for different strain states can be obtained by placing unknown $x$ in some positions and determining
it from $J=1$ or $J=1+O(x^3)$. Important point here is that some components of the strain tensor are of higher order of smallness. They should be kept when elastic energy is evaluated in atomistic simulations. However, when substituted in
Eq.~\eqref{tag514-psi}, the second order terms can be neglected because they produce higher than second-order terms in expression for energy. In particular, for  strain Eq.~\eqref{matr-she1}, $\beps= \frac{1}{2} \gamma (\m \n+\n\m)$
contributes to the  energy. Then, when neglecting second-order terms in distortions, incompressibility constraint in Eq. ~\eqref{tag508d} reduces to $\epsilon_0=0$. Decomposing small strain
\begin{eqnarray}
\beps=\frac{1}{3}\epsilon_0 \I + \beps_{dev}
\label{deceps}
\end{eqnarray}
into spherical and deviatoric parts, Eq.~\eqref{tag514-psi} specifies to
\begin{eqnarray}
&& \psi_{j=1} - \psi  (0)  
=  \frac{1}{2} \beps_{dev} : \B:\beps_{dev}.
   \label{tag514-psi-1}
\end{eqnarray}
It is clear that not all components of $\B$ contribute to the energy and can be found by calculating $ \psi_{j=1}$. It was mentioned in ~\citep{Vekilovetal-15} that shear moduli only can be determined, but as we will show explicitly, not only.    To find explicit expression for the contributing tensor, we utilize  the  volumetric
($\; {\fg V} \, = \, 1/3 \; {\fg I} \, {\fg I} \;$) and deviatoric
$\, \left( {\fg D} \right) \,$ parts of the symmetrizing fourth-order unit tensor
$\; {\I^4_s} \, = \, {\fg V} + {\fg D} \;$,  $\;\;$
$\, V_{ijkl} \, = \, 1/3 \, \delta_{ij} \, \delta_{kl} \,$, $\;\;$
$\, D_{ijkl} \, = \, 1/2 \, \left(\delta_{ik} \, \delta_{jl} +
\delta_{il} \, \delta_{jk} - 2/3 \, \delta_{ij} \, \delta_{kl} \right) \,$. The tensors
$\; {\fg V} \;$ and $\; {\fg D} \;$ possess the following properties:
\par
\noindent
$\, {\fg V} : {\fg V} \, = \, {\fg V} \, $; $\qquad$
$\, {\fg D} : {\fg D} \, = \, {\fg D} \, $; $\qquad$
$\, {\fg V} : {\fg D} \, = \, {\fg D} : {\fg V} \, = \, 0 \, $; $\qquad$
\\
$\, {\fg V} : {\fg A} \, = \, 1/3 \left( {\fg I} :
{\fg A} \right) {\fg I} \, $; $\qquad$
$\, {\fg D} : {\fg A} \, = \,  {\fg A}_{dev} \, $; $\qquad$
$\, {\fg A} \, = \, {\fg V} : {\fg A} + {\fg D} : {\fg A} \,$ ,
\par
\noindent
i.e., they allow one to separate the  volumetric and deviatoric parts of a symmetric tensor $ {\fg A} $.
Matrix presentation of tensors $\I^4$, $\I^4_s$, $\I^4_{as}$, ${\fg V}$, and ${\fg D}$ is given in Appendix B.
Then we can present  Eq.~\eqref{tag514-psi-1} as
\begin{eqnarray}
&& \psi_{j=1} - \psi  (0)  =
  \frac{1}{2} \beps : \fg D:   {\B}: \fg D :\beps= \frac{1}{2} \beps : \B_{j=1}: \beps; \qquad \B_{j=1}:= \fg D:  {\B}: \fg D
    ,
   \label{tag514-psi-2}
\end{eqnarray}
i.e., $\psi_{j=1}$ depends on the deviatoric projection $ \B_{j=1}$ of the tensor $\B$. In the explicit form,
\begin{eqnarray}
&&
\B_{j=1}=
\begin{pmatrix}
 \B_{1-3} & \B_{1-6}  \\
 \B_{1-6}^T &  \B_{4-6}  \\
   \end{pmatrix}
; \qquad
\B_{4-6}  = \begin{pmatrix}
B_{44}& B_{45} & B_{46} \\
 B_{45} &  B_{55}  & B_{56} \\
  B_{46} & B_{56} &  B_{66}\\
 \end{pmatrix},
   \label{DBD}
\end{eqnarray}
with
\bey
&&\B_{1-3}= \frac{1}{9}\begin{pmatrix}
\hat{B}_{11}& \hat{B}_{12} & \hat{B}_{13} \\
\hat{B}_{12} &  B_{22}  & \hat{B}_{23} \\
\hat{B}_{13} & \hat{B}_{23} &  \hat{B}_{33}\\
 \end{pmatrix};
\label{upleft}
 \eey
 \bey
&&
\hat{B}_{11}=4 {B_{11}}-4 {B_{12}}-4 {B_{13}}+{B_{22}}+2 {B_{23}}+{B_{33}};
\nonumber\\
&&  \hat{B}_{12}=   -2 {B_{11}}+5 {B_{12}}-{B_{13}}-2 {B_{22}}-{B_{23}}+{B_{33}}; \quad
\nonumber\\
&& \hat{B}_{13}=   -2 {B_{11}}-{B_{12}}+5 {B_{13}}+{B_{22}}-{B_{23}}-2 {B_{33}} ; \quad
\nonumber\\
&& \hat{B}_{22}=    {B_{11}}-4 {B_{12}}+2 {B_{13}}+4 {B_{22}}-4 {B_{23}}+{B_{33}}  ;
\nonumber\\
&&\hat{B}_{23}=    {B_{11}}-{B_{12}}-{B_{13}}-2 {B_{22}}+5 {B_{23}}-2 {B_{33}}  ; \quad
\nonumber\\
&&\hat{B}_{33}=    {B_{11}}+2 {B_{12}}-4 {B_{13}}+{B_{22}}-4 {B_{23}}+4 {B_{33}};
\label{upleft-c}
\eey
\begin{eqnarray}
&&\B_{1-6}
=
\begin{pmatrix}
\hat{B}_{14}& \hat{B}_{15} & \hat{B}_{16} \\
 \hat{B}_{24} &  \hat{B}_{25}  & \hat{B}_{26} \\
 \hat{B}_{34} & \hat{B}_{35} &  \hat{B}_{36}\\
 \end{pmatrix}
=
\nonumber\\
&& \frac{1}{3}\left(
\begin{array}{ccc}
 2  B_{14}- B_{24}- B_{34}  &    2  B_{15}- B_{25}- B_{35}  &   2  B_{16}- B_{26}- B_{36}  \\
  - B_{14}+2  B_{24}- B_{34}  &  - B_{15}+2  B_{25}- B_{35}  &    - B_{16}+2  B_{26}- B_{36}  \\
   - B_{14}- B_{24}+2  B_{34}  &    - B_{15}- B_{25}+2  B_{35}  &    - B_{16}- B_{26}+2  B_{36}  \\
\end{array}
\right).
   \label{B16}
\end{eqnarray}
Matrix $ \B_{j=1}$ has the block structure: the $3 \times 3$ upper left block $\B_{1-3}$ contains all $B_{ij}$ for $i, j=1,\, 2,\, 3$ (which participate in the definition of the bulk modulus $K$);
the $6 \times 6$ upper right block $\B_{1-6}$ contains all $B_{ij}$ for $i =1,\,2,\, 3$ and $ j=4,\,5,\, 6$ ;
and the $3 \times 3$ lower right  block $\B_{4-6}$ contains all $B_{ij}$ for $i, j=4,\,5,\,6$. Evidently,   $\B_{4-6}$  is not affected by the deviatoric projection and can be found by determining energy under different combinations of all shear strains supplemented by second-order normal strains making them isochoric.

This is a very important result that for isochoric deformation  and up to the second order in distortion $\be$ approximation,
the elastic energy $\psi_{j=1} - \psi  (0)$ differs from the Gibbs potential $G_{j=1}- \psi  (0)$ by $\be$-independent term $p$ only and it represents quadratic form in small strains with the effective elastic moduli $\B_{j=1}$.
Then the effective elastic moduli $\B_{j=1}$ can be determined
from the atomistic simulations, by calculating the elastic energy $\psi_{j=1} - \psi  (0)$ for different volume-preserving distortions $\be$ and then  approximating results by quadratic function Eq.~\eqref{psi-wr}.

The question arises: since $\B_{j=1}$ does not include all components of the tensor $\B$, how then to determine the remaining components with using  Eq.~\eqref{tag514} for Gibbs energy.
The answer is straightforward: they can be found from  the consistency conditions for $\B$ introduced in Section \ref{Consistency}. Let us do some accounting while using Voigt presentation of the tensors $\B$ and $\B_{j=1}$ as symmetric $6 \times 6$ matrices.  For the lowest (triclinic) symmetry, matrix $\B$ has total $36$ components, including $6$ diagonal  and $30/2=15$ independent off-diagonal components, i.e., total $21$ independent components. Since  $\B_{j=1}$  is contracted from both sides by $\beps_{dev}$ with $5$ independent components, it is equivalent to $5 \times 5$ matrix,   $25$ components, including $5$ diagonal  and $20/2=10$ independent off-diagonal components, i.e., total $15$ independent components.

 Thus, one needs $15$ calculations for $15$ independent isochoric distortions to find equations for $15$ components of $\B_{j=1}$ and consequently $\B$. Missing $6$ components of the $\B$ tensor can be found from $6$ consistency conditions (\ref{tag510-1bd}) that come from measurement of $6$ components of deformation gradient
$\F_* (p)$ in Eq.~\eqref{Esall-1s}. This is illustrated in more detail in Appendix B.

For higher symmetry, there are less missing component of the tensor $\B$ and less components of deformation gradient
$\F_* (p)$ in Eq.~\eqref{Esall-1s}. For example, for orthorhombic, tetragonal (classes $4mm$, $\bar{4}2m$, $42m$, and $4/mmm$), hexagonal, and cubic lattices $\B_{1-6}= \fg 0$, i.e., 3 missing equations disappear.
For monoclinic and orthorhombic lattices, 3 missing equations for matrix $\B_{1-3}$ can be substituted with 3 equations for diagonal deformation gradient $\F_* (p)$. For trigonal, tetragonal, and hexagonal lattices, two equations for diagonal deformation gradient $\F_* (p)$ supplement two independent equations from the system Eqs.~\eqref{Meq1} for strains different from $a \F_* (p)$ for arbitrary $a$.
For an isotropic material,
\begin{eqnarray}
\B= 3K \fg V + 2 \mu \fg D \rightarrow \B_{j=1}= 2 \mu \fg D,
 \label{Biso}
\end{eqnarray}
  where $\mu$ is the shear modulus, and we took into account  that $\fg V : \fg D=\fg 0$. For cubic and isotropic materials, one consistency condition, e.g., for bulk moduli, substitutes one missing equation.
Next, we can present
Eq.~\eqref{tag514} for the Gibbs energy as
\begin{eqnarray}
  && G-\psi  (0)- p= \frac{1}{2} \beps: \B:\beps = \frac{1}{2} \beps: \I^4_s: \B: \I^4_s : \beps
 =
  \frac{1}{2} \beps: ({\fg V} + {\fg D}) : \B: ({\fg V} + {\fg D}) : \beps
  =
 \label{tag514-dec}\\
&&    \frac{1}{2} \beps: (    {\fg V} : \B: {\fg V}  +  {\fg D} : \B: {\fg D}  + 2{\fg V} : \B: {\fg D}   ) : \beps
= \frac{1}{2} \beps: (    K \I\I  + \frac{2}{3}\I \I: \B: {\fg D}   ) : \beps  +  \psi_{j=1} - \psi  (0) ,
\nonumber
\end{eqnarray}
where we took into account $\frac{1}{9}\I \I: \B: \I \I=K \I\I$ (see definition of $K$ in  Eq.~\eqref{K-B}).

Also,  Eq.~\eqref{tag515} can be elaborated as
\begin{equation}
\s +p\I=  \frac{\p G}{\p \beps }=  \frac{\p G}{\p \epsilon_0 }\I + \frac{\p G}{\p \beps_{dev} }=  \frac{\p G}{\p \epsilon_0 }\I + \frac{\p G-\psi_{j=1}}{\p \beps_{dev} } + \frac{\p \psi_{j=1}}{\p \beps_{dev} } \neq  \frac{\p G}{\p \epsilon_0 }\I + \frac{\p \psi_{j=1}}{\p \beps_{dev} }
      \label{tag515-sig}
\end{equation}
because of cross ${\fg V} - {\fg D}$ term in Eq.~\eqref{tag514-dec}, i.e., there is no sense to use $ \psi_{j=1} $ for presentation of stresses. However, based on Eq.~\eqref{tag514-psi-2}
\begin{equation}
{\B}_{j=1} =  \frac{\p ^2 \psi_{j=1} }{\p \beps \p \beps},
      \label{tag515-el}
\end{equation}
which is part of Eq.~\eqref{tag515}.

\subsubsection{Hydrostatic loading and isotropic deformation through energy minimization\label{minimization}}

In many papers, e.g., ~\citep{Mehl-etal-1990,Cohen-etal-1997,Steinle-Neumann-etal-PRB-99} for tetragonal and hexagonal crystals, energy is calculated as function of volume for parameter $c/a$ that minimizes an energy. But they do not describe what does this minimization mean in terms of stress state. Here, we will show for an arbitrary crystal, that energy minimization at fixed volume leads either to pure hydrostatic stress state or to isotropic deformation with respect to the current state. The power balance is
\bey
\s: \bd dt = -p  d{\vep}_0 +  \s_{dev}:\bd_{dev} dt = J^{-1} d\psi,
 \label{power}
\eey
where we considered Eq.~\eqref{sph-dev}.  For fixed volume one has $d\vep_0=0$. If energy is minimized with respect to all strains (or lattice parameters)  at fixed volume, then $d\psi=0$ and Eq.~\eqref{power} can be satisfied  at $ \s_{dev}=\fg 0$ or $\bd_{dev}=\fg 0 $ only.

It is stated in ~\citep{marcus2002importance} that the equilibrium $c/a$ at prescribed $p$ does not correspond to the energy minimum. However, as it was just proven, $c/a$ corresponding to the pure hydrostatic loading does minimize energy at fixed  volume $d\vep_0=0$.  It is also stated in ~\citep{marcus2002importance} that the equilibrium $c/a$ at prescribed $p$  corresponds to the Gibbs energy minimum. This can be proven in the general case.  Substituting $\psi= G-pJ$ in   Eq.~\eqref{power}, we obtain
  \bey
 -p  J^{-1} dJ +  \s_{dev}:\bd_{dev} dt = J^{-1}( dG - p dJ -J dp)  \quad \rightarrow \quad  \s_{dev}:\bd_{dev} dt  + dp = J^{-1} dG.
 \label{powerG}
\eey
Thus, for constant $p$ and strains minimizing $G$, i.e., for $ dp = dG=0$, Eq.~\eqref{powerG} results in $ \s_{dev}=0$ or $\bd_{dev}=\fg 0 $, i.e., either in hydrostatic loading or isotropic  deformation.

\subsection{Simplification for isotropic materials and cubic crystals\label{isotropic}}

For isotropic materials and cubic crystals, the intermediate configuration under hydrostatic pressure  is produced by {\it isotropic deformation}. Then Eqs. (\ref{tag500}) and (\ref{Esall}) simplify to
\bey
& &\F_*=a \I \; \rightarrow \quad  \F_0= a \F \qquad
J_*=a^3; \qquad J_0=a^3 J; \quad \E_*=\frac{1}{2}(a^2-1)\I;
\nonumber\\
& & \E_0=a^2 \E +\frac{1}{2}(a^2-1)\I; \; \rightarrow \; \E=( \E_0-\frac{1}{2}(a^2-1)\I)/a^2.
\label{Esall-1p}
\eey
To derive relationships between $\C$ and $\C_0$ for isotropic materials and cubic crystals, we evaluate using Eq.~\eqref{Esall-1p}
\begin{eqnarray}
  \frac{\partial   \E_0}{ \partial \E}= a^2 \fg I^4_s;
    \qquad
  \frac{\partial   E_{0ij}}{ \partial E_{kl}}= a^2 I^4_{s,ijkl}=
  \frac{1}{2} a^2  \left(\delta_{ki} \delta_{lj} + \delta_{li}  \delta_{kj}   \right).
    \label{dE0dEcom-p}
\end{eqnarray}
These equations are a particular case of Eq.~\eqref{dE0dE}.
Next, using Eqs.~\eqref{tag505} and ~\eqref{Esall-1p}, we obtain for the second Piola-Kirchhoff stress in the intermediate  configuration:
\begin{eqnarray}
     &&
        \T =\frac{1}{a}\T_0=  \frac{1}{a} \frac{\p\psi_0}{\p\E_0} ;
     \qquad
      T_{ij}=\frac{1}{a}T_{0,ij}= \frac{1}{a}   \frac{\p\psi_0}{\p E_{0,ij}}.
           \label{C-C0-is}
\end{eqnarray}
Based on Eq.~\eqref{C-C0}, relationship between elastic moduli in the intermediate and stress-free configurations
reduces to
 \begin{eqnarray}
   \C (\E)=  a \C_0 (\E_0);
   \quad
    C_{ijkl}(E_{mn}) = a  C_{0,ijkl} (E_{0,mn}).
    \label{C-C0-is-w}
\end{eqnarray}
Eqs.~\eqref{tag501CE}  and ~\eqref{tag501CE-co} for $\E_0=\E_*$ simplify to
\begin{eqnarray}
 && \C_0  \left( \E_* \right)= \C_0^4  + \frac{1}{2}(a^2-1) \C_0^6 : \fg \I +  \frac{1}{8}(a^2-1)^2 ( \C_0^8 : \fg \I ): \fg \I +
 \nonumber\\
 &&
1/(3!)  \frac{1}{8}(a^2-1)^3((\C_0^{10}: \fg \I ): \fg \I ) :\fg \I+ ...\, ;
    \label{tag501CE-isp}
\end{eqnarray}
\begin{eqnarray}
&&	C_{0,ijkl}  \left( E_{*ab} \right)= C_{0,ijkl}^4 +  \frac{1}{2}(a^2-1) C_{0,ijklff}^6 + \frac{1}{8}(a^2-1)^2 C_{0,ijklffss}^8  +
	\nonumber\\
&	&1/(3!)  \frac{1}{8}(a^2-1)^3  C_{0,ijklffssmm}^{10} + ...\,  .
 \label{tag501CE-coisp}
 \end{eqnarray}
Eq.~\eqref{tag505s-p}  with the help of Eq.~\eqref{Esall-1p} reduces to the traditional equation of state
\begin{eqnarray}
 -p \I = \frac{1}{a} \frac{\p  {\psi_0}}{\p  {\E_0}}\mid_{\E_0=\frac{1}{2}(a^2-1)\I} ;
 \qquad
 -p  \delta_{ij} = \frac{1}{a}  \frac{\p  \psi_0}{\p  E_{0,kl}}\mid_{E_{0,ab}=\frac{1}{2}(a^2-1) \delta_{ab}} \, .
         \label{tag505s1-isop}
\end{eqnarray}
Consistency condition  (\ref{Esall-1s-la}) for elastic compliances reduces to
\begin{equation}
 \la  (  p) :  \I =-  \frac{1}{a}\frac{d a (p)}{d p} \I =- \frac{d \ln a (p)}{d p} \I \quad \rightarrow \quad
 \lambda_{ijkk}=- \frac{1}{a} \frac{d a (p)}{d p}  \delta_{ij} =- \frac{d \ln a (p)}{d p}  \delta_{ij}.
\label{Esall-1s-la-p}
\end{equation}
Since for cubic crystals $\lambda_{11kk}=\lambda_{22kk}=\lambda_{33kk}$ and $ \lambda_{1122}= \lambda_{1133}$, then Eq. (\ref{Esall-1s-la-p})
specifies to
\begin{equation}
 \lambda_{1111}+2 \lambda_{1122}=-    \frac{1}{a}\frac{d a (p)}{d p} =- \frac{d \ln a (p)}{d p} .
 \label{Esall-1s-la-1a-is}
\end{equation}
For cubic system, due to symmetry, bulk compliance is
\begin{equation}
 k=\lambda_{iikk}=3(\lambda_{1111}+2 \lambda_{1122}).
 \label{k-c}
\end{equation}
Then Eq. (\ref{Esall-1s-la-1a-is}) can be transformed to
\begin{equation}
k= - 3\frac{d \ln a (p)}{d p}=- \frac{d \ln a^3 (p)}{d p}=- \frac{d \ln J (p)}{d p},
 \label{ll-1s-la-1a-is}
\end{equation}
i.e., it coincides with constraint related to the   bulk compliance.
We eliminated subscript $R$ for $k$, because for cubic system Reuss and Voigt approximations for bulk modulus/compressibility coincide.

Consistency condition  (\ref{tag510-1bd}) for elastic moduli $\B$ reduces to
\begin{eqnarray}
&&-\dot{p}  \I= - \B (p): \I   \frac{1}{a}\frac{d a (p)}{d p}=  - \B (p): \I \frac{d \ln a (p)}{d p} \quad \rightarrow \quad
\nonumber\\
&&\delta_{ij}=- B_{ijkk}  \frac{1}{a}\frac{d p}{d a}  =- B_{ijkk} \frac{d p}{d \ln a} .
      \label{tag510-1bd-is}
\end{eqnarray}
Since for cubic crystals $B_{11kk}=B_{22kk}=B_{33kk}$ and $B_{1122}=B_{1133 }$, then Eq. (\ref{tag510-1bd-is})
specifies to
\begin{eqnarray}
&&
B_{1111}+ 2B_{1122} = \frac{d p}{d \ln a}
 .
      \label{1bd-is}
\end{eqnarray}
For cubic system, due to symmetry, the bulk modulus is
\begin{equation}
 K=\frac{1}{9}B_{iikk}=\frac{1}{3}(B_{1111}+ 2B_{1122}).
 \label{K-c}
\end{equation}
Then Eq. (\ref{1bd-is}) can be transformed to
\begin{eqnarray}
&&
K = \frac{1}{3}\frac{d p}{d \ln a}=\frac{d p}{d \ln a^3}=\frac{d p}{d \ln J},
       \label{K-c-C}
\end{eqnarray}
i.e., it coincides with constraint related to the  bulk modulus.

\section{THE PRINCIPLE OF SUPERPOSITION FOR DEFECTS AND INELASTICITY IN NONLINEAR ELASTICITY\label{superposition}}

One of the particular cases of application of Eq.~\eqref{b10-int-1}  (or Eq.~\eqref{b10-int-p-a}) is when stress $ \s (t)$ (pressure $p$) is homogeneous (i.e., satisfies the equilibrium equation) and there is a relatively small perturbation of strains that can be described within small strain approximation. Then all nonlinearities are included in a homogeneous solution, which automatically satisfies the equilibrium equations.
Then the boundary-value problem should be solved for small-strain linear elastic constitutive relationships between increments of stress $\Delta \s= \s (t+\Delta t)-\s (t)$ and strain $  \beps $ and rotations $ \om $.  If, as it is done in traditional linear elasticity, the effect of small rotations is neglected, then the constitutive
equations for the linearized incremental problem coincide with those for linear elasticity. In particular, one can apply
known analytical solutions for
defects (e.g., cracks, point defects, dislocations, disclinations, inclusions of different phases and twins)  and their evolution by just changing traditional elastic moduli $\C_0$ with stress-dependent elastic moduli $\B$.    Eshelby solution for ellipsoidal inclusion is another example, which is broadly used for the description of phase transformations,
twinning, cracks, and effective elastic properties of polycrystalline and  multiphase materials.

The principle of superposition can be used while combining different dislocations, inclusions, grains, etc.
Also, the small-strain incremental inelastic problem can be included in a standard way similar to the linear elastic material,
like elastoplasticity, creep, phase transformation, damage, etc.

One of the possible limitations on stress $ \s (t)$ for analytical solutions is that it should preserve the full Voigt symmetry, like in linear elasticity, because analytical solutions were not found for $B_{ijkl} \neq B_{klij}$.
In addition, the non-existence of analytical solutions for low-symmetry lattices also limits the type of applied stress $\s$.

 As an example, let consider a homogenous isotropic body under pressure $p$ as the intermediate reference configuration. Utilizing dislocation theory within linear elasticity approximation,
the normal and shear stress fields of the
superdislocation (which could also approximate a dislocation pile-up) in the polar coordinate $r,\,\phi$ are ~\citep{hirt+lothe-1992}:
\bey
&&
\sigma_{rr} = \sigma_{\phi\phi} = -p +\frac{ \mu_*  N
|{\fg b_*}|}{2\left(\pi \left(1  -  \nu(p) \right)\right)}  \frac{\sin  \phi}{r}
=-p+ \frac{ \mu(p) (3K(p)+2 \mu(p)) N
|{\fg b_*}|}{\left(\pi \left(3K(p)+4 \mu(p) \right)\right)}  \frac{\sin  \phi}{r}
;
\nonumber\\
&&
\tau  =  \frac{ \mu(p)  N
|{\fg b_*}|}{2\left(\pi \left(1  -  \nu(p) \right)\right)}  \frac{\cos  \phi}{r}
 =  \frac{ \mu(p)   (3K(p)+2 \mu(p)) N
|{\fg b_*}|}{\left(\pi \left(3K(p)+4 \mu(p) \right)\right)}  \frac{\cos  \phi}{r}
 ,
\label{nano-2}
\eey
where $\fg b_*$ is the Burgers vector in the intermediate configuration under pressure $p$, $ N $ is the number if dislocations, and $\nu$ is the Poison ratio. Generally,  $\fg b_*=\F_* \cdot \fg b$, but for isotropic material  $\fg b_*=a \fg b$.
Pressure-dependent moduli $K(p)$ and $ \mu(p)$ can be found in terms of $\B(p)$ using equations for the effective elastic properties
 ~\eqref{bound-1} and  ~\eqref{EOS-con-a} for crystals of arbitrary symmetry.
Similarly, these expressions can be generalized for single crystals, using analytical solutions for anisotropic elasticity from ~\citep{hirt+lothe-1992} and can be used to determine the interaction between dislocations, dislocation and other defects and inclusions.

  As an example, let us qualitatively analyze the effect of dislocation pileup on the drastic reduction of phase transformation pressure due to
  large plastic shear. For example, pressure reduced from 52.5 to 6.7 GPa for transformation from hexagonal nanocrystalline BN  to superhard  wurtzitic BN  ~\citep{Ji-Levitasetal-2012} and from 70 to 0.7  GPa for transformation from graphite to cubic diamond  ~\citep{Gaoetal-19}; see also review  ~\citep{Levitas-MT-19} for other examples. As the main hypothesis for  explaining  such a transformation pressure reduction, strong stress concentration at the tip of the plastic-strain-induced dislocation pileups is considered.
  Both normal and shear stresses at the tip of the pileup significantly contribute to the total stresses and reach the lattice instability
criterion under complex loading (see \citep{Levitasetal-Instab-17,Levitasetal-PRL-17,zarkevich2018lattice})
and produce local nucleation of the high-pressure phase.
A simple analytical model in  ~\citep{Levitas-PRB-04,Levitas-chapter-04} based on Eq.~\eqref{nano-2} but with pressure-independent elastic moduli was
used for analytical estimates, which in fact correspond well to more complex phase-field modeling  ~\citep{Levitas-Mahdi-Nanoscale-14,Javanbakht-Levitas-PRB-16,Javanbakht-Levitas-JMS-18}, which also did not include pressure dependence of the elastic moduli.

Let us demonstrate that allowing for pressure dependence of the elastic moduli in Eq.~\eqref{nano-2} can lead to some useful qualitative analysis. First, due to $\fg b_*=a \fg b$, stress concentration under pressure reduces by a factor of $a$. Then, phase transformation starts with crystal lattice instability at the dislocation pileup, either due to elastic instability or phonon instability. For
elastic shear instability, $\mu(p) \rightarrow 0$, and the effect of stress concentrations is getting small. In contrast, for phonon instabilities, shear modulus should not reduce and may even grow; thus, promoting  such transformations by dislocation pileups will be much more effective. In more detail, for shear instability, if applied pressure $p$ is much below the instability pressure, $\mu$ can be large enough, and the heterogeneous pressure distribution due to pileup with large $N$ can cause local instabilities and phase transformations. In this case, the principle of superposition is, of course, a big assumption, which still could work beyond some distance from the tip.  Since screw dislocation does not generate pressure, its contribution to lattice instability is much lower even for constant $\mu$. Then transformation may start quite close to pressure, for which $\mu(p)=0$, which makes the concentration of shear stresses small as well. This approach can be applied for single crystals, using equations for dislocations in anisotropic elasticity presented in ~\citep{hirt+lothe-1992}.

{ \section{SUMMARY \label{Summarizing-remarks}}}

In the paper, a general nonlinear theory for the elasticity of pre-stressed single crystals is systematically presented. Various types of elastic moduli are defined, their importance is determined, and relationships between them are presented.

{ \subsection{Pre-stressing with arbitrary stress $\s_*$ \label{Summary-arb-str}}}
Complete information about the nonlinear elasticity of a crystal under arbitrary pressure is contained in the Taylor expansion of the elastic energy Eq.~\eqref{tag504-exp} to a high enough degree with respect to the stress-free state. This expansion defines
$n^{th}$-rank  elastic constants $\C_0^n$, which possess symmetry of the undeformed crystal. These are the only existing elastic constants; all other elastic moduli are functions of initial stress or pressure,  $\E_0$,  or $\E$, and the deformation gradient $\F_*$.
They have symmetry with respect to exchanging all pairs of indices and within each pair.

The second-order elastic moduli in the reference $\C_0 \left( \E_0 \right)$ and intermediate $\C \left( \E \right)$ configurations are defined in a natural way in Eq.~\eqref{tag501} via the first or second derivatives of the second Piola-Kirchhoff stress or elastic energy, respectively, with respect to Lagrangian strains in corresponding configurations. They also can be found from the linear relationship ~\eqref{b1}  $\dot{\T}=\C:\dot{\E}$. They represent tangent moduli in stress-strain $\T_0-\E_0$ or $\T-\E$ curves at any chosen strain $\E_0$ or $\E_0$.
Moduli $\C_0 \left( \E_0 \right)$ and $\C \left( \E \right)$ do not keep symmetry of the initial nondeformed lattice and for general $\E_0$  or $\E$ have symmetry of a trigonal crystal, i.e., all 21 unequal components. They have complete Voigt symmetry and
are related to each other by  Eq.~\eqref{C-C0}.  Also, $\C_0 \left( \E_0 \right)$ is related to elastic constants $\C_0^n$ via Eq.~\eqref{tag501CE}.

When the intermediate configuration coincides with the current one, elastic moduli are designated as $\bar{\C}$, see Eq.~\eqref{barC}.
These moduli also relate the Truesdell objective rate of the Cauchy stress and the deformation rate,
$ \stackrel{\nabla}\s_{Tr}  =\bar{ \C} :\bd$.
For finite deviation of the intermediate configuration from the current one, one may need higher order elastic moduli in the intermediate configuration to evaluate energy and stresses.
That is why the choice of the intermediate configuration that coincides with the current one is the most convenient and popular.

Elastic moduli $\B$ can be introduced from the relationship ~\eqref{mb7}
between the Jaumann objective derivative  $  \stackrel{\nabla}\s_{J}=\boldsymbol{B}:\bd$ and the deformation rate. They are connected to $\bar{\C}$ by Eq.~\eqref{mB}.
The tensor $B_{ijkl}$ is symmetric with respect to permutations $i\leftrightarrow j$ and $k \leftrightarrow l$, but generally is not symmetric in $(i,j) \leftrightarrow (k,l)$.
 The finite difference version of this rate equation ~\eqref{b10-int}, which also includes small rotations,
is broadly  used in numerical algorithms for the solution of boundary-value problems in many finite-element programs.
This incremental equation was traditionally utilized for the definition of $\B$ moduli.  Because of Eq. ~\eqref{b10-int}, the moduli $\B$ are the main elastic moduli tensor used for evaluation of crystal lattice instability under homogenous strains and stresses
      ~\citep{wallace-72,Grimvall-etal-2012,de2017ideal,Pokluda-etal-2015,wang1993crystal}.

The bulk modulus $K$ and compressibility $k$ under general stress-strain states  can be defined by derivatives
$- \frac{\p p}{\p \vep_0}$ and $- \frac{\p \vep_0}{\p p}$, respectively, under various constraints on the strain or stress states, which should be very clearly stated. For one of them, $K_V=1/k_V$ (Eq.~\eqref{K-B}), we fix deviatoric strain, and for another one, $K_R=1/k_R$ (Eq.~\eqref{k-S}), we fix deviatoric stress. Generally, the bulk modulus and compressibility  can be determined for arbitrary strain or stress states (which contribute to the definition of $\B$).  For $K_V=1/k_V$ small strain increment should be pure volumetric (isotropic) only, and
for $K_R=1/k_R$ small stress increment should be pure hydrostatic pressure only. If the entire loading corresponds to isotropic straining or hydrostatic loading, then two different energy functions $\psi (J)$ can be defined, and bulk modulus for both loadings can be in addition defined in terms of the second derivative of the energy with respect to $J$ or volume $V$ (Eqs.~\eqref{K-nondev} and  ~\eqref{K-hydros}). While there is no any averaging in the definition of the bulk modulus and compressibility for a single crystal, the expression for $K_V=1/k_V$ in terms of components of $\B$ coincides with the Voigt average for polycrystal, and  expression for $k_R=1/K_R$ in terms of components $\la=\B^{-1}$ coincides with the Reuss average. Neglecting differences between two types of bulk moduli/compressibility may lead to quantitative misinterpretation of experiments and qualitative contradictions, which are illustrated in Section \ref{Consist-bulk}.

Elastic moduli $\tilde{\C}$ are introduced in Eq.~\eqref{tag504a-Ta} as coefficients of the quadratic elastic energy in terms of small distortions. The symmetrized moduli $\tilde{\C}^s$ in Eq.~\eqref{motion} also directly appear in equations of motion and wave propagation.  Because of this,
the moduli $\tilde{\C}^s$ are    used for the analysis of the crystal lattice instability with respect to the propagation of small perturbations.
The moduli $\tilde{\C}$ are expressed in terms of $\C$ in Eq.~\eqref{tag504a-Ta} and in terms of $\B$ in Eq.~\eqref{mB-t}.  The tensor $\tilde{C}_{ijkl}$ is not symmetric in exchange between $l$ and $j$ as well as $k$ and $i$ but is symmetric in exchange between pairs $ij$ and $kl$.

After decomposing $\be$ into small strains $\beps$ and rotations $\om$, elastic energy   in Eq.~\eqref{504a-TaC} contains three types of "moduli". The tensor ${\C}^{\eps \eps}$ characterizing contribution of strains  possesses the full Voigt symmetry. The tensor $C^{ \ome\eps}_{ijkl}$, characterizing coupling between strains and rotations, is symmetric in $k \leftrightarrow l$ and antisymmetric in $i \leftrightarrow j$. The tensor $ C^{ \ome\ome}_{ijkl}$, characterizing contribution of small rotations,  is antisymmetric in $i \leftrightarrow j$ and  $k \leftrightarrow l$, and invariant under exchange of pairs $(ij) \leftrightarrow (lk)$,
 $(kl) \leftrightarrow (ji)$, and  $(ij) \leftrightarrow (kl)$.
 Importantly, as it follows from  Eq. \eqref{noteq},  none of the traditional relationships  that are expected from small strain theory are  valid for pre-stressed materials.

{ \subsection{Crystals under initial hydrostatic pressure \label{Summary-pressure}}}
All definitions remain the same, and all final equations for different elastic moduli under pressure can be obtained as a particular
case of the general equations. However, some new aspects start playing a role, and new relationships can be found.  It is clear that the hydrostatic pressure does not change the symmetry of the stress-free lattice, and elastic moduli have much fewer independent components.
In particular, new consistency conditions for moduli $\B$ (Eq. \eqref{tag510-1bd}) and corresponding compliances (Eq. \eqref{Esall-1s-la}) are derived based on the existing  generalized equation of state Eq. \eqref{Esall-1s} under hydrostatic loading.
This is done for the data obtained for single crytal and  polycrystalline samples. Relationships between bulk and shear moduli of the isotropic
polycrystal and moduli $\B$ and compliances $\la$ under pressure are presented in Eq. \eqref{EOS-con-a} in the Voigt and Reuss approximations.  They can be used in the Hill approximation.

Moduli $\B$ under hydrostatic pressure possess new properties. Thus, the tensor $\B$ has full Voigt symmetry
and is related to moduli $\tilde{\C}$ by relationship $\tilde{C}_{ijkl}+\tilde{C}_{ilkj}= B_{ijkl}+B_{ilkj}$  (Eq. \eqref{mB-co-t-p-s}).
The tensor $\tilde{C}_{ijkl}$  is not symmetric with respect to $i \leftrightarrow j$  and $k \leftrightarrow l$ and does not possess the Voigt  symmetry, however, it is symmetric with respect to $ (i,j)\leftrightarrow (k,l)$. At the same time, $  B_{ijkl}$ is not symmetric with respect to $j \leftrightarrow l$ and $i \leftrightarrow k$. Because of Eq. \eqref{mB-co-t-p-s}, symmetrized $\B$ moduli also participate  in equations of motion and wave propagation \eqref{motion-p} and \eqref{wave-p}, and, consequently, in the  crystal lattice instability with respect to propagation of small perturbations.

{ For small distortions in the intermediate configuration, it is shown that the energy
is  invariant with respect to small rigid-body rotations, although it explicitly contains
small rotations. The  simple-shear example explicitly shows the effect of  small rotations (and initial stress) on the
energy and determination of the elastic moduli based on energy. If neglected, there are
errors in the determination of the elastic moduli.
}

The Gibbs energy or enthalpy is a quadratic form in strains with moduli $\B$ (Eq. \eqref{tag514}) and can be used as a potential for the Cauchy stress increment. { However, there is no any proof that the higher-order elastic
moduli  derived from the enthalpy or Gibbs energy  are of any use for stress-strain relationships. } It is also proven that the elastic energy for small isochoric strains is a quadratic form with the deviatoric projection $ \B_{j=1}$ of moduli $\B$ (Eq. \eqref{tag514-psi-2}). The rest of the components of $\B$ can be determined from the consistency conditions \eqref{tag510-1bd}. Simplifications for isotropic materials and cubic crystals are presented in Section \ref{isotropic}.

{ It is proven that   hydrostatic loading or isotropic deformation can be obtained by   energy minimization at fixed volume or by the Gibbs energy minimization at a fixed pressure.
}

The principle of superposition for defects and inelasticity in nonlinear elasticity with application to superdislocation and promotion of phase transformations by dislocation pileups are considered in Section \ref{superposition}.

{ Obtained results are compared with known ones. In some cases, this leads to the strict justification of them and expanding for the more general case, and in other cases, this shows various existing errors.
}
\\
\\
	
\numberwithin{equation}{section}
\renewcommand{\theequation}{A\arabic{equation}}
{ \section{Appendices}}

{\bf Appendix A:  Tensor notations}

Vectors and
tensors are denoted in boldface type, e.g.,  ${\fg A}= A_{ij}\e_i \e_j$, where $A_{ij}$ are components in the Cartesian system with unit basis vectors $\e_i$  and summation over the repeated indices is assumed.
{ Expressions $\e_i \e_j$ and $\e_i \e_k \e_t \e_d$ designate the direct or dyadic product of vectors, which represent second- and fourth-rank tensors, respectively.  }
Let  {$\, {\fg A} \, {\fg \cdot} \,
{\fg B} \,  = \, A_{ik} \, B_{kj} \e_i \e_j \,$  and $\, {\fg A} \, {\fg :} \,
{\fg B} \,  =\, tr ({\fg A} \, {\fg \cdot} \,
{\fg B}) \, = \, A_{ij} \, B_{ji}\,$} be  the contraction (or scalar product) of tensors
over one and two nearest indices, { where $tr$ is the trace operation (sum of the diagonal components), and $A_{ik} \, B_{kj}$ is the matrix product.} In the equations, first ${\fg \cdot}$ is performed, and then $\fg :$, e. g.,
$\fg A: \fg B \cdot \fg K= \fg A: (\fg B \cdot \fg K)$.
{ The direct (or dyadic) product of two tensors $\fg K$ and $\fg M$ is the  tensor $\fg K \fg M$ of rank equal to the sum of  the two initial ranks. In particular, for the second-rank tensors $\fg K=K_{ij} \e_i \e_j $ and $\fg M=M_{kl}\e_k \e_l$, one has $\fg K \fg M= K_{ij} M_{kl}\e_i \e_j \e_k \e_l$.
}

Superscripts $ - 1 $ and $T$ denote
inverse operation and transposition, respectively, $ := $ means equals per definition;  subscripts
$ s $ and  $as $ designate symmetrization and antisymmetrization of the second-rank tensors.

For the scalar product with the higher-order tensors,
it is convenient to introduce $\stackrel{n}*$,
{  which
designates contraction of the closest basis vector of the second-rank tensor with the $n^{th}$ closest basis vector of a  higher than second-rank tensor from the right or left.
Thus,
\begin{equation}
	(\tikzmarknode{A}{\fg A}   \stackrel{4}{*}
	\tikzmarknode{C}{\fg C}   \stackrel{2}{*}
	\tikzmarknode{E}{\fg E})  \stackrel{3}{*}
	\tikzmarknode{B}{\fg B} =
	(   A_{ij} \e_i \e_j \stackrel{4}*  C_{klmn}\e_k \e_l \e_m \e_n \stackrel{2}* E_{st} \e_s \e_t) \stackrel{3}*  B_{cd} \e_c \e_d= A_{ij} C_{kcsj} E_{st} B_{cd}  \e_i \e_k \e_t \e_d.
	\label{stackrel}
\end{equation}

\begin{tikzpicture}[remember picture,overlay]
	\foreach\i in {A,E,B}
	\pic at (\i) {vlines};
	\pic at (C) {vlines={4}}; 
	\draw (A) to[underb={0.05   }{0.15}{0.07}] (C);
	\draw (C) to[underb={0.05}{ -0.05   }{0.14}]  (E);
	\draw (C) to[underb={-0.05   }{ -0.05   }{0.21}] (B);
\end{tikzpicture}
\noindent
 Operations with the smaller $n$ are performed first, starting with $\cd $, which is equivalent to $ \stackrel{1}*$; parentheses  can be used to avoid ambiguity, if necessary.
For vivid and fast operation in index-free notations, it is convenient to designate basis vectors as  vertical bars and arcs connecting contracted basis vectors \cite{levitas-book96}.
All manipulations $\stackrel{n}*$ below can be checked using these arcs or in index form, or can be ignored because corresponding expressions in the component form are always given.

Without giving a general definition, in some particular cases, $\stackrel{n}*$ can be applied to a second-rank tensor. For example, for a symmetric tensor $\E$, it is convenient to present
\begin{equation}
	\tikzmarknode{F1}\F_*^T \cdot
	\tikzmarknode{E1}\E \cdot
	\tikzmarknode{F2}\F_*=
(	\tikzmarknode{F3}\F_*^T \stackrel{2}{*}
	\tikzmarknode{F4}\F_*^T \cdot
	\tikzmarknode{E2}\E )^T=\F_*^T \stackrel{2}{*}
	\F_*^T \cdot \E,
	\label{I4E-0}
\end{equation}

\begin{tikzpicture}[remember picture,overlay]
	\foreach\i in {F1,E1,F2,F3,F4,E2}
	\pic at (\i) {vlines};
	\draw (F1) to[underb={0.02    }{-0.02}{0.07}]  (E1);
	\draw (E1) to[underb={0.05}{ -0.05    }{0.07}]  (F2);
	\draw (F3) to[underb={0.05    }{ 0.05}{0.07}] (E2);
	\draw (F4) to[underb={0.05}{-0.05}{0.14}]  (E2);
\end{tikzpicture}

\noindent
see Eq. ~\eqref{I4E} and its application. Here, we took into account a symmetry of $\F_*^T{\cd} \E {\cd} \F_*$.
Similarly, in Eq. ~\eqref{tag505C-C0}
\begin{eqnarray}
	\tikzmarknode{C3}\C_0  \fg: \left(
	\tikzmarknode{F9}\F_*^T \stackrel{2}*
	\tikzmarknode{F10}\F_*^T {\cd}
	\tikzmarknode{I1}{\fg I^4_s} \right)
	=
	\left(
	\tikzmarknode{C4}\C_0  \cd
	\tikzmarknode{F11}\F_*^T \stackrel{2}*
	\tikzmarknode{F12}\F_*^T \right) \fg :
	\tikzmarknode{I2}{\fg I^4_s}
	= \C_0  \cd \F_*^T \stackrel{2}* \F_*^T \, ;
\label{pic1}
\end{eqnarray}

\begin{tikzpicture}[remember picture,overlay]
	\foreach\i in {F9,F10,F11,F12}
	\pic at (\i) {vlines};
	\foreach\j in {C3,C4,I1,I2}
	\pic at (\j) {vlines={4}};
	\draw (C3) to[underb={0.15    }{-0.05}{0.07}]  (F9);
	\draw (C3) to[underb={0.05}{ -0.05    }{0.21}]  (F10);
	\draw (F9) to[underb={0.05    }{-0.05}{0.14}] (I1);
	\draw (F10) to[underb={0.05}{-0.15}{0.07}]  (I1);
	\draw (C4) to[underb={0.15    }{-0.05}{0.07}]  (F11);
	\draw (C4) to[underb={0.05}{ -0.05    }{0.21}]  (F12);
	\draw (F11) to[underb={0.05    }{-0.05}{0.14}] (I2);
	\draw (F12) to[underb={0.05}{-0.15}{0.07}]  (I2);
\end{tikzpicture}
\begin{eqnarray}
	\tikzmarknode{F1}{{\F}_*}    \stackrel{2}*
	\tikzmarknode{F2}{\F}_* \cd
	\tikzmarknode{C1}\C_0  \cd
	\tikzmarknode{F3}\F_*^T \stackrel{2}*
	\tikzmarknode{F4}\F_*^T =
	\tikzmarknode{F5}\F_*  \stackrel{4}*
	\tikzmarknode{F6}\F_*  \stackrel{3}*
	\tikzmarknode{F7}{\F}_*    \stackrel{2}*
	\tikzmarknode{F8}{\F}_* \cd
	\tikzmarknode{C2}\C_0 .
	\label{tag505C-C0-0}
\end{eqnarray}

\begin{tikzpicture}[remember picture,overlay]
	\foreach\i in {F1,F2,F3,F4,F5,F6,F7,F8}
	\pic at (\i) {vlines};
	\foreach\j in {C1,C2}
	\pic at (\j) {vlines={4}};
	\draw (F1) to[underb={0.05    }{-0.05}{0.14}]  (C1);
	\draw (F2) to[underb={0.05}{ -0.15    }{0.07}]  (C1);
	\draw (C1) to[underb={0.05    }{-0.05}{0.14}] (F4);
	\draw (C1) to[underb={0.15}{-0.05}{0.07}]  (F3);
	\draw (F5) to[underb={0.05    }{0.15}{0.28}]  (C2);
	\draw (F6) to[underb={0.05}{ 0.05    }{0.21}]  (C2);
	\draw (F7) to[underb={0.05    }{-0.05}{0.14}] (C2);
	\draw (F8) to[underb={0.05}{-0.15}{0.07}]  (C2);
\end{tikzpicture}
In  Eq. ~\eqref{pic1} we used symmetry of the expression in the parentheses with respect to indices 3 and 4;   in  Eq. ~\eqref{tag505C-C0-0} we utilized that $\C_0$ is multiplied by the same tensors $\F$. }

The second-rank unit tensor is $\fg I= \delta_{ij}  \e_{i}\e_{j} $, where $\delta_{ij}$ is the Kronecker delta. Let us define the following isotropic  fourth-rank  tensors that will be used in the paper:
\begin{eqnarray}
{\I^4}=\delta_{il}\delta_{jk}\e_{i}\e_{j}\e_{k}\e_{l}= \e_{i}\e_{j}\e_{j}\e_{i}; \qquad
      \I^4_t=\delta_{ik}\delta_{jl}\e_{i}\e_{j}\e_{k}\e_{l}=\e_{i}\e_{j}\e_{i}\e_{j}.
     \label{bb1}
\end{eqnarray}
For any second-order tensor $\boldsymbol{A}$,
\begin{eqnarray}
    \I^4:\boldsymbol{A}= \boldsymbol{A}:\I^4=\boldsymbol{A}; \qquad
  \I^4_t:\boldsymbol{A}= \boldsymbol{A}: \I^4_t=\boldsymbol{A}^{T},
    \label{bb2}
\end{eqnarray}
i.e.,  $   \I^4$ is the fourth-order identity tensor and $ \I^4_t$ is the transposing tensor.
Then the fourth order symmetrizing tensor, defined as $\I^4_s = \frac{1}{2}({\I^4}+\I^4_t)$,
and the anti-symmetrizing tensor, defined as $\I^4_{as} = \frac{1}{2}({\I^4}-\I^4_t)$
 have the following properties:
\begin{eqnarray}
    \I^4_s:\boldsymbol{A}= \boldsymbol{A}:\I^4_s= \boldsymbol{A}_{s}; \qquad
    \I^4_{as}:\boldsymbol{A}= \boldsymbol{A}:\I^4_{as}= \boldsymbol{A}_{as},
    \label{bb3}
\end{eqnarray}
where $\boldsymbol{A}_s$ is the symmetric part of $\boldsymbol{A}$
and $\boldsymbol{A}_{as}$ is the antisymmetric part of $\boldsymbol{A}$.
\\
\\
{\bf Appendix B: Derivation of some equations}

{\bf 1. Derivation of Eq.~\eqref{b5}.}

Using $\T=J\F^{-1}\cd\s\cd\F^{-1T}$  and expression for $\dot{\E}$ in Eq.~\eqref{b1} gives
\begin{eqnarray}
    \dot{\overline{J\F^{-1}\cd\s\cd\F^{-1T}}} = \C : \left( \F^{T} \cd \bd \cd \F \right) = \left(\C \cd\F^{T}\stackrel{2}*\F^{T}  \right):\bd .
    \label{b2}
\end{eqnarray}
In component  notations,
\begin{eqnarray}
    \left(\C \cd\F^{T}\stackrel{2}*\F^{T}  \right):\bd=  {C }_{ijpq}{F^{T}}_{qr}{F^{T}}_{ps}d_{sr}\e_{i}\e_{j}
    =  {C }_{ijpq}{F}_{rq}{F}_{sp}d_{sr}\e_{i}\e_{j}.
\end{eqnarray}
Expanding the  LHS of Eq.~\eqref{b2} gives
\begin{eqnarray}
    &&J\F^{-1}\cd\dot{\s}\cd\F^{-1T} + J\dot{\F}:\F^{-1}\left(\F^{-1}\cd\s\cd\F^{-1T} \right)- J\F^{-1}\cd\dot{\F}\cd\left( \F^{-1}\cd\s\cd\F^{-1T} \right)
    \nonumber\\
    &&- J\left( \F^{-1}\cd\s\cd\F^{-1T} \right)\cd\dot{\F^{T}}\cd\F^{-1T} = \left( \C \cd\F^{T}\stackrel{2}*\F^{T} \right):\bd ;
    \label{b3}
\end{eqnarray}
\begin{eqnarray}
	&&JF^{-1}_{il}\dot{\sigma}_{lk}  F^{-1}_{jk} +
J\dot{F}_{ab} F^{-1}_{ba}\left(F^{-1}_{il} \sigma_{lk}  F^{-1}_{jk} \right)
-
JF^{-1}_{in} \dot{F}_{np} \left( F^{-1}_{pl}\ \sigma_{lk}\cd F^{-1}_{jk} \right)
	\nonumber\\
	&&- J\left(F^{-1}_{il} \sigma_{lk}  F^{-1}_{ak} \right) \dot{F}_{ba}  F^{-1}_{jb} = {C }_{ijpq}{F}_{rq}{F}_{sp}d_{sr}.
\end{eqnarray}
The time-derivative of $\F\cd\F^{-1}=\I$ gives $\dot{\F}\cd\F^{-1} + \F\cd\dot{\F}^{-1}=0$, which can be resolved
\begin{eqnarray}
    \dot{\F}^{-1} = -\F^{-1}\cd\dot{\F}\cd\F^{-1}, \quad \dot{\F}^{-1T} = -\F^{-1T}\cd\dot{\F}^{T}\cd\F^{-1T};
    \label{b4}
\end{eqnarray}
\begin{eqnarray}
	\dot{F}_{ij}^{-1} = -F_{ik}^{-1}\dot{F}_{kl} F^{-1}_{lj}, \quad \dot{F}^{-1}_{ji} = -F^{-1}_{ki}\dot{F}_{lk} F^{-1}_{jl}.
  \label{b4C}
  \end{eqnarray}
Eq.~\eqref{b4} was utilized in Eq.~\eqref{b3}.
Operating by $\frac{1}{J}\F$ from the left and by $\F^{T}$ from the right on each term in Eq.~\eqref{b3} and also using Eq.~\eqref{b4} gives Eq.~\eqref{b5}.
\\\\
{\bf 2. Derivation of Eqs.~\eqref{mb7} and ~\eqref{b10}.}

Using the fourth-order symmetrizing tensor $\I_s^4$, $\s\cd\bd$ and $\bd\cd\s$ can be presented as
\begin{eqnarray}
    &&\s\cd\bd=\s\cd(\I_s^{4}:\bd)=(\s\cd\I_s^{4}):\bd;
    \nonumber\\
    &&\bd\cd\s= (\bd:\I_s^{4})\cd\s=\bd:(\I_s^4\cd\s)=(\I_s^{4}\cd\s)^{T}:\bd.
    \label{b7}
\end{eqnarray}
The transposition of the  fourth-order tensor $\I_s^4\cd\s$ means permutation of pair basis vectors (1,2) and (3,4):
\begin{eqnarray}
\I_s^4\cd\s= I_{s,ijkm}^4 \sigma_{ml}\e_{i}\e_{j}\e_{k}\e_{l}= \frac{1}{2} (\sigma_{il}\delta_{jk} + \sigma_{jl} \delta_{ik}) \e_{i}\e_{j}\e_{k}\e_{l}; \qquad
\\
    (\I_s^4\cd\s)^{T}=(\I_s^4\cd\s)_{ijkl}\e_{k}\e_{l}\e_{i}\e_{j} =
    \frac{1}{2} (\sigma_{il}\delta_{jk} + \sigma_{jl} \delta_{ik}) \e_{k}\e_{l}\e_{i}\e_{j}
    =
      \frac{1}{2} (\sigma_{kj}\delta_{li} + \sigma_{jl} \delta_{ik})
 \e_{i}\e_{j}\e_{k}\e_{l}   .
\nonumber
\end{eqnarray}
Using Eq.~\eqref{b7} in Eq.~\eqref{b6} gives Eq.~\eqref{mb7}.

Similarly, using the fourth-order anti-symmetrizing tensor $\I_{as}^4$, the tensors  $\bw\cd\s$ and $\s\cd\bw^{T}$ can be transformed to the form
\begin{eqnarray}
    \bw\cd\s= (\bw:\I_{as}^{4})\cd\s= \bw:(\I_{as}^{4}\cd\s)= (\I_{as}^{4}\cd\s)^{T}:\bw ;
    \label{b8}
\end{eqnarray}
\begin{eqnarray}
    \s\cd\bw^{T}= -\s\cd\bw= -\s\cd(\I_{as}^{4}:\bw)=-(\s\cd\I_{as}^{4}):\bw .
    \label{b9}
\end{eqnarray}
Using Eqs.~\eqref{b8} and~\eqref{b9} in Eq.~\eqref{b7}, the time-derivative of the Cauchy stress $\dot{\s}$ is obtained in Eq.~\eqref{b10}.
\\
\\
{\bf 3. Derivation of Eq.~\eqref{504a-Ta}. }

 Decomposing $\be$ into $\beps$ and $\om$ in Eq.~\eqref{Esall-1}, we obtain the more detailed expression
\begin{eqnarray}
& &\s_* :  \be^T \cd \be =  \s_* : (\beps + \om^T) \cdot (\beps + \om)=   \s_* : \beps  \cdot \beps
+   \s_* : (\om^T\cdot \beps + \beps \cdot \om) + \s_* : \om^T \cdot \om
=
\nonumber\\
& &\beps \cdot \s_* : \beps
+  2\om \cdot \s_* :  \beps+\om \cdot \s_* : \om^T,
  \label{tag4a-T}
\end{eqnarray}
where we took into account the symmetry of $\s_*$ and that $\om^T\cdot \beps =(\beps \cdot \om)^T$.
Next,
\begin{eqnarray}
 &&\beps \cdot \s_* : \beps=   \beps :   \fg I^4_s  \cd \s_* :  \fg I^4_s : \beps;
 \qquad
 \om \cdot \s_* :  \beps =   \om \fg  :   {\C}^{ \om \eps} \fg :  \beps;
 \qquad
  {\C}^{ \om \eps}:=  \fg I^4_{as}  \cd \s_* :  \fg I^4_s  ; \qquad
  \nonumber\\
& &\om \cdot \s_* : \om^T = \om :   {\C}^{ \om \om}:  \om^T;
\qquad {\C}^{ \om \om}:= \fg I^4_{as}   \cd \s_* : \fg I^4_{as}.
  \label{tag04a-Ta}
\end{eqnarray}
In Eq.~\eqref{tag04a-Ta} we took into account the symmetry of $\beps$ and antisymmetry of $\om$.
Substituting Eq.~\eqref{tag04a-Ta}  in the expression for the  energy Eq.~\eqref{tag504a-T}, we obtain Eq.~\eqref{504a-Ta}.
\\
\\
{\bf 4. Derivation of Eq.~\eqref{504a-TaCsh} for simple shear. }

For $\F- \fg I= \be=\gamma \m \n $, we evaluate in addition to Eq.~\eqref{tag409sg}
\begin{eqnarray}
 &&\beps = \frac{1}{2} \gamma (\m \n +\n\m)   =
 \frac{1}{2} \gamma   \begin{pmatrix}
  0 & 1 & 0 \\
  1 & 0 & 0 \\
  0 & 0 & 0 \\
 \end{pmatrix}; \quad\om=\frac{1}{2} \gamma (\m \n -\n\m)  =
\frac{1}{2}   \gamma   \begin{pmatrix}
  0 & 1 & 0 \\
  -1 & 0 & 0 \\
  0 & 0 & 0 \\
 \end{pmatrix};
   \quad
    \nonumber\\
&&  \be^T \cdot \be =  \gamma^2  \n\n
 =
  \gamma^2   \begin{pmatrix}
  0 & 0 & 0 \\
  0 & 1 & 0 \\
  0 & 0 & 0 \\
 \end{pmatrix};
 \quad
\beps \cd \om=  \frac{1}{4}  \gamma^2 (\n\n - \m\m) =
 \frac{1}{4} \gamma^2   \begin{pmatrix}
  -1 & 0 & 0 \\
  0 & 1 & 0 \\
  0 & 0 & 0 \\
 \end{pmatrix};
\nonumber\\
&&
\beps \cd \beps= \om^T \cd \om= \frac{1}{4}  \gamma^2 (\m\m + \n\n) =
 \frac{1}{4} \gamma^2   \begin{pmatrix}
  1 & 0 & 0 \\
  0 & 1 & 0 \\
  0 & 0 & 0 \\
 \end{pmatrix}.
 \nonumber\\
    \label{tag409sg-mod}
\end{eqnarray}
Let us evaluate
\begin{eqnarray}
& &\beps \fg : \C \fg : \beps = \frac{1}{4} \gamma^2  (\m \n +\n\m) \fg : \C \fg :  (\m \n +\n\m) =
\nonumber\\
& &\frac{1}{4} \gamma^2 (C_{1212}+C_{2121}+C_{2112}+C_{1221})= \gamma^2 C_{1212};
\nonumber\\
& &  \beps \fg : \C^{\eps \eps} \fg : \beps = \gamma^2 C_{1212}^{\eps \eps};
\label{shear-1}
\end{eqnarray}
\begin{eqnarray}
& &\s_* \fg : \beps \cd \beps= \frac{1}{4} \s_*  \fg : \gamma^2 (\m\m + \n\n) = \frac{1}{4}  \gamma^2 ( \sigma_{*11}+ \sigma_{*22})
= \gamma^2 \left( C_{1212}^{\eps \eps}-C_{1212}\right) ; \qquad
\nonumber\\
& &C_{1212}^{\eps \eps}= C_{1212} + \frac{1}{4}   ( \sigma_{*11}+ \sigma_{*22});
\label{shear-2}
\end{eqnarray}
\begin{eqnarray}
& &  \om \fg  :   {\C}^{ \om \eps} \fg :  \beps = \frac{1}{4} \gamma^2  (\m \n -\n\m)  \fg  :   {\C}^{ \om \eps} \fg :  (\m \n +\n\m) =
\nonumber\\
&&
  \frac{1}{4}\gamma^2 (C_{2121}^{ \ome \eps}-C_{1221}^{ \ome \eps}  +C_{2112}^{ \ome \eps} -C_{1212}^{ \ome \eps} )=\gamma^2 C_{2121}^{ \ome \eps} ;
 \label{shear-3}
\end{eqnarray}
\begin{eqnarray}
& &\s_* \fg : \beps \cd \om= \frac{1}{4} \s_*  \fg : \gamma^2 (\n\n - \m\m) = \frac{1}{4}  \gamma^2 ( \sigma_{*22}- \sigma_{*11})
= \gamma^2 C_{2121}^{ \ome \eps} ; \qquad
\nonumber\\
& &C_{2121}^{ \ome \eps}= \frac{1}{4}   ( \sigma_{*22}- \sigma_{*11});
\label{shear-4}
\end{eqnarray}
\begin{eqnarray}
& &  \om \fg  :   {\C}^{ \om \om} \fg :  \om^T = \frac{1}{4} \gamma^2  (\m \n -\n\m)  \fg  :   {\C}^{ \om \om} \fg :  (\n \m - \m\n) =
 \nonumber\\
& &\frac{1}{4} \gamma^2 (C_{1221}^{ \om \om} -C_{2121}^{ \om \om} +C_{2112}^{ \om \om} -C_{1212}^{ \om \om} )=\gamma^2 C_{1221}^{ \om \om} ;
 \label{shear-5}
\end{eqnarray}
\begin{eqnarray}
\s_* \fg : \om^T \cd \om=
\s_* \fg : \beps \cd \beps= \frac{1}{4}  \gamma^2 ( \sigma_{*11}+ \sigma_{*22})
= \gamma^2 C_{1221}^{ \ome \ome} ; \quad C_{1221}^{\ome \ome}= C_{1221}^{\eps \eps}=\frac{1}{4}   ( \sigma_{*11}+ \sigma_{*22}).
\label{shear-6}
\end{eqnarray}
We took into account the symmetry and antisymmetry properties of all $\C$-tensors. Note that the same expressions for
$C_{1212}^{\eps \eps}$, $C_{2121}^{\ome \eps}$, and $C_{1221}^{\ome \ome}$ can be obtained directly from
Eq.~\eqref{504a-TaC}. Combining all terms, we obtain from Eq.~\eqref{504a-TaC} for the energy Eq.~\eqref{504a-TaCsh}.
\\
\\
{\bf 5. Matrix presentation of tensors $\I^4$, $\I^4_s$, $\I^4_{as}$, ${\fg V}$, and ${\fg D}$.}

To get a better feeling of the structure of all these tensors, we present first the forth-rank identity tensor in explicit form as both $9\times 9$ matrix and in Voigt designations as $6\times 6$ matrix:
\bey
\I^4=\left(
\begin{array}{ccc}
 \left(
\begin{array}{ccc}
 1 & 0 & 0 \\
 0 & 0 & 0 \\
 0 & 0 & 0 \\
\end{array}
\right) & \left(
\begin{array}{ccc}
 0 & 1 & 0 \\
 0 & 0 & 0 \\
 0 & 0 & 0 \\
\end{array}
\right) & \left(
\begin{array}{ccc}
 0 & 0 & 1 \\
 0 & 0 & 0 \\
 0 & 0 & 0 \\
\end{array}
\right) \\
 \left(
\begin{array}{ccc}
 0 & 0 & 0 \\
 1 & 0 & 0 \\
 0 & 0 & 0 \\
\end{array}
\right) & \left(
\begin{array}{ccc}
 0 & 0 & 0 \\
 0 & 1 & 0 \\
 0 & 0 & 0 \\
\end{array}
\right) & \left(
\begin{array}{ccc}
 0 & 0 & 0 \\
 0 & 0 & 1 \\
 0 & 0 & 0 \\
\end{array}
\right) \\
 \left(
\begin{array}{ccc}
 0 & 0 & 0 \\
 0 & 0 & 0 \\
 1 & 0 & 0 \\
\end{array}
\right) & \left(
\begin{array}{ccc}
 0 & 0 & 0 \\
 0 & 0 & 0 \\
 0 & 1 & 0 \\
\end{array}
\right) & \left(
\begin{array}{ccc}
 0 & 0 & 0 \\
 0 & 0 & 0 \\
 0 & 0 & 1 \\
\end{array}
\right) \\
\end{array}
\right);
\qquad
\tilde{\I}^4=\left(
\begin{array}{cccccc}
 1 & 0 & 0 & 0 & 0 & 0 \\
 0 & 1 & 0 & 0 & 0 & 0 \\
 0 & 0 & 1 & 0 & 0 & 0 \\
 0 & 0 & 0 & 1 & 0 & 0 \\
 0 & 0 & 0 & 0 & 1 & 0 \\
 0 & 0 & 0 & 0 & 0 & 1 \\
\end{array}
\right).
\label{unit4}
\eey
Advantage of the Voigt notations is evident. Similar symmetrizing  $\I^4_{s}$ and  anti-symmetrizing  $\I^4_{as} $ parts of $\I^4$ are
\bey
\I^4_{s}=\left(
\begin{array}{ccc}
 \left(
\begin{array}{ccc}
 1 & 0 & 0 \\
 0 & 0 & 0 \\
 0 & 0 & 0 \\
\end{array}
\right) & \left(
\begin{array}{ccc}
 0 & \frac{1}{2} & 0 \\
 \frac{1}{2} & 0 & 0 \\
 0 & 0 & 0 \\
\end{array}
\right) & \left(
\begin{array}{ccc}
 0 & 0 & \frac{1}{2} \\
 0 & 0 & 0 \\
 \frac{1}{2} & 0 & 0 \\
\end{array}
\right) \\
 \left(
\begin{array}{ccc}
 0 & \frac{1}{2} & 0 \\
 \frac{1}{2} & 0 & 0 \\
 0 & 0 & 0 \\
\end{array}
\right) & \left(
\begin{array}{ccc}
 0 & 0 & 0 \\
 0 & 1 & 0 \\
 0 & 0 & 0 \\
\end{array}
\right) & \left(
\begin{array}{ccc}
 0 & 0 & 0 \\
 0 & 0 & \frac{1}{2} \\
 0 & \frac{1}{2} & 0 \\
\end{array}
\right) \\
 \left(
\begin{array}{ccc}
 0 & 0 & \frac{1}{2} \\
 0 & 0 & 0 \\
 \frac{1}{2} & 0 & 0 \\
\end{array}
\right) & \left(
\begin{array}{ccc}
 0 & 0 & 0 \\
 0 & 0 & \frac{1}{2} \\
 0 & \frac{1}{2} & 0 \\
\end{array}
\right) & \left(
\begin{array}{ccc}
 0 & 0 & 0 \\
 0 & 0 & 0 \\
 0 & 0 & 1 \\
\end{array}
\right) \\
\end{array}
\right)
\label{sym4}
\eey
and
\bey
\I^4_{as}=\frac{1}{2}\left(
\begin{array}{ccc}
 \left(
\begin{array}{ccc}
 0 &\;\;\, 0 & \;\;\,0 \\
 0 & \;\;\,0 &\;\;\, 0 \\
 0 &\;\;\, 0 & \;\;\,0 \\
\end{array}
\right) & \left(
\begin{array}{ccc}
 \;\;\, 0 &1 & \;\;\,0 \\
 -1 & 0 & \;\;\,0 \\
\;\;\, 0 & 0 & \;\;\, 0 \\
\end{array}
\right) & \left(
\begin{array}{ccc}
\;\;\, 0 & \;\;\,0 & 1 \\
 \;\;\,0 &\;\;\, 0 & 0 \\
 -1 &\;\;\, 0 & 0 \\
\end{array}
\right) \\
 \left(
\begin{array}{ccc}
 0 & -1 &\;\;\, 0 \\
1 & \;\;\, 0 & \;\;\,0 \\
 0 & \;\;\,0 &\;\;\, 0 \\
\end{array}
\right) & \left(
\begin{array}{ccc}
\;\;\, 0 & 0 & \;\;\,0 \\
 \;\;\,0 & 0 &\;\;\, 0 \\
 \;\;\,0 & 0 & \;\;\,0 \\
\end{array}
\right) & \left(
\begin{array}{ccc}
 \;\;\,0 & \;\;\,0 & 0 \\
 \;\;\,0 & \;\;\,0 & 1 \\
 \;\;\,0 & - 1& 0 \\
\end{array}
\right) \\
 \left(
\begin{array}{ccc}
 0 & \;\;\,0 & -1 \\
 0 & \;\;\,0 & \;\;\,0 \\
1 & \;\;\,0 & \;\;\, 0 \\
\end{array}
\right) & \left(
\begin{array}{ccc}
\;\;\, 0 & 0 & \;\;\, 0 \\
\;\;\, 0 & 0 & -1 \\
 \;\;\, 0 & 1 & \;\;\, 0 \\
\end{array}
\right) & \left(
\begin{array}{ccc}
\;\;\, 0 & \;\;\,0 & 0 \\
\;\;\, 0 &\;\;\, 0 & 0 \\
\;\;\, 0 &\;\;\, 0 & 0 \\
\end{array}
\right) \\
\end{array}
\right).
\label{antisym4}
\eey
The symmetrizing tensor $\I^4_{s}$ has the same presentation in the Voigt notations like $\I^4$ (because the Voigt notations are applicable to the symmetric second-rank tensors  and fourth-rank tensors with $A_{ijkl}=A_{ijlk}=A_{jikl}$). The antisymmetric and antisymmetrizing tensors cannot be presented in the Voigt notations. For spherical and deviatoric parts of  $\I^4_{s}$  we have
\bey
\fg V=
\frac{1}{3} \left(
\begin{array}{cccccc}
 1 & 1 & 1 & 0 & 0 & 0 \\
 1 & 1 & 1 & 0 & 0 & 0 \\
 1 & 1 & 1 & 0 & 0 & 0 \\
 0 & 0 & 0 & 0 & 0 & 0 \\
 0 & 0 & 0 & 0 & 0 & 0 \\
 0 & 0 & 0 & 0 & 0 & 0 \\
\end{array}
\right);
\qquad
\fg D=
\left(
\begin{array}{cccccc}
\;\;\, \frac{2}{3} & -\frac{1}{3} & -\frac{1}{3} & 0 & 0 & 0 \\
 -\frac{1}{3} &\;\;\, \frac{2}{3} & -\frac{1}{3} & 0 & 0 & 0 \\
 -\frac{1}{3} & -\frac{1}{3} &\;\;\, \frac{2}{3} & 0 & 0 & 0 \\
 \;\;\,0 &\;\;\, 0 & \;\;\,0 & 1 & 0 & 0 \\
 \;\;\,0 &\;\;\, 0 &\;\;\, 0 & 0 & 1 & 0 \\
\;\;\, 0 & \;\;\,0 &\;\;\, 0 & 0 & 0 & 1 \\
\end{array}
\right).
\label{sph-dev-a}
\eey
\\
\\
{\bf  6. Equations for matrix $\B$ in terms of matrix $\B_{j=1}$.}

Let us assume that we know all components of the matrix  $\B_{j=1}$ and we need to find component of matrix $\B$. We can do this for blocks $\B_{1-3}$ and $\B_{1-6}$ separately from the linear   Eqs.~\eqref{upleft-c} and  ~\eqref{B16}:
\begin{eqnarray}
 \left(
\begin{array}{cccccc}
\;\;\, 4 & -4 & -4 & \;\;\,1 & \;\;\,2 & \;\;\,1 \\
 -2 & \;\;\,5 & -1 & -2 & -1 & \;\;\,1 \\
 -2 & -1 & \;\;\,5 & \;\;\,1 & -1 & -2 \\
\;\;\, 1 & -4 &\;\;\, 2 &\;\;\, 4 & -4 &\;\;\, 1 \\
\;\;\, 1 & -1 & -1 & -2 & \;\;\,5 & -2 \\
 \;\;\,1 &\;\;\, 2 & -4 &\;\;\, 1 & -4 &\;\;\, 4 \\
\end{array}
\right)
\left(\begin{array}{c}B_{11}\\  B_{12}\\ B_{13}\\ B_{22}\\ B_{23}\\ B_{33}\\
 \\
\end{array}
\right)
=
\left(\begin{array}{c}\hat{B}_{11}\\  \hat{B}_{12}\\ \hat{B}_{13}\\ \hat{B}_{22}\\ \hat{B}_{23}\\ \hat{B}_{33}\\
 \\
\end{array}
\right)
  \label{Meq1}
 \eey
and
\begin{eqnarray}
\frac{1}{3} \left(
\begin{array}{ccccccccc}
\;\;\, 2 &\;\;\, 0 &\;\;\, 0 & -1 & \;\;\,0 &\;\;\, 0 & -1 & \;\;\,0 &\;\;\, 0 \\
\;\;\, 0 &\;\;\, 2 & \;\;\,0 &\;\;\, 0 & -1 &\;\;\, 0 & \;\;\,0 & -1 & \;\;\,0 \\
\;\;\, 0 &\;\;\, 0 & \;\;\,2 &\;\;\, 0 &\;\;\, 0 & -1 & \;\;\,0 & \;\;\,0 & -1 \\
 -1 & \;\;\,0 & \;\;\,0 &\;\;\, 2 &\;\;\, 0 &\;\;\, 0 & -1 & \;\;\,0 &\;\;\, 0 \\
\;\;\, 0 & -1 &\;\;\, 0 &\;\;\, 0 & \;\;\,2 & \;\;\,0 & \;\;\,0 & -1 & \;\;\,0 \\
\;\;\, 0 &\;\;\, 0 & -1 & \;\;\,0 &\;\;\, 0 &\;\;\, 2 & \;\;\,0 & \;\;\,0 & -1 \\
 -1 &\;\;\, 0 & \;\;\,0 & -1 &\;\;\, 0 & \;\;\,0 & \;\;\,2 & \;\;\,0 & \;\;\,0 \\
\;\;\, 0 & -1 & \;\;\,0 &\;\;\, 0 & -1 &\;\;\, 0 &\;\;\, 0 & \;\;\,2 & \;\;\,0 \\
 \;\;\, 0 & \;\;\,0 & -1 &\;\;\, 0 & \;\;\,0 & -1 &\;\;\, 0 & \;\;\,0 &\;\;\, 2 \\
\end{array}
\right)
\left(\begin{array}{c}B_{14}\\  B_{15}\\ B_{16}\\ B_{24}\\ B_{25}\\ B_{26}\\B_{34}\\ B_{35}\\ B_{36}\\
 \\
\end{array}
\right)
=
\left(\begin{array}{c}\hat{B}_{14}\\  \hat{B}_{15}\\ \hat{B}_{16}\\ \hat{B}_{24}\\ \hat{B}_{25}\\ \hat{B}_{26}\\ \hat{B}_{34}\\ \hat{B}_{35}\\ \hat{B}_{36}\\
 \\
\end{array}
\right)
  .
  \label{Meq2}
 \eey
The rank of matrices of coefficients in  Eqs.~\eqref{Meq1} and  ~\eqref{Meq2} is 3 and 6 respectively, i.e., there are 3 missing linear independent equations in each of them, total 6, as we found above.
\\
\\
{\bf 7. Note for transition from Eq. ~\eqref{tag504a} to Eq. ~\eqref{tag514-psi}.}

While our proof of Eqs. ~\eqref{tag514} and ~\eqref{tag514-psi}  is concise and strict, it is useful to show how the very different expression ~\eqref{tag504a} for elastic energy with rotations, which does not contain $\B$, transforms to Eq.~\eqref{tag514-psi}.
It follows from Eq.~\eqref{tag508d}
\begin{eqnarray}
    J= 1 +\I : \beps + \frac{1}{2} (\I : \beps)^2- \frac{1}{2}  \left( \beps : \beps + \om : \om \right)= 1
    \rightarrow
    \nonumber\\
 \I : \beps =- \frac{1}{2} (\I : \beps)^2+ \frac{1}{2}  \left( \beps : \beps + \om : \om \right)  .
    \label{tag508d-1}
\end{eqnarray}
Substitution of the expression ~\eqref{tag508d-1} for the first-order term $\I : \beps$ (which does not contribute to moduli in Eq.~\eqref{tag504a}) in terms of the second-order terms  eliminate rotations and changes elastic moduli to $\B_{j=1}$.
\\
\\
{\bf Appendix C: Invariance under superposed rigid-body rotations in the current configuration}

Potentially concerning point in Eqs.~\eqref{tag504a-T}, ~\eqref{tag504a-Ta}, and ~\eqref{504a-Ta} is that  $ \E $  and, consequently, free energy depend on the small rotation $ \om $, which would violate the principle of material objectivity. To elaborate this, consider transformation of the relevant tensors under superposition of the rigid-body rotation ${\fg  r}^*= \Q \cd {\fg  r}$ in the current configuration, where  $\Q$ is the orthogonal tensor, i.e., $\Q \cd \Q^{T}=\Q^{T}   \cd\Q =\I$ . We obtain
\begin{eqnarray}
 && \F^*= \Q \cd \F; \quad  {\F^{T}}^*= {{\F}^T} \cd {\Q^T};
 \quad   \E^*= \frac{1}{2} (({\F^T} \cd \Q^{T}) \cd (\Q \cd  \F)-\I)= \E;  \quad \C^*=\C;
\nonumber\\
&&\be^*= \Q \cd  \F -\I; \quad  \beps^*=(  \Q \cd  \F)_s -\I; \quad \om^*=(  \Q \cd  \F)_a; 
 \label{tag408}\\
&
&\be^{T*} \cd \be^*=  \F^T \cd \F +\I - 2 ( \Q \cd \F)_s;
\quad \beps^* \cd \beps^*=  ( ( \Q \cd  \F)_s -\I)^2; \quad \om^{T*}\cd \om^*=-( \Q \cd \F)_a^2;
\nonumber
\end{eqnarray}
Any parameters defined in configurations $  \Omega_0  $ and $  \Omega_*  $ are independent of the rigid-body rotations in $  \Omega $.
Thus, as it is known, $ \E $ is independent of the rigid-body rotations in $  \Omega $, but all its parts
($ \be $, $ \beps $, and $ \om $) depend on $\Q$.
That is why in Eq.~\eqref{tag504a-T} the term $\E=\beps +\frac{1}{2} \be^T \cd \be $
is independent of arbitrary rotations, while the term $ \be:\C:\be^T =\beps:\C:\beps  $
is not.  However, when we substitute in equations $ \E $ with its linearized part, $ \beps $, we should consider small rotations and invariance of the equations with respect to small rotations only. For small rotations, $\Q\simeq \I + \ph$, where $\ph=-\ph^T$ is the small antisymmetric rotation tensor. Then
\begin{eqnarray}
  &&
 \F^* = \Q \cd \F =(\I + \ph) \cd (\I+ \beps+ \om) \simeq  \I + \beps + \om + \ph;
  \quad
    \nonumber\\
&& \be^* = \beps + \om + \ph; \quad \beps^*=\beps; \quad \om^*= \om + \ph,
    \label{tag408a}
\end{eqnarray}
where all higher order products are neglected. Thus, $ \beps $  is invariant with respect to  small rigid-body rotations as well as  the term $ \be:\C:\be^T =\beps:\C:\beps  $, and entire
Eq.~\eqref{tag504a-T}.
Since Eqs. ~\eqref{tag504a-Ta} and ~\eqref{504a-Ta} represent  strict algebraic transformation of  Eq.~\eqref{tag504a-T}, they are also invariant with respect to small rigid-body rotations in $\Omega$, despite they contain explicitly small rotations $\om$.
\\\\
{\bf Appendix D: Simple shear under hydrostatic pressure}

For the initial hydrostatic loading, the following simplifications can be made in the results of Section \ref{sec-shear}.
First, we evaluate using Eqs. (\ref{tag409sg})  that
\begin{equation}
\be^T : \be =2 \beps : \beps= 2\om^T : \om= \gamma^2 ;
\qquad    \beps :\I =0,
    \label{tag508b-p}
\end{equation}
and from Eq. (\ref{tag508c}) that
\begin{equation}
\I : \E= \beps : \I +\frac{1}{2} \be^T : \be = \frac{1}{2}  \gamma^2.
    \label{tag508c-p}
\end{equation}
Then from Eq. (\ref{tag504a}) or as a particular case of Eq. (\ref{504a-TaCsh}), the expression for the elastic energy is
\begin{eqnarray}
& &  \psi=-p\underbrace{[\cancel{\beps :\I}  + \frac{1}{2} \left( \beps : \beps  + \om^T: \om \right)]}_{\I : \E} +\frac{1}{2} \beps:\C:\beps= \frac{1}{2} \underbrace{( C_{1212} -p )}_{C_{\psi}} \gamma^2 =
    \nonumber\\
  & &
\frac{1}{2} ( C_{1212}^{\eps \eps}+C^{ \ome\ome}_{1221} ) \gamma^2 ;   \qquad  C_{1212}^{\eps \eps}=   C_{1212} -\frac{1}{2} p ; \qquad C^{ \ome\ome}_{1221}= -\frac{1}{2} p.
    \label{tag504a-p}
\end{eqnarray}
Note that from Eq.~\eqref{mB-co-sh}
\begin{eqnarray}
    B_{1212} = {C}_{1212}  - p = C_{\psi}=  C_{1212}^{\eps\eps} + C^{ \ome\ome}_{1221}\neq C_{1212}^{\eps\eps} ;
      \label{mB-co-sh-p}
     \end{eqnarray}
     \begin{eqnarray}
      \psi= \frac{1}{2}B_{1212} \gamma^2 .
      \label{psi-sh}
     \end{eqnarray}
Eq.~\eqref{psi-sh} is a particular case of  Eq.~\eqref{tag514-psi}, since $det (\I+ \gamma \m \n)=1$ and simple shear
is an  isochoric distortion.  It is clear that  the expression for the shear modulus $C_{1212}$ acquires a correction of $-p$ due to initial pressure, half of which comes from strains $\beps : \beps$ and half from rotations $\om^T: \om$.

However, for rotation-free shear, $\be=\beps = \frac{1}{2} \gamma (\m \n +\n\m)$ and $\om=\fg 0$,  the term with   $C^{ \ome\ome}_{1212}$ disappears from the expression for $\psi$ and one obtains
\begin{eqnarray}
 & &\psi =\psi  (0)    + \frac{1}{2}
\underbrace{(C_{1212}- \frac{1}{2}   p)}_{C_{1212}^{\eps\eps}}\gamma^2
 .
 \label{TaCsh-p}
  \end{eqnarray}
  Thus, $  B_{1212}  \neq  C_{1212}^{\eps\eps} $. Since $det (\I+  \frac{1}{2} \gamma (\m \n +\n\m))\neq 1$ and rotation-free shear is not an isochoric strain, this inequality demonstrates that for non-isochoric strain or distortion Eq.~\eqref{tag514-psi} is not valid.

The above example shows explicitly the effect of initial pressure and small rotations on the energy and determination of the elastic moduli based on energy. If neglected, they produce error in determination of elastic moduli.
\\\\
{\bf Appendix E: Comparison with the previous  approaches}

{
In a highly cited paper ~\citep{Mehl-etal-1990}, elastic moduli  have been determined from the quadratic approximation of the elastic energy for different simple strain states. For cubic crystals, volume-preserving strains were used, i.e., moduli $\B$ were determined based on Eq.~\eqref{tag514-psi-2}.
However, for tetragonal crystals, some of the strain states (in particular, with the only nonzero terms $\epsilon_{11}=\epsilon_{22}$ or $\epsilon_{33}$) are not isochoric, and Eq.~\eqref{tag504a-c} with pressure-corrections of elastic moduli should be used.
Thus, elastic moduli $C_{33}$ and combination $C_{11}+ C_{12}$ have been determined with neglected pressure-corrections, which affected each of these moduli and $C_{13}$.
Note that while it is mentioned in ~\citep{Mehl-etal-1990} that the bulk modulus and other moduli depend on volume,
actual calculations were performed for the volume close to the equilibrium one, i.e., at zero pressure. However, in
~\citep{Stixrude-Cohen-1995} these equations have been applied at high pressures, but  pressure correction was not mentioned.

In ~\citep{Fastetal-95,Soderlindetal-96} for hexagonal and in ~\citep{Marcus-01} for tetragonal crystals   none of  strain states were isochoric, thus, if applied under applied pressure,  all moduli would require pressure corrections according to Eq.~\eqref{tag504a-c} for $\C$ and Eq.~\eqref{b6-c-p} for $\B$.
Note that while energy was calculated in ~\citep{Soderlindetal-96} for three different volumes, it is not clear how elastic moduli were evaluated and why they are formally equivalent to $\B_{ijkl}$.

It was suggested for the hydrostatic state in the intermediate configuration in ~\citep{barsch1968second} for cubic crystals  to use
the $n^{th}$ derivatives of the function  $U+p(J-1)$, close to the enthalpy (instead of internal $U$ or Helmholtz $\psi$ energies)
with respect to the Lagrangian strain tensor to determine "effective" elastic moduli. It is claimed that these effective elastic moduli describe under pressure the same phenomena as normal elastic constants under normal pressure, especially the stress-strain curves and elastic wave propagation. From the point of view of determining elastic moduli,  the function $U+p(J-1)$ is equivalent to the Gibbs energy at the fixed temperature or enthalpy for the adiabatic process.
The same statement for the Gibbs energy is repeated in ~\citep{Vekilovetal-15}
and calculated from the Gibbs energy with DFT in  ~\citep{Vekilovetal-15,Mosyaginetal-08,Krasilnikovetal-12}  contain  pressure corrections of the type $\alpha p$ with some integer $\alpha$. 
However, there is not any proof that these higher-order elastic moduli
are of any use for stress-strain relationships or some other purposes.
For the second-order elasticity only, such   proof was given in  Section \ref{Gibbs energy}.
 Also, as it is mentioned in  ~\citep{wallace-70}, while the initial stress $\s_*$ (or $p$) contributes to the linear fourth-order  propagation matrix in the wave propagation equation, it does not  contribute to the nonlinear sixth-order  propagation matrix.
Thus, the higher-order elastic moduli from ~\citep{barsch1968second,Vekilovetal-15,Mosyaginetal-08,Krasilnikovetal-12}
 cannot be directly applied in nonlinear stress-strain $\s-\beps$ and  wave propagation equations.
Nevertheless, higher-order effective moduli can be transformed back to the traditional moduli based on $\U$, $\psi$ or just elastic energy using equations derived in  ~\citep{barsch1968second} and re-derived for neglected rotations in ~\citep{Vekilovetal-15} using a slightly different method.
In ~\citep{Vekilovetal-15} second to fourth-order elastic moduli of bcc tungsten  were calculated up to 600 GPa using DFT.
The same approach for the higher-order elasticity with application to different materials was applied as well in ~\citep{Mosyaginetal-08,Krasilnikovetal-12}; however, presentation is very confusing. It is explicitly written that the deformation gradient, Lagrangian strain, and all energies  are evaluated with respect to undeformed state rather than state under pressure  $p$. The only hint that the reference configuration is  in fact, under pressure  $p$ can be found in the expansion for the free energy that starts with the term $- p E_{ii}$.

It was suggested in  ~\citep{marcus2002importance} for body-centered tetragonal and hexagonal closed packed crystals to use the second derivatives of the Gibbs energy $G$ (instead of Helmholtz energy) with respect to the
strain tensor to determine "correct" elastic moduli. { The treatment in  ~\citep{marcus2002importance} is  not  systematic from the viewpoint  of the continuum mechanics (e.g., using energy and defining elastic moduli in terms of Eulerian strain,
which differs from all papers cited here and was not justified). Still,  these moduli coincide
with $\B$  derived in ~\citep{Barron-Klein-65}}. Note that in ~\citep{marcus2002importance}
 $B_{ijkl}$ are expressed in terms of coefficients
$C^{\eps \eps}_{ijkl} $ in  Eq.~\eqref{504a-TaC} instead of $\bar{C}_{ijkl}$ defined in Eq.~\eqref{b5com} and Eq.~\eqref{tag501C}, that is why they differ  from  expressions in
~\citep{wallace-67} and Eq.~\eqref{b6-c-p}.

However, in the later paper ~\citep{Marcus-Qiu-09}, a non-justified expression for stress $\sigma_{ij}=J^{-1}\frac{\p\psi}{\p \epsilon_{ij}}$ was used, which resulted in "new" elastic coefficients connecting  $\sigma_{ij}$ and $\epsilon_{ij}$, which was claimed as the main novelty. Based on strict derivations here and result in Eq.~\eqref{noteq-p}, this is the incorrect equation. 
The surprise   in ~\citep{Marcus-Qiu-09} why the classical works ~\citep{Barron-Klein-65,wallace-67} could not arrive at the same "new" result is easy to resolve: because in  ~\citep{Barron-Klein-65,wallace-67} strict finite strain theory is consistently linearized instead of using intuitive small-strain theory in ~\citep{Marcus-Qiu-09}.

Also,
it is stated in ~\citep{marcus2002importance} that the bulk modulus $K$, unlike the $\B$, is a second derivative of energy $\psi$, rather than of $G$. 
This is true if one considers the derivative with respect to volume,
 see Eqs.~\eqref{K-nondev} and ~\eqref{K-hydros}.  However, $p$ is not part of the second Piola-Kirchhoff stress, and $V$ is not part of Lagrangian strain, as it should be for the definition of moduli $\C$. At the same time,
 we derived in  Eq.~\eqref{K-B-1-0} that $K_V=\frac{\p ^2 G }{\p \epsilon_0^2}$, i.e., the bulk modulus at isotropic strain increment  is the second derivative of the Gibbs energy with respect to small volumetric strain.

Another statement in ~\citep{marcus2002importance} was that elastic moduli determined in ~\citep{Steinle-Neumann-etal-PRB-99} as the second derivatives of the elastic energy $\psi$ with respect to small strains at pressure $p$,  are $\C$-moduli instead of $\B$-moduli, and should be pressure-corrected.
 However, as it was mentioned in reply ~\citep{steinle2004comment}, isochoric strains were used in all simulations in ~\citep{Steinle-Neumann-etal-PRB-99}, which according to Eq.~\eqref{tag514-psi-2}
 results in dependence of energy on the deviatoric projection $\B_{j=1}:= \fg D:  {\B}: \fg D$ of  $\B$-moduli.
 In reply-to-reply by  ~\citep{Marcus-Qiu-04rep}, which is correct in practically all points, they agreed with such a justification from ~\citep{steinle2004comment}.

It was stated in ~\citep{Vekilovetal-15} that, as it follows from ~\citep{Barron-Klein-65}, that for isochoric strains up to the second-order term, the elastic moduli $\B$ can be determined as the second derivatives of the elastic energy without pressure corrections.
As we see from Eq.~\eqref{tag515-el}, this is true, but for ${\B}_{j=1}$; however, isochoric strains were never mentioned in  ~\citep{Barron-Klein-65}. In  ~\citep{Cohen-etal-1997}, pressure-dependence of the elastic moduli  for Fe, Xe, and Si was calculated based on evaluations of the elastic energy for isochoric strains up to the second-order term. There was no justification that these results in $\B$ moduli, but based on Eq.~\eqref{tag514-psi-2}, this is true. Isochoric strains were probably used because the bulk modulus and (for hexagonal lattices) ratio $c/a$ were calculated independently from the simulated equation of state and $c/a (p)$, which play a similar role as the consistency conditions Eq. (\ref{tag510-1bd}). Thus, results in  ~\citep{Cohen-etal-1997} (as well as in ~\citep{Steinle-Neumann-etal-PRB-99}) are correct. Responding to critics from ~\citep{marcus2002importance}, it was demonstrated in ~\citep{steinle2004comment} using the correct expression for energy from ~\citep{Barron-Klein-65} that for three isochoric strains, energy indeed involves moduli $\B$ without any pressure corrections. Similar examples are given in ~\citep{Grimvall-etal-2012}. However, general proof with all limitations, like we performed in Section \ref{isochoric}, was absent in the literature.
Also, small rotations were not considered in ~\citep{Cohen-etal-1997,Steinle-Neumann-etal-PRB-99,marcus2002importance,steinle2004comment,Grimvall-etal-2012}.

On the other hand,  it is  stated in ~\citep{sin2002ab}, based on the expression for energy similar to Eq. (\ref{tag504a}) but without rotations, that pressure contributes to the energy and, consequently, to elastic moduli, even for  isochoric strains, and elastic moduli for Ta in  ~\citep{Gulseren-Cohen-02} require corrections. Thus, for cubic lattices under isochoric strains, quadratic in strain $\Tilde\beps$ terms in free energy
 include pressure correction for $  C_{44}$ and $  C_{11}- C_{12}$. The confusion appeared because in ~\citep{sin2002ab} $\C$ moduli were discussed, and their pressure corrections were calculated correctly; however, in  ~\citep{Gulseren-Cohen-02} $\B$ moduli were discussed, and they also were calculated correctly.

In ~\citep{sin2002ab}, elastic moduli $ \C$ were determined from the expression for energy (\ref{tag504a}) with allowing for all required terms but rotations $  \om$. While it is not stated explicitly, but it follows from the equations in  ~\citep{sin2002ab} that mechanical instability is determined based on the function similar to the increment of the Gibbs energy, and, consequently, effective elastic moduli $\B$. 
These approach and loadings were applied in ~\citep{Hao-08} for finding pressure-dependent elastic moduli $ \C$ for $\omega-$Zr at $0 K$.
\\\\
{\bf Acknowledgements.}
Support from
NSF (CMMI-1943710 and DMR-1904830),  ONR
(N00014-19-1-2082),  and Iowa State University (Vance Coffman
Faculty Chair Professorship) is greatly appreciated.
Author thanks his PhD student Achyut Dhar for very careful reading this paper, raising many questions and concerns, and finding numerous errors/misprints, especially in equations. The author also thanks his another PhD student, Raghunandan Pratoori, for the technical help in manuscript
preparation.



\end{document}